\global\def\draftcontrol{0}

   \def\versionno{WF, CP, ms -- draft   }

\catcode`\@=11

\expandafter\ifx\csname draftcontrol\endcsname\relax\global\def\draftcontrol{0}
\fi

{\count255=\time\divide\count255 by 60
\xdef\hourmin{\number\count255}
\multiply\count255 by-60\advance\count255 by\time
\xdef\hourmin{\hourmin:\ifnum\count255<10 0\fi\the\count255}}
\def\draftdate{\number\month/\number\day/\number\year\ \ \ \hourmin }

\newcommand\makepapertitle{\par
  \begingroup
    \renewcommand\thefootnote{\@fnsymbol\c@footnote}%
    \def\@makefnmark{\rlap{\@textsuperscript{\normalfont\@thefnmark}}}%
    \long\def\@makefntext##1{\parindent 1em\noindent
            \hb@xt@1.8em{%
                \hss\@textsuperscript{\normalfont\@thefnmark}}##1}%
     \newpage
     \global\@topnum\z@   
     \@makepapertitle
     \thispagestyle{empty}\@thanks
  \endgroup
  \setcounter{footnote}{0}%
  \global\let\thanks\relax
  \global\let\makepapertitle\relax
  \global\let\@makepapertitle\relax
  \global\let\@thanks\@empty
  \global\let\@author\@empty
  \global\let\@date\@empty
  \global\let\@title\@empty
  \global\let\title\relax
  \global\let\author\relax
  \global\let\date\relax
  \global\let\and\relax
  \def\version{\let\version\@version\@gobble}
}
\def\@makepapertitle{%
  \newpage
   \ifnum\draftcontrol=1 {}
   \version\versionno
   \vskip 3em%
   \else
   \hfill\hbox to 3cm {\parbox{4cm}{\@pubnum}\hss}%
   \vskip 3em%
   \fi
   \begin{center}%
   \let \footnote \thanks
     {\LARGE {\@title}}%
     \vskip 1.5em%
     {\normalsize
       \lineskip .5em%
       \begin{tabular}[t]{c}%
         \@author
       \end{tabular}\par}%
     \vskip 1.5em%
     {\@bstract}%
     \end{center}%
     \vskip 1.5em
     \@date%
   \par
}

\gdef\@pubnum{}
\def\pubnum#1{%
  \gdef\@pubnum{#1}}

\gdef\@bstract{}
\def\Abstract#1{%
  \gdef\@bstract{%
   \parbox{\textwidth-0pc}{%
   \centerline{\bf Abstract}\penalty1000%
\kern.2cm%
\noindent
\renewcommand\baselinestretch{1.0}%
{#1}}}
}

\def\ps@paper{\let\@mkboth\@gobbletwo%
     \ifnum\draftcontrol=1
    \def\@oddfoot{\hbox to \textwidth{\tiny \versionno \hfil\tiny\draftdate}%
    \hskip -\textwidth \hbox to \textwidth{\hfil\rm\thepage\hfil}}%
     \else\def\@oddfoot{\hbox to \textwidth{\hfil\rm\thepage\hfil}}
     \fi
     \let\@evenfoot\@oddfoot
}

\def\body{\clearpage
          \pagestyle{paper}
    }

\def\@version#1{\ifnum\draftcontrol=1
\typeout{}\typeout{#1}\typeout{}
\vskip3mm\centerline{\hbox{\fbox{\normalsize{\tt DRAFT -- #1 -- }
                   {\draftdate}}}}\vskip3mm
\fi}
\let\version\@version
\long\def\eqlabel#1{\ifnum\draftcontrol=1
                    \tag@false  
                    \tag*{(\theequation) \hbox to -0.2cm{\hspace{0cm}\small{#1}\hss}}
                    \refstepcounter{equation}
                    \edef\@currentlabel{\theequation}
                    \ltx@label{#1}          
                    \else
                    \label{#1}
                    \fi
                    }
\let\st@bibitem\@bibitem
\let\st@lbibitem\@lbibitem
\ifnum\draftcontrol=1
  \def\@bibitem#1{%
    \st@bibitem{#1}\a@@label{#1}\ignorespaces}
  \def\@lbibitem[#1]#2{%
    \st@lbibitem[#1]{#2}\a@@label{#2}\ignorespaces}
  \def\a@@label#1{%
    \gdef\a@lab{\smash{\normalfont\small#1}}
    \ifvmode
      \if@inlabel
        \global\setbox\@labels\hbox{%
          \llap{\a@lab\let\a@lab\relax
                \kern\@totalleftmargin\kern\marginparsep}%
          \box\@labels}%
      \fi
    \fi}
\fi

\documentclass[10pt]{article}

\usepackage{amsmath,amssymb,array,calc,epsfig,bm}
\usepackage{cite}
\usepackage[
      colorlinks=true,
      linkcolor=red,  
      urlcolor=blue,    
      filecolor=red,     
      linkcolor=blue,
      citecolor = red,
      pdfstartview=FitV,
      pdftitle={},
        pdfauthor={Paolo Benincasa},
        pdfsubject={},
        pdfkeywords={},
        pdfpagemode=None,
        bookmarksopen=true    
      ]{hyperref}
      
\usepackage{graphicx}
\usepackage{ccaption}
\usepackage{mathpazo}
\usepackage{graphics}
\usepackage{mathtools}
\usepackage{stmaryrd}
\usepackage{pifont}
\usepackage{colortbl}
\usepackage{multirow,bigdelim}
\usepackage{arydshln}
\usepackage{tikz, pict2e}
\usepackage{geometry}
\usepackage{lmodern}
\usepackage{wrapfig}
\usepackage{float}
\usepackage{placeins}
\usetikzlibrary{calc,backgrounds}
\usetikzlibrary{arrows,shapes,positioning,shadows,trees}
\usetikzlibrary{mindmap}
\usetikzlibrary{decorations.pathreplacing}
\usetikzlibrary{decorations.pathmorphing}
\usetikzlibrary{decorations.markings}
\usetikzlibrary{matrix}

\tikzstyle arrowstyle=[scale=1]
\tikzstyle directed=[postaction={decorate,decoration={markings,
    mark=at position .5 with {\arrow[arrowstyle]{stealth}}}}]
\tikzstyle reverse directed=[postaction={decorate,decoration={markings,
    mark=at position .5 with {\arrowreversed[arrowstyle]{stealth};}}}]

\ifnum\draftcontrol=1
\fi

\renewcommand\baselinestretch{1.25}
\renewcommand\section{\@startsection {section}{1}{\z@}%
                                   {-3.5ex \@plus -1ex \@minus -.2ex}%
                                   {2.3ex \@plus.2ex}%
                                   {\normalfont\large\bfseries}}
\renewcommand\subsection{\@startsection{subsection}{2}{\z@}%
                                   {-3.25ex\@plus -1ex \@minus -.2ex}%
                                   {1.5ex \@plus .2ex}%
                                   {\normalfont\normalsize\bfseries}}
\renewcommand\subsubsection{\@startsection{subsubsection}{3}{\z@}%
                                   {-3.25ex\@plus -1ex \@minus -.2ex}%
                                   {1.5ex \@plus .2ex}%
                                   {\normalfont\normalsize\it}}
\renewcommand\paragraph{\@startsection{paragraph}{4}{\z@}%
                                   {-3.25ex\@plus -1ex \@minus -.2ex}%
                                   {1.5ex \@plus .2ex}%
                                   {\normalfont\normalsize\bf}}

\numberwithin{equation}{section}

\def\revise#1       {\raisebox{-0em}{\rule{3pt}{1em}}%
                     \marginpar{\raisebox{.5em}{\vrule width3pt\
                     \vrule width0pt height 0pt depth0.5em
                     \hbox to 0cm{\hspace{0cm}{%
                     \parbox[t]{4em}{\raggedright\footnotesize{#1}}}\hss}}}}

\def\sqr#1#2{{\vcenter{\vbox{\hrule height.#2pt
 \hbox{\vrule width.#2pt height#1pt \kern#1pt
 \vrule width.#2pt}\hrule height.#2pt}}}}

\def\r{\rho}

\def\aa1{\phi}
\def\cc1{\psi}

\catcode`\@=12

\begin{document}


\title{{\bf Cosmological Polytopes and the} \\ \vspace{.2cm} {\bf Wavefuncton of the Universe for Light States}}

\pubnum{%
arXiv:xxxx.xxxxx}
\date{September 2019}

\author{
\scshape Paolo Benincasa${}^{\dagger}$ \\[0.4cm]
\ttfamily ${}^{\dagger}$ Niels Bohr International Academy and Discovery Center, \\
\ttfamily University of Copenhagen, The Niels Bohr Institute,\\
\ttfamily Blegdamsvej 17, DK-2100, Copenhagen, Denmark\\
\small \ttfamily pablowellinhouse@anche.no \\[0.2cm]
}

\Abstract{We extend the investigation of the structure of the late-time wavefunction of the universe to a class of toy models of scalars with time-dependent masses and polynomial couplings, which contains general massive scalars in FRW cosmologies. We associate a universal integrand to each Feynman diagram contributing to the wavefunction of the universe. For certain (light) masses, such an integrand satisfies recursion relations involving certain differential operators, connecting states with different masses and having, as a seed, the massless scalar (which describes a conformally coupled scalar in FRW cosmologies as a special case). We show that it is a {\it degenerate limit} of the canonical form of a generalisation of the cosmological polytopes describing the subclass of these models with massless scalars. Intriguingly, the flat-space scattering amplitude appears as a higher codimension face of this generalisation of the cosmological polytope: this is the reflection of the fact that it is contained in the leading term in the Laurent expansion as the total energy is taken to zero, with the codimension of the face providing the order of the total energy pole. The same connection between the other faces  and the Laurent expansion coefficients holds for the other singularities of the wavefunction of the universe, all of them connectable to flat-space processes. This new construction makes manifest the origin of the multiple poles in the universal integrand of the wavefunction, which is exactly obtained in a degenerate limit, where some of the singularities of the canonical form of the polytope collapse onto each other. Finally, we consider the mass as a perturbative coupling as well, showing that the contribution to the wavefunction coming from graphs with mass two-point couplings can be identified with a degenerate limit of the canonical form of the cosmological polytope, if the perturbative expansion is done around the massless (conformally coupled) state; or as double degenerate limit of the canonical form of the extension of the cosmological polytopes introduced in the present paper, if the perturbative expansion is done around minimally coupled states.}

\makepapertitle

\body

\version\versionno

\tableofcontents

\section{Introduction}\label{sec:Intro}

Physics at accessible high energies is extremely constrained by the unitarity of time evolution as well as Lorentz invariance and the locality of the interactions: these basic principles fix all the possible three-particle couplings \cite{Benincasa:2007xk, McGady:2013sga, Arkani-Hamed:2017jhn}, Yang's theorem \cite{Arkani-Hamed:2017jhn}, the consistency of the interactions among massless particles with spin less or equal to two \cite{Benincasa:2007xk, Benincasa:2011pg, McGady:2013sga, Arkani-Hamed:2017jhn}, the impossibility of having interaction involving a finite number of massless particles with spin higher than 2 \cite{Weinberg:1964ew, Benincasa:2007xk, Benincasa:2011pg, McGady:2013sga}, as well as the charge conservation for interactions mediated by massless spin-$1$ particles, the equivalence principle \cite{Weinberg:1964ew} and the uniqueness of the graviton \cite{Benincasa:2007xk}.

The imprint of locality and unitarity in the relevant quantum mechanical observables, {\it i.e.} the scattering amplitudes, is given by their sufficiently analytic structure with at most poles and branch cuts, with locality fixing the location of such singularities at those points of kinematic space where the square of the sum of two or more momenta vanishes, while unitarity reflects into the fact that when such singularities are approached, the scattering amplitudes factorise into lower point ones.

However, while Lorentz invariance is broken at cosmological scales, the phase of accelerated expansion the universe is undergoing \cite{Riess:1998cb, Perlmutter:1998np}, makes impossible even in principle to have a well-defined quantum mechanical observable. However, for cosmologies in which the universe opens up to become infinitely large and flat at sufficiently late times -- which indeed is not ours, due precisely to the current accelerated expansion --, it is possible to define spatial correlation functions, or, equivalently, the wavefunction of the universe whose squared modulus provides the probability distribution through which the spatial correlations can be computed. They are static quantities which depends only on data living at the future spatial boundary of the universe. Having now both Lorentz invariance and unitarity as approximated concepts (the former is broken, while the latter is {\it hidden} because the time evolution has been integrated out), the features listed above are not bounded to hold. And, indeed, important differences appear, {\it e.g.} in cosmological settings we no longer have cluster decompositions globally, but it can hold only in each branch in which the wavefunction of the universe separates via a branched diffusion process as the universe expands\cite{Starobinsky:1982ee}. The lack of global cluster decomposition manifests itself even in the structure of the two-point function for massless scalars, which grows logarithmically at large distances, as well as in the ultrametric structure of the wavefunction of the universe \cite{Anninos:2011kh}. All these features are tied to the tree-like structure of the cosmological bulk \cite{Winitzki:2001np, Harlow:2011az}.

Thus, it is fair and necessary to ask whether there exists a cosmological counterpart of the list of constraints which hold for flat-space scattering, and which are the fundamental principles behind it. Said differently, we need to understand what are the invariant properties that the wavefunction of the universe ought to satisfy in order to come from a consistent causal evolution in cosmological space-times. Despite the existence of a number of consistency conditions for inflationary correlation functions \cite{Maldacena:2002vr, Seery:2008ax, Leblond:2010yq, Creminelli:2011rh, Creminelli:2012ed, Senatore:2012wy, Assassi:2012zq, Goldberger:2013rsa, Hinterbichler:2013dpa, Pimentel:2013gza, Creminelli:2013mca, Bordin:2016ruc, Bordin:2017ozj, Finelli:2017fml, Pajer:2019jhb}, yet no general rules are known for cosmological observables, and very little is known about the structure of the wavefunction of the universe \cite{Anninos:2014lwa, Konstantinidis:2016nio}.

In order to address this class of questions, we need to collect more theoretical data: this is an important zero-th order step for gaining a deeper understanding of the general analytic structure of cosmological observables and how physics is encoded into it. One distinctive feature that we have already learnt for observables with Bunch-Davies condition in the infinite past, is that the lack of time translation invariance shows up as a dependence on the sum of the energies, {\it i.e.} the length of the momenta, of all the states in the correlation function/wavefunction of the universe. 

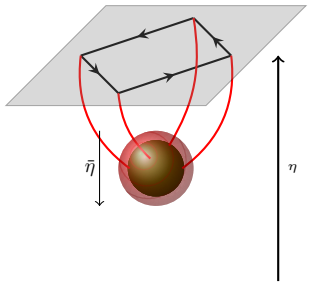
\begin{wrapfigure}{l}{4.5cm}
 \begin{tikzpicture}[node distance=2cm, cross/.style={cross out, draw, inner sep=0pt, outer sep=0pt}]
  \begin{scope}[parallelogram/.style={trapezium, draw, fill=gray, opacity=.3, minimum width=4cm, trapezium left angle=45, trapezium right angle=135}]
   \def\xa{1}
   \def\ya{3}
   \def\r{.35} 
   \def\rb{.15} 
   \coordinate (Ct) at (\xa,\ya); 
   \coordinate (TC) at ($(Ct)+(0,-2)$);   
   \coordinate (S) at ($(Ct)+(0,-3.5)$);  
   \coordinate (A1) at ($(TC)+(-1,0)$);
   \coordinate (A2) at ($(TC)+(-.5,-.5)$);
   \coordinate (A3) at ($(TC)+(1,0)$);
   \coordinate (A4) at ($(TC)+(.5,.5)$);
   \pgfmathsetmacro\Axi{\r*cos(180)}
   \pgfmathsetmacro\Ayi{\r*sin(180)}
   \coordinate (As) at ($(S)+(\Axi,\Ayi)$);
   \pgfmathsetmacro\Bxi{\r*cos(0)}
   \pgfmathsetmacro\Byi{\r*sin(0)}
   \coordinate (Bs) at ($(S)+(\Bxi,\Byi)$);
   \pgfmathsetmacro\Cxi{\r*cos(60)}
   \pgfmathsetmacro\Cyi{\r*sin(60)}
   \coordinate (Cs) at ($(S)+(\Cxi,\Cyi)$);
   \pgfmathsetmacro\Dxi{\rb*cos(120)}
   \pgfmathsetmacro\Dyi{\rb*sin(120)}
   \coordinate (Ds) at ($(S)+(\Dxi,\Dyi)$);
   \coordinate (BR) at ($(Ct)+({3/2},-5)$);
   \coordinate (TR) at ($(Ct)+({3/2},-2)$);
   \coordinate (tb) at ($(BR)+(0.125,0)$);
   \coordinate (tt) at ($(TR)+(0.125,0)$);
   \node [shade, shading=ball,circle,ball color=green!70!black,minimum size=.75cm] (Ampl) at (S) {};
   \draw[-, directed, black, thick] (A1) -- (A2);
   \draw[-, directed, black, thick] (A2) -- (A3);
   \draw[-, directed, black, thick] (A3) -- (A4);
   \draw[-, directed, black, thick] (A4) -- (A1);   
   \draw[-, red, thick] (As) edge [bend left] (A1);
   \draw[-, red, thick] (Ds) edge [bend left=20] (A2);   
   \draw[-, red, thick] (Bs) edge [bend right] (A3);   
   \draw[-, red, thick] (Cs) edge [bend right=20] (A4);   
   \draw [->, thick] (tb) -- (tt);
   \coordinate [label=right:{\tiny $\eta$}] (t) at ($(tb)!0.5!(tt)$);
   \node[parallelogram, trapezium stretches=false,minimum height=1cm] at (TC) {};
   \coordinate (LT) at ($(Ct)+(0,-6.5)$);
   \node [shade, shading=ball,circle,ball color=red!90!black,opacity=.25, minimum size=1cm] (Ampl2) at (S) {};
   \coordinate (tc) at ($(S)-(.75,0)$);
   \coordinate (tc1) at ($(tc)+(0,.5)$);
   \coordinate (tc2) at ($(tc)-(0,.5)$);
   \draw[->] (tc1) -- (tc2);
   \node[scale=.75] (tcl) at ($(tc)-(.125,0)$) {$\bar{\eta}$};
  \end{scope}
 \end{tikzpicture}
\end{wrapfigure}
Outside of the physical sheet, they develop a singularity in such a sum, which can be reached upon analytic continuation. At this point in energy space, the process shows energy conservation and, thus, is time-translation invariant as well as Lorentz invariant, and it reduces to the high energy limit of the flat-space scattering amplitudes -- this is a fact which can be understood by realising that the point in energy space $\sum_{j=1}^n E_j\,\longrightarrow\,0$  dominates as the interaction are taken at early times, with the late-time boundary which effectively becomes infinitely far away and disappears, restoring the conditions which characterise a scattering process in flat-space \cite{Maldacena:2011nz, Arkani-Hamed:2015bza}. It is quite remarkable how the wavefunction of the universe and the spatial correlation functions in a static Bunch-Davies vacuum encode the flat-space scattering amplitudes. This relation between cosmological and flat-space observables has deep implications for the analytic structure of the formers, which are not yet fully understood: there should be an imprint of all the theorems and properties holding for the flat-space S-matrix in the wavefunction of the universe. For example, it has to factorise in a codimension-two surface of the energy space, reflecting the factorisation properties in flat-space. 

\noindent
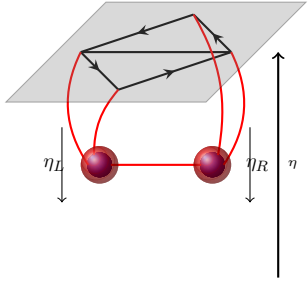
\begin{wrapfigure}{l}{4.5cm}
 \begin{tikzpicture}[node distance=2cm, cross/.style={cross out, draw, inner sep=0pt, outer sep=0pt}]
  \begin{scope}[parallelogram/.style={trapezium, draw, fill=gray, opacity=.3, minimum width=4cm, trapezium left angle=45, trapezium right angle=135}]
   \def\xa{1}
   \def\ya{3}
   \def\r{.15} 
   \def\rb{.15} 
   \coordinate (Ct) at (\xa,\ya); 
   \coordinate (TC) at ($(Ct)+(0,-2)$);   
   \coordinate (S) at ($(Ct)+(0,-3.5)$);
   \coordinate (Sl) at ($(S)-(.75,0)$);
   \coordinate (Sr) at ($(S)+(.75,0)$); 
   \coordinate (A1) at ($(TC)+(-1,0)$);
   \coordinate (A2) at ($(TC)+(-.5,-.5)$);
   \coordinate (A3) at ($(TC)+(1,0)$);
   \coordinate (A4) at ($(TC)+(.5,.5)$);
   \pgfmathsetmacro\Axi{\r*cos(180)}
   \pgfmathsetmacro\Ayi{\r*sin(180)}
   \coordinate (As) at ($(Sl)+(\Axi,\Ayi)$);
   \pgfmathsetmacro\Bxi{\r*cos(0)}
   \pgfmathsetmacro\Byi{\r*sin(0)}
   \coordinate (Bs) at ($(Sr)+(\Bxi,\Byi)$);m
   \pgfmathsetmacro\Cxi{\r*cos(60)}
   \pgfmathsetmacro\Cyi{\r*sin(60)}
   \coordinate (Cs) at ($(Sr)+(\Cxi,\Cyi)$);
   \pgfmathsetmacro\Dxi{\rb*cos(120)}
   \pgfmathsetmacro\Dyi{\rb*sin(120)}
   \coordinate (Ds) at ($(Sl)+(\Dxi,\Dyi)$);
   \coordinate (Es) at ($(Sl)+(\Bxi,\Byi)$);
   \coordinate (Fs) at ($(Sr)+(\Axi,\Ayi)$);   
   \coordinate (BR) at ($(Ct)+({3/2},-5)$);
   \coordinate (TR) at ($(Ct)+({3/2},-2)$);
   \coordinate (tb) at ($(BR)+(0.125,0)$);
   \coordinate (tt) at ($(TR)+(0.125,0)$);
   \node [shade, shading=ball,circle,ball color=blue,minimum size=.2cm] (AmplL) at (Sl) {};
   \node [shade, shading=ball,circle,ball color=blue,minimum size=.2cm] (AmplR) at (Sr) {};   
   \draw[-, directed, black, thick] (A1) -- (A2);
   \draw[-, directed, black, thick] (A2) -- (A3);
   \draw[-, directed, black, thick] (A3) -- (A4);
   \draw[-, directed, black, thick] (A4) -- (A1);   
   \draw[-, black, thick] (A1) -- (A3);
   \draw[-, red, thick] (As) edge [bend left] (A1);
   \draw[-, red, thick] (Ds) edge [bend left=20] (A2);   
   \draw[-, red, thick] (Bs) edge [bend right] (A3);   
   \draw[-, red, thick] (Cs) edge [bend right=20] (A4);
   \draw[-, red, thick] (Es) -- (Fs);   
   \draw [->, thick] (tb) -- (tt);
   \coordinate [label=right:{\tiny $\eta$}] (t) at ($(tb)!0.5!(tt)$);
   \node[parallelogram, trapezium stretches=false,minimum height=1cm] at (TC) {};
   \coordinate (LT) at ($(Ct)+(0,-6.5)$);
   \node [shade, shading=ball,circle,ball color=red!90!black,opacity=.25, minimum size=.5cm] (AmplL2) at (Sl) {};
   \coordinate (tc) at ($(Sl)-(.5,0)$);
   \coordinate (tc1) at ($(tc)+(0,.5)$);
   \coordinate (tc2) at ($(tc)-(0,.5)$);
   \draw[->] (tc1) -- (tc2);
   \node[scale=.75] (tcl) at ($(tc)-(.1,0)$) {$\eta_L$};
   \node [shade, shading=ball,circle,ball color=red!90!black,opacity=.25, minimum size=.5cm] (AmplR2) at (Sr) {};
   \coordinate (tc) at ($(Sr)+(.5,0)$);
   \coordinate (tc1) at ($(tc)+(0,.5)$);
   \coordinate (tc2) at ($(tc)-(0,.5)$);
   \draw[->] (tc1) -- (tc2);
   \node[scale=.75] (tcl) at ($(tc)+(.1,0)$) {$\eta_R$};
  \end{scope}
 \end{tikzpicture}
\end{wrapfigure}
This fact has been used, together with the requirement that Bunch-Davies observables should not have singularities in the physical sheet as well as conformal symmetry, to compute the four-point correlation functions with external conformally-coupled or massless scalars and internal massive states in de Sitter space-time and the inflationary three-point functions which can be obtained from the former by evaluating one of the external states on the time dependent background \cite{Arkani-Hamed:2018kmz}.

Even more surprisingly, for a large class of toy models described by a massless scalar state in flat-space with time-dependent polynomial interactions, which, upon a specific choice for the time-dependence of the couplings, contains the conformally-coupled scalars with polynomial interactions in FRW cosmologies \cite{Arkani-Hamed:2017fdk}, it is possible to reconstruct the Bunch-Davies perturbative wavefunction at all order in perturbation theory from the knowledge of the flat-space scattering amplitudes and the requirement of the absence of unphysical singularities \cite{Benincasa:2018ssx}. Despite this latter result cannot be completely general, but it may hold for a larger class of toy models, it suggests that the flat-space physics constrains the wavefunction of the universe more than what one would have ever expected. In the case of \cite{Benincasa:2018ssx}, it reflects into the fact that the coefficients of all the singularities can be interpreted in terms of flat-space processes or, anyhow, expressed in terms of them.

These features are made manifest in the formulation of the wavefunction in terms of {\it cosmological polytopes} introduced in \cite{Arkani-Hamed:2017fdk}. They are combinatorial-geometrical objects with their own first principle definition, characterised by a differential form, called {\it canonical form}\footnote{For an extensive study of positive geometries and canonical forms, see \cite{Arkani-Hamed:2017tmz}.}, whose coefficient has all the properties that we ascribe to the wavefunction of the universe. In particular, their boundaries are lower-dimensional polytopes which encode the residues of the wavefunction poles, with the hyperplanes identifying them being related to the poles themselves. Thus, there is a codimension-one boundary, named {\it scattering facet}, which is related to the total energy pole and encodes the relevant flat-space scattering amplitude. Amazingly, the vertex structure of such a facet makes the cutting rules manifest, allowing us for a novel combinatorial-geometrical proof of them, while its dual makes Lorentz invariance manifest \cite{Arkani-Hamed:2018ahb}.

The works  \cite{Arkani-Hamed:2018kmz} and  \cite{Arkani-Hamed:2017fdk, Arkani-Hamed:2018ahb, Benincasa:2018ssx} provide two different but complementary approaches for understanding the general rules behind cosmological processes, both of which take the perspective of not considering explicitly the time evolution: in the former the correlation functions are determined from symmetries and the knowledge of their singularities, in a very S-matrix-like fashion; the latter instead consider a totally new mathematical formulation with the rules we are looking for which should emerge from its first principles\footnote{Also this approach takes a lesson coming from the most recent developments in the context of flat-space scattering amplitudes, which have a combinatorial characterisation in a number of cases, see \cite{ArkaniHamed:2012nw, Benincasa:2016awv} and \cite{Arkani-Hamed:2013jha, Arkani-Hamed:2017mur, Frost:2018djd, Salvatori:2018aha, Banerjee:2018tun, Raman:2019utu}.}. However, in both the approaches we still need more theoretical data in order to grasp the fundamental properties ruling the cosmological observables. 

In this paper, we will extend the exploration of the detailed structure of the perturbative wavefunction of the universe developed in \cite{Arkani-Hamed:2017fdk, Arkani-Hamed:2018ahb, Benincasa:2018ssx}, focusing on a class of toy models of scalars with time-dependent masses and time-dependent polynomial couplings, which contains massive scalars with polynomial interactions in FRW cosmologies, upon a specific choice of the time-dependence of the mass and couplings. In Section \ref{sec:Rev}, after having introduced the model and discussed its generalities, we restrict to a subclass  for which the time-dependent mass is inversely proportional to the (conformal) time and we define a set of differential operators mapping free flat-space massless\footnote{The model is formulated as a scalar in flat-space, with the cosmology encoded into the time-dependence of the mass and of the coupling. Thus, when we refer to the states, we will always use the flat-space wording, unless otherwise specified.} scalars ({\it i.e.} conformally coupled scalars in FRW cosmologies) to states with generic masses. This allows us to focus on the wavefunction of the universe with external massless scalars and prove a novel set of recursion relations which relates wavefunction with internal states with different masses and involve certain differential operators. In Section \ref{sec:EWG}, we exploit these recursion relations for a class of values of the masses, for which the iterated recursion relations has the structure of a differential operator acting on the wavefunction with all the internal states being massless ({\it i.e.} conformally coupled). In these cases the masses on a given edge $e$ of the graph representing a certain contribution to the wavefunction, can be labelled by an integer $l_e$, and the wavefunction is represented by edge-weighted graphs with the integer $l_e$ being the weight of the edge $e$. While the combinatorial rules proven in \cite{Arkani-Hamed:2017fdk} for computing the seed of our new recursion relations together with the differential operators, allows us to compute the contribution to the wavefunction from a given graph, we also provide a combinatorial rule to predict the order of the poles in the wavefunction. Section \ref{sec:Pert} is devoted to generalise the discussion of the previous section. In this case we treat the mass perturbatively. In Section \ref{sec:CPl1} we discuss a generalisation of the cosmological polytopes, whose canonical form encode the wavefunction of the universe for $l\,=\,1$, which contains the minimally coupled scalars in FRW cosmologies. We discuss in detail its face structure. The wavefunction of the universe turns out to be a degenerate limit of the canonical form of these polytopes, and the flat-space amplitude is returned by a higher codimension face. For these wavefunctions, the flat-space amplitudes are given by the coefficient of the leading term in its Laurent expansion when the total energy goes to zero. This is beautifully reflected in the polytope picture by the fact that the scattering face has now higher codimension, with the codimension giving the degree of the pole. We conclude the section commenting on the polytope description of the perturbative mass expansion, whose contribution can be obtained as a degenerate limit of a certain subclass of the standard cosmological polytopes. The degenerate limit allows to obtain a rational function with multiple poles from the canonical forms of the polytopes which are characterised by having logarithmic singularities only. This is a very similar phenomenon to what happens in the case of the halohedron \cite{Salvatori:2018aha}. Finally, Section \ref{sec:Concl} contains our conclusion and outlook.


\section{The wavefunction of the universe for massive scalars}\label{sec:Rev}

We consider a class of toy models of a scalar $\phi$ in a $(d+1)$-dimensional flat space-time with a time-dependent mass $\mu(\eta)$ as well as time-dependent polynomial couplings:
\begin{equation}\eqlabel{eq:Sact}
 S\:=\:\int d^dx\,\int_{-\infty}^{0}d\eta\,
  \left\{
   \frac{1}{2}\left(\partial\phi\right)^2-\frac{1}{2}\mu^2(\eta)\phi^2-\sum_{k\ge3}\lambda_{k}(\eta)\phi^{k}
  \right\}.
\end{equation}
Such a model describes a massive scalar $\phi$ in FRW cosmologies with polynomial self-interactions, for the following choices for the mass $\mu(\eta)$ and the couplings $\lambda_{k}(\eta)$
\begin{equation}\eqlabel{eq:ml}
  \mu^2(\eta)\:=\:m^2a^2(\eta)+2d\left(\xi-\frac{d-1}{4d}\right)\left[\partial_{\eta}\left(\frac{\dot{a}}{a}\right)+\frac{d-1}{2}\left(\frac{\dot{a}}{a}\right)^2\right],\quad
  \phantom{\ldots}\\
  \lambda_{k}(\eta)\:=\:\lambda_{k}\left[a(\eta)\right]^{2+\frac{(d-1)(2-k)}{2}}
\end{equation}
where $\dot{\phantom{a}}$ indicates the derivative with respect to the conformal time $\eta$, $a(\eta)$ is the time-dependent warp factor for FRW cosmologies:
\begin{equation}\eqlabel{eq:FRW}
 ds^2\:=\:a^2(\eta)\left[-d\eta+dx_idx^i\right],
\end{equation}
and $\xi$ is a parameter such that for $\xi\,=(d-1)/4d$ and $m\,=\,0$ the model reduces to the one of a conformally coupled scalar with (non-conformal) polynomial interactions\footnote{There is also another special case for which the model reduces to a massless scalar with time-dependent interactions, but without requiring that the parameter $\xi$ has the conformal value. Setting $m\,=\,0$, it corresponds to a specific choice of cosmology, such that:
$$
 \partial_{\eta}\left(\frac{\dot{a}}{a}\right)+\frac{d-1}{2}\left(\frac{\dot{a}}{a}\right)^2\,=\,0.
$$
For $d\,=\,1$, the warp factor is an exponential $a(\eta)\,=\,a_0\,e^{A_0\eta}$, which vanishes in the far past if $A_0\,\in\,\mathbb{R}_{+}$. For $d\,>\,1$, the solution blows up as $\eta\,\longrightarrow\,-\infty$. Finally notice that, in a cosmology $a(\eta)\,=\,a_0e^{A_0\eta}$ ($A_0\,\in\,\mathbb{R}_+$) and with $d\,>\,1$ (and still $m\,=\,0$), it is described by a scalar with a constant mass in a flat space-time -- while for $m\,\neq\,0$, the time-dependent mass of the flat-space scalar increases as the universe expands.}, which has been discussed in \cite{Arkani-Hamed:2017fdk, Arkani-Hamed:2018ahb,Benincasa:2018ssx}; for $\xi\,=\,0$ the scalar becomes minimally coupled. The mode functions are determined by the following differential equation
\begin{equation}\eqlabel{eq:mfeq}
 \ddot{\phi}_{\circ}(\eta)+\left(E^2+\mu^2(\eta)\right)\phi_{\circ}(\eta)\:=\:0,\hspace{1.5cm} E\:\equiv\:|\overrightarrow{p}|,
\end{equation}
with the condition that it vanishes in the far past, as $\eta\,\longrightarrow\,-\infty$. A solution for such an equation is not known for an arbitrary time-dependent mass $\mu(\eta)$, but it can be studied if it is considered perturbatively. For the time being, let us focus on the specific choice $\mu(\eta)\,=\,\mu_{\alpha}^2\eta^{-2}$, which corresponds to cosmologies with $a(\eta)\,=\,(-\eta)^{-\alpha}$ ($\alpha\,\in\,\mathbb{R}$), where $\mu^2_{\alpha}\,\equiv\,2d\alpha(\xi-(d-1)/4d)(1+(d-1)\alpha/2)$ for $m\,=\,0$, and it includes also the case $m\,\neq\,0$ for $\alpha\,=\,1$ ({\it i.e.} in de Sitter), with $\mu^2_1\,=\,m^2+d(\xi-(d-1)/4d)(1+d)$. In this case, the solution of the mode equation \eqref{eq:mfeq} ensuring the correct oscillating behaviour in the far past $\phi_{\circ}(\eta)\:\overset{\eta\,\longrightarrow\,-\infty}{\sim}\:e^{iE\eta}$ is given in terms of Hankel functions of the second type
\begin{equation}\eqlabel{eq:mm}
 \phi_{\circ}^{\mbox{\tiny $(\nu)$}}\:=\:\sqrt{-E\eta}H_{\nu}^{\mbox{\tiny $(2)$}}(-E\eta)\:\overset{\eta\,\longrightarrow\,-\infty}{\sim}\:e^{iE\eta},\hspace{1.5cm}\nu\,\equiv\,\sqrt{\frac{1}{4}-\mu_{\alpha}^2}.
\end{equation}
Notice from \eqref{eq:mm} that $\nu$ can be either real or purely imaginary depending on whether $\mu^2_{\alpha}$ is respectively smaller or greater than $1/4$. More explicitly, the order parameter $\nu$ writes
\begin{equation}\eqlabel{eq:cups}
 \begin{array}{ccl}
  \mbox{Conformal coupling } {\displaystyle\xi\,=\,\frac{d-1}{4d}}\phantom{\ldots} & \mbox{Minimal coupling } {\displaystyle\xi\,=\,0}\phantom{\ldots} & {} 			     \\ 
  {}								                   &						                       & {} 			     \\
  {\displaystyle\nu\,=\,\sqrt{\frac{1}{4}-m^2}}			                   & {\displaystyle\nu\,=\,\sqrt{\frac{d^2}{4}-m^2}}                    & (\alpha\,=\,1,\;m) \\
  {}										   & 								       & {}   			     \\
  {\displaystyle\nu\,=\,\frac{1}{2}}						   & {\displaystyle\nu\,=\,\frac{1}{2}+\frac{d-1}{2}\alpha}            & (\alpha,\;m\,=\,0)	     \\
 \end{array}
\end{equation}
and it can be either imaginary ($\nu\,=\,i\zeta$, $\zeta\,\in\,\mathbb{R}$) or real for $\alpha\,=\,1$ -- they are respectively the principal and complementary series in de Sitter --, while it is only real for generic $\alpha$. In this last case, $\nu\,\in\,\mathbb{Z}_{\frac{1}{2}}$ if $(d-1)\alpha\,=\,2l$ ($l\,\in\,\mathbb{Z}$). Furthermore, as for the case of a generic function $\mu(\eta)$, the mode equation \eqref{eq:mfeq} for generic $\alpha$ and $m$ cannot be solved exactly, but it can be treated considering $m$ perturbatively, which will be analysed in Section \ref{sec:Pert}.

\begin{figure}[t]
 \centering
 \begin{tikzpicture}[line join = round, line cap = round, ball/.style = {circle, draw, align=center, anchor=north, inner sep=0}, scale=1.5, transform shape]
  \begin{scope}
   \def\cx{1.75}
   \def\cy{3}
   \def\r{.6}
   \pgfmathsetmacro\Axi{\cx+\r*cos(135)}
   \pgfmathsetmacro\Ayi{\cy+\r*sin(135)}
   \pgfmathsetmacro\Axf{\Axi+cos(135)}
   \pgfmathsetmacro\Ayf{\Ayi+sin(135)}
   \coordinate (pAi) at (\Axi,\Ayi);
   \coordinate (pAf) at (\Axf,\Ayf);
   \pgfmathsetmacro\Bxi{\cx+\r*cos(45)}
   \pgfmathsetmacro\Byi{\cy+\r*sin(45)}
   \pgfmathsetmacro\Bxf{\Bxi+cos(45)}
   \pgfmathsetmacro\Byf{\Byi+sin(45)}
   \coordinate (pBi) at (\Bxi,\Byi);
   \coordinate (pBf) at (\Bxf,\Byf);
   \pgfmathsetmacro\Cxi{\cx+\r*cos(-45)}
   \pgfmathsetmacro\Cyi{\cy+\r*sin(-45)}
   \pgfmathsetmacro\Cxf{\Cxi+cos(-45)}
   \pgfmathsetmacro\Cyf{\Cyi+sin(-45)}
   \coordinate (pCi) at (\Cxi,\Cyi);
   \coordinate (pCf) at (\Cxf,\Cyf);
   \pgfmathsetmacro\Dxi{\cx+\r*cos(-135)}
   \pgfmathsetmacro\Dyi{\cy+\r*sin(-135)}
   \pgfmathsetmacro\Dxf{\Dxi+cos(-135)}
   \pgfmathsetmacro\Dyf{\Dyi+sin(-135)}
   \coordinate (pDi) at (\Dxi,\Dyi);
   \coordinate (pDf) at (\Dxf,\Dyf);
   \pgfmathsetmacro\Exi{\cx+\r*cos(90)}
   \pgfmathsetmacro\Eyi{\cy+\r*sin(90)}
   \pgfmathsetmacro\Exf{\Exi+cos(90)}
   \pgfmathsetmacro\Eyf{\Eyi+sin(90)}
   \coordinate (pEi) at (\Exi,\Eyi);
   \coordinate (pEf) at (\Exf,\Eyf);
   \coordinate (ti) at ($(pDf)-(.25,0)$);
   \coordinate (tf) at ($(pAf)-(.25,0)$);
   \draw[->] (ti) -- (tf);
   \coordinate[label=left:{\tiny $\displaystyle\eta$}] (t) at ($(ti)!0.5!(tf)$);
   \coordinate (t0) at ($(pBf)+(.1,0)$);
   \coordinate (tinf) at ($(pCf)+(.1,0)$);
   \node[scale=.5, right=.0125 of t0] (t0l) {\tiny $\displaystyle\eta\,=\,0$};
   \node[scale=.5, right=.0125 of tinf] (tinfl) {\tiny $\displaystyle\eta\,=\,-\infty$};
   \draw[-] ($(pAf)-(.1,0)$) -- (t0);
   \draw[-] ($(pDf)-(.1,0)$) -- ($(pCf)+(.1,0)$);
   \coordinate (d2) at ($(pAf)!0.25!(pBf)$);
   \coordinate (d3) at ($(pAf)!0.5!(pBf)$);
   \coordinate (d4) at ($(pAf)!0.75!(pBf)$);
   \node[above=.01cm of pAf, scale=.625] (d1l) {$\displaystyle\overrightarrow{p}_1$};
   \node[above=.01cm of d2, scale=.625] (d2l) {$\displaystyle\overrightarrow{p}_2$};
   \node[above=.01cm of d3, scale=.625] (d3l) {$\displaystyle\overrightarrow{p}_3$};
   \node[above=.01cm of d4, scale=.625] (d4l) {$\displaystyle\overrightarrow{p}_4$};
   \node[above=.01cm of pBf, scale=.625] (d5l) {$\displaystyle\overrightarrow{p}_5$};
   \def\rb{.55}
   \pgfmathsetmacro\sax{\cx+\rb*cos(180)}
   \pgfmathsetmacro\say{\cy+\rb*sin(180)}
   \coordinate[label=below:{\scalebox{0.5}{$x_1$}}] (s1) at (\sax,\say);
   \pgfmathsetmacro\sbx{\cx+\rb*cos(135)}
   \pgfmathsetmacro\sby{\cy+\rb*sin(135)}
   \coordinate (s2) at (\sbx,\sby);
   \pgfmathsetmacro\scx{\cx+\rb*cos(90)}
   \pgfmathsetmacro\scy{\cy+\rb*sin(90)}
   \coordinate (s3) at (\scx,\scy);
   \pgfmathsetmacro\sdx{\cx+\rb*cos(45)}
   \pgfmathsetmacro\sdy{\cy+\rb*sin(45)}
   \coordinate (s4) at (\sdx,\sdy);
   \pgfmathsetmacro\sex{\cx+\rb*cos(0)}
   \pgfmathsetmacro\sey{\cy+\rb*sin(0)}
   \coordinate[label=below:{\scalebox{0.5}{$x_3$}}] (s5) at (\sex,\sey);
   \coordinate[label=below:{\scalebox{0.5}{$x_2$}}] (sc) at (\cx,\cy);
   \draw (s1) edge [bend left] (pAf);
   \draw (s1) edge [bend left] (d2);
   \draw (s3) -- (d3);
   \draw (s5) edge [bend right] (d4);
   \draw (s5) edge [bend right] (pBf);
   \draw [fill] (s1) circle (1pt);
   \draw [fill] (sc) circle (1pt);
   \draw (s3) -- (sc);
   \draw [fill] (s5) circle (1pt);
   \draw[-,thick] (s1) -- (sc) -- (s5);
  \end{scope}
  \begin{scope}[shift={(4.5,0)}, transform shape]
   \def\cx{1.75}
   \def\cy{3}
   \def\r{.6}
   \pgfmathsetmacro\Axi{\cx+\r*cos(135)}
   \pgfmathsetmacro\Ayi{\cy+\r*sin(135)}
   \pgfmathsetmacro\Axf{\Axi+cos(135)}
   \pgfmathsetmacro\Ayf{\Ayi+sin(135)}
   \coordinate (pAi) at (\Axi,\Ayi);
   \coordinate (pAf) at (\Axf,\Ayf);
   \pgfmathsetmacro\Bxi{\cx+\r*cos(45)}
   \pgfmathsetmacro\Byi{\cy+\r*sin(45)}
   \pgfmathsetmacro\Bxf{\Bxi+cos(45)}
   \pgfmathsetmacro\Byf{\Byi+sin(45)}
   \coordinate (pBi) at (\Bxi,\Byi);
   \coordinate (pBf) at (\Bxf,\Byf);
   \pgfmathsetmacro\Cxi{\cx+\r*cos(-45)}
   \pgfmathsetmacro\Cyi{\cy+\r*sin(-45)}
   \pgfmathsetmacro\Cxf{\Cxi+cos(-45)}
   \pgfmathsetmacro\Cyf{\Cyi+sin(-45)}
   \coordinate (pCi) at (\Cxi,\Cyi);
   \coordinate (pCf) at (\Cxf,\Cyf);
   \pgfmathsetmacro\Dxi{\cx+\r*cos(-135)}
   \pgfmathsetmacro\Dyi{\cy+\r*sin(-135)}
   \pgfmathsetmacro\Dxf{\Dxi+cos(-135)}
   \pgfmathsetmacro\Dyf{\Dyi+sin(-135)}
   \coordinate (pDi) at (\Dxi,\Dyi);
   \coordinate (pDf) at (\Dxf,\Dyf);
   \pgfmathsetmacro\Exi{\cx+\r*cos(90)}
   \pgfmathsetmacro\Eyi{\cy+\r*sin(90)}
   \pgfmathsetmacro\Exf{\Exi+cos(90)}
   \pgfmathsetmacro\Eyf{\Eyi+sin(90)}
   \coordinate (pEi) at (\Exi,\Eyi);
   \coordinate (pEf) at (\Exf,\Eyf);
   \coordinate (ti) at ($(pDf)-(.25,0)$);
   \coordinate (tf) at ($(pAf)-(.25,0)$);
   \def\rb{.55}
   \pgfmathsetmacro\sax{\cx+\rb*cos(180)}
   \pgfmathsetmacro\say{\cy+\rb*sin(180)}
   \coordinate[label=below:{\scalebox{0.5}{$x_1$}}] (s1) at (\sax,\say);
   \pgfmathsetmacro\sbx{\cx+\rb*cos(135)}
   \pgfmathsetmacro\sby{\cy+\rb*sin(135)}
   \coordinate (s2) at (\sbx,\sby);
   \pgfmathsetmacro\scx{\cx+\rb*cos(90)}
   \pgfmathsetmacro\scy{\cy+\rb*sin(90)}
   \coordinate (s3) at (\scx,\scy);
   \pgfmathsetmacro\sdx{\cx+\rb*cos(45)}
   \pgfmathsetmacro\sdy{\cy+\rb*sin(45)}
   \coordinate (s4) at (\sdx,\sdy);
   \pgfmathsetmacro\sex{\cx+\rb*cos(0)}
   \pgfmathsetmacro\sey{\cy+\rb*sin(0)}
   \coordinate[label=below:{\scalebox{0.5}{$x_3$}}] (s5) at (\sex,\sey);
   \coordinate[label=below:{\scalebox{0.5}{$x_2$}}] (sc) at (\cx,\cy);
   \draw [fill] (s1) circle (1pt);
   \draw [fill] (sc) circle (1pt);
   \draw [fill] (s5) circle (1pt);
   \draw[-,thick] (s1) -- (sc) -- (s5);
  \end{scope}
 \end{tikzpicture}
 \caption{Example of a Feynman graph contribution to the wavefunction of the universe (left) and its associated reduced graph (right), which is obtained from the former by suppressing the external lines.}
 \label{Fig:G}
\end{figure}
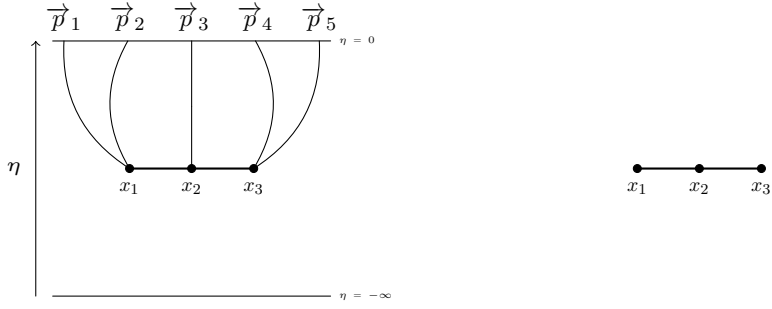

As usual, the perturbative wavefunction can be computed via Feynman graphs whose vertices are associated to the time-dependent couplings $\lambda_{k}(\eta)$, the external edges to the bulk-to-boundary propagators, which are given by the solution of \eqref{eq:mfeq} satisfying the Bunch-Davies boundary condition, and the internal edges are associated to a bulk-to-bulk propagator, which shows three terms, two encoding the time-ordered Feynman propagators and the third one fixed by the condition that the fluctuations have to vanish at the boundary:
\begin{equation}\eqlabel{eq:WF}
 \tilde{\psi}_{\mathcal{G}}\:=\:\int_{-\infty}^{0}\prod_{v\in\mathcal{V}}\left[d\eta_{\nu}\,V_v\phi_{\circ}^{\mbox{\tiny $(v)$}}\right]\prod_{e\in\mathcal{E}}G_e\left(\eta_{v_e},\,\eta_{v'_e}\right),
\end{equation}
where $\mathcal{V}$ and $\mathcal{E}$ are the sets of, respectively, the vertices and the internal edges of the graph $\mathcal{G}$, $V_v\:\equiv\:i\lambda_{k}(\eta_v)$, is the interaction associated to the vertex $v$, $\displaystyle \phi_{\circ}^{\mbox{\tiny $(v)$}}\,\equiv\,\prod_{j\in v}\phi_{\circ}(-E_j\eta_v)$ are the free solution associated to the external states, and $G_e$ is the propagator which is given by
\begin{equation}\eqlabel{eq:ge}
 \begin{split}
 & G_e(\eta_{v_e},\eta_{v'_e})\:=\:\frac{1}{2y_e}
   \left[
    \bar{\phi}_{\circ}(-y_e\eta_{v_e})\phi_{\circ}(-y_e\eta_{v'_e})\vartheta(\eta_{v_e}-\eta_{v'_e})+\phi_{\circ}(-y_e\eta_{v_e})\bar{\phi}_{\circ}(-y_e\eta_{v'_e})\vartheta(\eta_{v'_e}-\eta_{v_e})-
   \right.\\
 & \left.\hspace{3cm}
   -\frac{\bar{\phi}_{\circ}(0)}{\phi_{\circ}(0)}\phi_{\circ}(-y_e\eta_{v_e})\phi_{\circ}(-y_e\eta_{v'_e})
   \right],
 \end{split}
\end{equation}
$E_j$ and $y_e$ being the modulus of the momentum of an external state $j$ and of the momentum running along the edge $e$, respectively, while $\bar{\phi}_{\circ}$ identifies the complex conjugate of the mode function $\phi_{\circ}$.
Finally, considering the time-dependent coupling constants $\lambda_k(\eta)$ in Fourier space
\begin{equation}\eqlabel{eq:Fcc} 
 \lambda(\eta)\:=\:\int_{-\infty}^{+\infty}d\varepsilon\,e^{i\varepsilon\eta}\tilde{\lambda}_k(\varepsilon),
\end{equation}
the perturbative wavefunction can be written as 
\begin{equation}\eqlabel{eq:PWF2a}
 \tilde{\psi}_{\mathcal{G}}\:=\:
  \int_{-\infty}^{+\infty}\prod_{v\in\mathcal{V}}
   \left[d\varepsilon_v\,\tilde{\lambda}_k(\varepsilon_v)\right]\psi_n(\left\{\varepsilon_v\right\}),
\end{equation}
where
\begin{equation}\eqlabel{eq:PWF2b}
 \psi_{\mathcal{G}}(\left\{\varepsilon_v\right\})\:\equiv\:
  \int_{-\infty}^{0}\prod_{v\in\mathcal{V}}\left[d\eta_v e^{i\varepsilon_v\eta_v}\phi_{\circ}^{\mbox{\tiny $(v)$}}(\eta_v)\right]
   \prod_{e\in\mathcal{E}}G_e\left(\eta_{v_e},\eta_{v'_e}\right).
\end{equation}
Our analysis will mainly focus on the structure of \eqref{eq:PWF2b}, leaving the integrations \eqref{eq:PWF2a} as the very last step. From the perspective of the integrand $\psi_{\mathcal{G}}$, the time-dependence of the coupling constants is reflected in the presence of an additional massless external state at each vertex. For cosmologies $a(\eta)\,\propto\,(-\eta)^{-\alpha}$ ($\alpha\,\in\,\mathbb{R}_+$), where $\lambda_k(\eta)$ is given by \eqref{eq:ml}, the Fourier coupling constant $\tilde{\lambda}_k(\varepsilon)$ has support on the Heaviside step function $\vartheta(\varepsilon)$ and, consequently, it takes values just on the positive energy axis. More precisely:
\begin{equation}\eqlabel{eq:Fcc2} 
 \lambda_k(\eta)\:=\:\lambda_k\int_{-\infty}^{+\infty}d\varepsilon\,e^{i\varepsilon\eta}\varepsilon^{\gamma_k-1}\vartheta(\varepsilon)\:\equiv\:
                   \lambda_k\int_{0}^{+\infty}d\varepsilon\,e^{i\varepsilon\eta}\varepsilon^{\gamma_k-1}
\end{equation}
as long as $\gamma_k\,\equiv\,\alpha\left[2+(d-1)(2-k)/2\right]\,>\,0$. As we will discuss in more details later, for $\gamma_k\,<\,0$, $\tilde{\psi}_{\mathcal{G}}$ can be obtained by acting with a derivative operator on $\psi_{\mathcal{G}}$.

As a final remark, in the rest of the paper, unless stated explicitly, we will focus on cosmologies $a(\eta)\,\propto\,(-\eta)^{-\alpha}$, for which the (squared) time-dependent mass is $\mu^2(\eta)\,=\,\mu_{\alpha}^2\eta^{-2}$ and the mode functions are given in terms of Hankel functions, as discussed earlier.


\subsection{Boundary representations for the wavefunction of the universe}\label{subsec:BRca}

In order to get more insights into the structure of the wavefunction of the universe, the zero-th order step is to look for new ways of computing it. For cosmologies $a(\eta)\,\propto\,(-\eta)^{-\alpha}$, the data characterising the wavefunction of the universe are the moduli $E_j$ of the spatial momenta, the angles among the spatial momenta themselves which can be parametrised via the moduli $y_J$ of sums of momenta ($y_J\,\equiv\,\sum_{j\in J}\overrightarrow{p}_j$, $J$ being a subset of the external momenta), as well as the masses of the states involved, which can be encoded into the parameter $\nu_j$ defined as in \eqref{eq:cups}. So, given a graph $\mathcal{G}$, the related wavefunction will be denoted as $\psi_{\mathcal{G}}^{\mbox{\tiny $(\{\nu_j\};\{\nu_e\})$}}(\{E_j\},\{y_e\})$, with the energies $E_j$'s associated to the external states, $y_e$'s associated to the edge $e$, while $\nu_j$ and $\nu_e$ respectively encode masses of the external state $j$ and the internal state on the edge $e$ of $\mathcal{G}$.

In this section we first discuss how contributions $\psi_{\mathcal{G}}^{\mbox{\tiny $(\{\nu_j\}; \{\nu_e\})$}}$ to the wavefunction of the universe related to a graph $\mathcal{G}$ with generic external scalars, can be obtained by acting with certain operators on the contribution to the wavefunction $\psi_{\mathcal{G}}^{\mbox{\tiny $(\{1/2\}; \{\nu_e\})$}}\,\equiv\,\psi_{\mathcal{G}}^{\mbox{\tiny $(\{\nu_e\})$}}$ from the very same graph $\mathcal{G}$ but with external states with order $\nu\,=\,1/2$, corresponding to the case $\mu(\eta)\,=\,0$ with the mode functions that reduce to be exactly exponentials. Then we will show how the wavefunction satisfies a new set of recursion relations. As a final comment, we will consider either $\nu\,\in\,\mathbb{R}_+$, or $\nu\,\in\,i\zeta$ with $\zeta\,\in\,\mathbb{R}_+$ given that the bulk-to-bulk propagator $G^{\mbox{\tiny $(\nu)$}}$ is invariant under the sign flip of $\nu$: $G^{\mbox{\tiny $(-\nu)$}}\,=\,G^{\mbox{\tiny $(\nu)$}}$.


\subsubsection{General external scalars from $\nu\,=\,1/2$}

In this paper, we will focus on wavefunctions whose external states are given by just $\nu\,=\,1/2$, the reason being that {\it any state with a generic mass can be obtained by applying a suitable differential operator on the $\nu\,=\,1/2$ ones} -- an example of this fact was illustrated in \cite{Arkani-Hamed:2015bza, Arkani-Hamed:2018kmz} where weight-shifting operators were defined for mapping the wavefunction of the universe with  external conformally coupled scalars in de Sitter space, to a wavefunction with {\it all} external massless states. Here we will define operators which, acting on the wavefunction, changes $\nu\,=\,1/2$ to an arbitrary $\nu$, {\it i.e.} an arbitrary mass, for a single state. The first direct observation is that any mode function $\phi_{\circ}^{\mbox{\tiny $(\nu)$}}$ can be conveniently written as an operator $\hat{\mathcal{O}}_{\nu}(E)$ in energy space acting on $\phi_{\circ}^{\mbox{\tiny $(1/2)$}}\,\equiv\,e^{iE\eta}$. Concretely
\begin{equation}\eqlabel{eq:ExSop}
 \phi_{\circ}^{\mbox{\tiny $(\nu)$}}\:=\:\frac{1}{(-E\eta)^{\nu-\frac{1}{2}}}\hat{\mathcal{O}}_{\nu}(E)\,e^{iE\eta},\qquad
\end{equation} 
where
\begin{equation}\eqlabel{eq:ExSop2}
 \hat{\mathcal{O}}_{\nu}(E)\:\equiv\:\frac{E^{2\nu}}{\Gamma\left(\frac{1}{2}+\nu\right)\Gamma\left(\frac{1}{2}-\nu\right)}\int_0^{+\infty}dt\,t^{\nu-\frac{1}{2}}\int_0^{+\infty}ds\,s^{-\nu-\frac{1}{2}}
			    e^{-\left(Et+is\frac{\partial}{\partial E}\right)}.
\end{equation}
There is a class of values of $\nu$, and consequently of the masses, for which the expression \eqref{eq:ExSop2} simplifies. For $\nu\,=\,l+\frac{1}{2}$ ($l\,\in\,\mathbb{Z}_+$), we have
\begin{equation}\eqlabel{eq:Opl12}
 \hat{\mathcal{O}}_{l}(E)\:=\:\prod_{r=1}^{l}\left(E\frac{\partial}{\partial E}-(2r-1)\right),
\end{equation}

With the relation \eqref{eq:ExSop} among a generic mode function $\phi_{\circ}^{\mbox{\tiny $(\nu)$}}$ and $\phi_{\circ}^{\mbox{\tiny $(1/2)$}}\,\equiv\,e^{iE\eta}$ at hand, we can also deduce how the wavefunction $\psi_{\mathcal{G}}^{\mbox{\tiny $(\{\nu_j\}; \{\nu_e\})$}}$ can be obtained from $\psi_{\mathcal{G}}^{\mbox{\tiny $(\{\nu_e\})$}}$. Let us begin with considering  a general graph $\mathcal{G}$ with $n_v$ vertices and $n_e$ edges, and a rescaled wavefunction 
\begin{equation}\eqlabel{eq:WFres}
  \tilde{\psi}_{\mathcal{G}}^{\mbox{\tiny $(\{\nu_j\};\{\nu_e\})$}}\:\longrightarrow\:
   \prod_{j=1}^nE^{\frac{1}{2}-\nu_j}_j\prod_{e\in\mathcal{E}}y_e^{2\left(\frac{1}{2}-\nu_e\right)}\tilde{\psi}_{\mathcal{G}}^{\mbox{\tiny $(\{\nu_j\};\{\nu_e\})$}}
\end{equation}
whose explicit expression as a time integral is given by
\begin{equation}\eqlabel{eq:tWF}
  \tilde{\psi}_{\mathcal{G}}^{\mbox{\tiny $(\{\nu_j\};\{\nu_e\})$}}\:=\:
   \prod_{j\in v}\hat{\mathcal{O}}_{\nu_j}(E_j)
   \int_{-\infty}^{0}\prod_{v\in\mathcal{V}}\left[d\eta_v\,\frac{i\lambda_k(\eta_v)}{(-\eta_v)^{\nu_v-\frac{\rho_v}{2}}}\,e^{iX_v\eta_v}\right]\prod_{e\in\mathcal{E}}G_e\left(y_e;\,\eta_{v_e},\eta_{v'_e}\right)
\end{equation}
where $X_v\,\equiv\,\sum_{j\in v}E_j$, the propagators $G_e$ have been rescaled by $[(-y_e\eta_{v_e})(-y_e\eta_{v'_e})]^{\frac{1}{2}-\nu_e}$, $\rho_v$ is the sum of all the states (both internal and external) at the vertex $v$, and $\nu_v\,\equiv\,\sum_{j\in v}\nu_j + \sum_{e\in\mathcal{E}_v}\nu_e$. For each vertex $v$, the coupling constant and the other factors of $\eta_v$ can be consider all together in Fourier space:
\begin{equation}\eqlabel{eq:FCCt}
 \frac{i\lambda_k(\eta_v)}{\left(-\eta_v\right)^{\nu_v-\frac{\rho_v}{2}}}\:=\:
  i^{\beta_{k,\nu}}\left(i\lambda_k\right)\int_{0}^{\infty}d\varepsilon_v\,e^{i\varepsilon_v\eta_v}\varepsilon_v^{\beta_{k,\nu}-1},
\end{equation}
for $\mbox{Re}\left\{\beta_{k,\nu}\right\}\,>\,0$, where $\beta_{k,\nu}\,\equiv\,\gamma_k+\nu_v-\frac{\rho_v}{2}$. The power $\beta_{k,\nu}$ contains both the information about the cosmology and the type of interactions through the parameters $\alpha$ and $k$ in $\gamma_k$, and about the internal and external states at each vertex via $\nu_v$. It is possible to keep the two set of information separated, by writing two different Fourier spaces for $\mbox{Re}\{\gamma_k\}\,>\,0$ and $\mbox{Re}\{\nu_v-\rho_v/2\}\,>\,0$ separately -- then \eqref{eq:FCCt} results from the convolution theorem. Indeed, if either $\mbox{Re}\{\gamma_k\}\,<\,0$ or $\mbox{Re}\{\nu_v-\rho_v/2\}\,<\,0$ (or both), the related Fourier integral is substituted by a derivative operator. Thus, the wavefunction of the universe related to a generic graph $\mathcal{G}$ can be written as
\begin{equation}\eqlabel{eq:tWF2}
 \tilde{\psi}_{\mathcal{G}}\:=\: \prod_{j=1}^{n}\hat{\mathcal{O}}_{\nu_j}(E_j)
   \int_{0}^{+\infty}\prod_{v\in\mathcal{V}}\left[d\varepsilon_v\,\varepsilon^{\beta_{k,v}-1}\right]
   \underbrace{\int_{-\infty}^{0}\prod_{v\in\mathcal{V}}\left[d\eta_v\,i\,e^{i(X_v+\varepsilon_v)\eta_v}\right]\prod_{e\in\mathcal{E}}G_e\left(\eta_{v_e},\eta_{v'_e}\right)}_{\text{$\displaystyle \psi_{\mathcal{G}}^{\mbox{\tiny $(\{\nu_e\})$}}$}}
\end{equation}
{\it i.e.} the (rescaled) wavefunction $\tilde{\psi}_{\mathcal{G}}$  with arbitrary external states as well as its integrand $\psi_{\mathcal{G}}$ can be obtained by acting with the operators $\hat{\mathcal{O}}$ on the (rescaled) wavefunction $\tilde{\psi}_{\mathcal{G}}^{\mbox{\tiny $(\{\frac{1}{2}\};\{\nu_e\})$}}$ with just external conformally coupled scalars and its integrand $\psi_{\mathcal{G}}^{\mbox{\tiny $(\{\frac{1}{2}\};\{\nu_e\})$}}$ respectively. Notice that the formula \eqref{eq:tWF2} is valid as long as  $\mbox{Re}\left\{\beta_{k,\nu}\right\}\,>\,0$. For  $\mbox{Re}\left\{\beta_{k,\nu}\right\}\,\le\,0$, the integrations over $\varepsilon_v$ is substituted by a differential operator of order $\beta\,\in\,\mathbb{Z}_+$, with $\beta$ which is finally analytically continued to $-\beta_{k,\nu}$. Hence, we can write in full generality:
\begin{equation}\eqlabel{eq:tWF3}
 \psi'_{\mathcal{G}}\:=\:\left[\prod_{j=1}^{n}\hat{\mathcal{O}}_{\nu_j}(E_j)\right]\psi_{\mathcal{G}}^{\mbox{\tiny $(\{\frac{1}{2}\};\{\nu_e\})$}},
 \qquad
 \tilde{\psi}'_{\mathcal{G}}\:=\:\left[\prod_{j=1}^{n}\hat{\mathcal{O}}_{\nu_j}(E_j)\right]\tilde{\psi}_{\mathcal{G}}^{\mbox{\tiny $(\{\frac{1}{2}\};\{\nu_e\})$}},
\end{equation}
with $\tilde{\psi}_{\mathcal{G}}^{\mbox{\tiny $(\{\frac{1}{2}\};\{\nu_e\})$}}$ given by
\begin{equation}\eqlabel{eq:tWF3b}
 \tilde{\psi}_{\mathcal{G}}^{\mbox{\tiny $(\{\frac{1}{2}\};\{\nu_e\})$}}\:=\:\hat{\mathcal{W}}(x,X)\psi_{\mathcal{G}}^{\mbox{\tiny $(\{\frac{1}{2}\};\{\nu_e\})$}},
 \qquad
 \hat{\mathcal{W}}\:=\:
 \left\{
  \begin{array}{l}
   {\displaystyle \prod_{v\in\mathcal{V}}\int_{X_v}^{+\infty}dx_v\,(x_v-X_v)^{\beta_{k,v}-1}},\quad\mbox{for } \mbox{Re}\left\{\beta_{k,\nu}\right\}\,>\,0,\\
   \phantom{\ldots}\\
   {\displaystyle \left.\prod_{v\in\mathcal{V}}\left(i\frac{\partial}{\partial X_v}\right)^{\beta}\right|_{\beta\longrightarrow -\beta_{k,\nu}}},\hspace{1.25cm} \mbox{for } \mbox{Re}\left\{\beta_{k,\nu}\right\}\,\le\,0,
  \end{array}
 \right.
 .
\end{equation}
Notice that in \eqref{eq:tWF3} the integrated (rescaled) wavefunction $\tilde{\psi}'_{\mathcal{G}}$ with arbitrary scalars is obtained from the integrated (rescaled) wavefunction $\tilde{\psi}_{\mathcal{G}}^{\mbox{\tiny $(\{\frac{1}{2}\};\{\nu_e\})$}}$ with external $\nu\,=\,1/2$ states only. It is also possible to obtaining it by applying the operator $\hat{\mathcal{W}}$, written now as an integral over $\varepsilon_v$, acting on $\psi'_{\mathcal{G}}$.

Thus, the integrand $\psi'_{\mathcal{G}}$ and the integral $\tilde{\psi}'_{\mathcal{G}}$ for contact interactions acquire the following form respectively:
\begin{equation}\label{eq:CI1}
 \psi'_{\mathcal{G}}\:=\:\left[\prod_{j=1}^{n}\hat{\mathcal{O}}_{\nu_j}(E_j)\right]\frac{1}{X},
 \qquad
 \tilde{\psi}'_{\mathcal{G}}\:=\:\left[\prod_{j=1}^{n}\hat{\mathcal{O}}_{\nu_j}(E_j)\right]\hat{\mathcal{W}}\frac{1}{x},
\end{equation}
where $X$ is the sum of all the energies, and the product in the operator $\hat{\mathcal{W}}$ is given by a single term. For $k\,=\,3$ and $d\,=\,5$ as well as $k\,=\,4$ and $d\,=\,3$, $\gamma_k\,=\,0$ and the operator $\hat{\mathcal{W}}$ has an integral or derivative form depending only on whether $\mbox{Re}\{\nu\}-n/2$ is positive or negative. Indeed, if all the external states have $\nu_j\,=\,1/2$, then such an operator is just the identity.


\subsubsection{Recursive relations}\label{subsubsec:RR}

Let us now focus on the integrand $\psi_{\mathcal{G}}^{\mbox{\tiny $(\{\frac{1}{2}\};\{\nu_e\})$}}$ with external $\nu\,=\,1/2$ states only. For brevity, and because it will not give rise to any confusion, we will drop the $\{1/2\}$ in the upper index:
\begin{equation}\eqlabel{eq:WFint}
 \psi_{\mathcal{G}}^{\mbox{\tiny $(\{\nu_e\})$}}\:=\:\left(i\lambda_k\right)^{n_v}\int_{-\infty}^0\prod_{v\in\mathcal{V}}\left[d\eta_{v}\,e^{ix_v\eta_v}\right]\prod_{e\in\mathcal{E}}G^{\mbox{\tiny $(\nu_e)$}}\left(y_e;\,\eta_{v_e},\eta'_{v_e}\right),
\end{equation}
with the propagators $G_e$ which, as in \eqref{eq:tWF}, have been rescaled by $[(-y_e\eta_{v_e})(-y_e\eta_{v'_e})]^{\frac{1}{2}-\nu_e}$.

From this integral representation, we can consider the integral $\mathcal{I}$ obtained from \eqref{eq:WFint} by inserting the time-translation operator $\Delta$ in such a way that it acts on the full integrand of \eqref{eq:WFint}:
\begin{equation}\eqlabel{eq:rrWF}
 \mathcal{I}\:\equiv\:\left(i\lambda_k\right)^{n_v}\int_{-\infty}^0\prod_{v\in\mathcal{V}}d\eta_{v}\,\Delta\left[\prod_{v\in\mathcal{V}}e^{ix_v\eta_v}\prod_{e\in\mathcal{E}}G^{\mbox{\tiny $(\nu_e)$}}\left(y_e;\,\eta_{v_e},\eta'_{v_e}\right)\right],\qquad
 \Delta\:\equiv\:-i\sum_{v\in\mathcal{V}}\partial_{\eta_v}.
\end{equation}
It is straightforward to notice that this integral vanishes: because of $\Delta$, the integrand of $\mathcal{I}$ is a sum of total derivatives whose contribution from $-\infty$ vanishes because of the positive frequency external states, while from the boundary because the propagation vanishes there. Thus, allowing the total time translation operator acting before on the external states and then on the propagators, $\mathcal{I}$ can also be written as
\begin{equation}\eqlabel{eq:rrWF2}
 0\:=\:\mathcal{I}\:=\:
  \left(\sum_{v\in\mathcal{V}} x_v\right)\psi_{\mathcal{G}}^{\mbox{\tiny $(\{\nu_e\})$}}\:+\:
   \left(i\lambda_k\right)^{n_v}\sum_{e\in\mathcal{E}}\int_{-\infty}^0\prod_{v\in\mathcal{V}}\left[d\eta_{v}\,e^{ix_v\eta_v}\right]\Delta G_e^{\mbox{\tiny $(\nu_e)$}}\prod_{\bar{e}\in\mathcal{E}\setminus\left\{e\right\}}G_e^{\mbox{\tiny $(\nu_{\bar{e}})$}},
\end{equation}
where the notation has been shortened by writing $G_{e}^{\mbox{\tiny $(\nu_e)$}}\,\equiv\,G_e^{\mbox{\tiny $(\nu_{\bar{e}})$}}\left(y_e;\,\eta_{v_e},\eta'_{v_e}\right)$. Interestingly, the total time translation operator maps the propagator $G_e^{\mbox{\tiny $(\nu_e)$}}$ of a state $\nu_e$ into a propagator of a state $\nu_e-1$:
\begin{equation}\eqlabel{eq:rrWF3}
 \Delta G_{e}^{\mbox{\tiny $(\nu_e)$}}\:=\:-i\left(\eta_{v_e}+\eta_{v'_e}\right)2(\nu_e-1)y_e^2 G_{e}^{\mbox{\tiny $(\nu_e-1)$}} + y_e^2\eta_{v_e}\eta_{v'_e}\Delta G_{e}^{\mbox{\tiny $(\nu_e-1)$}}.
\end{equation}
Substituting \eqref{eq:rrWF3} into \eqref{eq:rrWF2}, the first term in \eqref{eq:rrWF3} gives rise to a first derivative operator dependent on the energies $x_{v_e}$ and $x_{v'_e}$ of the endpoints of the edge $e$ acting on the wavefunction $\psi_{\mathcal{G}}^{\mbox{\tiny $(\nu_e-1,\{\nu_{\bar{e}}\})$}}$ of a graph with the very same topology of $\mathcal{G}$ but with the edge $e$ now related to a state with order $\nu_e-1$. The second term of \eqref{eq:rrWF3} instead gives rise to a double derivative operator dependent again on the energies $x_{v_e}$ and $x_{v'_e}$ of the endpoints of the edge $e$, which now acts on the product of the total energy times $\psi_{\mathcal{G}}^{\mbox{\tiny $(\nu_e-1,\{\nu_{\bar{e}}\})$}}$. Hence, the wavefunction with arbitrary internal states satisfies the following recursion relation
\begin{equation}\eqlabel{eq:rrWFfin1}
  \left(\sum_{v\in\mathcal{V}} x_v\right)\psi_{\mathcal{G}}^{\mbox{\tiny $(\{\nu_e\})$}}\:=\:\sum_{e\in\mathcal{E}}
  \left[
   \left(\frac{\partial}{\partial x_{v_e}}+\frac{\partial}{\partial x_{v'_e}}\right)2\left(\nu_e-1\right)-
   \frac{\partial^2}{\partial x_{v_e} \partial x_{v'_e}}\left(\sum_{v\in\mathcal{V}} x_v\right)
  \right]
  \psi_{\mathcal{G}}^{\mbox{\tiny $(\nu_e-1,\{\nu_{\bar{e}}\})$}}
\end{equation}
which can be schematically visualised as
\begin{equation}\eqlabel{eq:rrWfin1pic}
  \begin{tikzpicture}[node distance=2cm, cross/.style={cross out, draw, inner sep=0pt, outer sep=0pt}, ball/.style = {circle, draw, align=center, anchor=north, inner sep=0}]
  \begin{scope}[shift={(2.125,0)}, scale={.9}, transform shape]
   \begin{scope}[shift={(-1,0)}, transform shape]
    \fill[shade,thick] (0,0) circle (.8);
    \node[text width=.18cm,color=black] at (0,0) (Fn) {$\displaystyle\psi_{\mbox{\tiny $\mathcal{G}$}}$};
    \node[below=.5cm of Fn] (lelhs) {$\displaystyle\left\{\nu_e\right\}$};
    \node[text width=.18cm,color=black] at (-2.8,-0.1) (sum) {$\displaystyle\left(\sum_{v\in\mathcal{V}}x_v\right)$};
    \node[ball,text width=.18cm,fill,color=black,label=left:$\mbox{\tiny $x_1$}$, scale=.75] at (-.8,0) (x1) {}; 
    \node[ball,text width=.18cm,fill,color=black,label=left:$\mbox{\tiny $x_2$}$, scale=.75] at ({.8*cos(150)},{.8*sin(150)}) (x2) {}; 
    \node[ball,text width=.18cm,fill,color=black,label={right:$\mbox{\tiny $x_{i-1}$}$}, scale=.75] at ({.8*cos(30)},{.8*sin(30)}) (xi1) {};
    \node[ball,text width=.18cm,fill,color=black,label={right:$\mbox{\tiny $x_i$}$}, scale=.75] at ({.8},{0}) (xi) {};   
    \node[ball,text width=.18cm,fill,color=black,label={right:$\mbox{\tiny $x_{i+1}$}$}, scale=.75] at ({.75*cos(-30)},{.75*sin(-30)}) (xi1b) {};  
    \node[ball,text width=.18cm,fill,color=black,label=left:$\mbox{\tiny $x_n$}$, scale=.75] at ({.8*cos(150)},{-.8*sin(150)}) (xn) {};   
   \end{scope}
   \node[color=black, scale=1.125] at (2.25,0) (eq) {$\displaystyle\quad=\:\sum_{e\in\mathcal{E}}\hat{\mathcal{O}}_{e}
							\hspace{-.125cm}
							\left[
 							 \begin{array}{l}
							  \phantom{ldots}\\
							  \phantom{ldots}\\
							  \phantom{ldots}
							 \end{array}
 							\right.$};
   \fill[shade,thick] (4,0) circle (.8);  
   \node[text width=.18cm,color=black] at (4,0) (Fl) {$\displaystyle\psi_{\mathcal{L}}$};
   \node[ball,text width=.18cm,fill,color=black, scale=.75] at (3.2,0) (x1) {}; 
   \node[ball,text width=.18cm,fill,color=black, scale=.75] at ({.8*cos(150)+4},{.8*sin(150)}) (x2) {}; 
   \node[ball,text width=.18cm,fill,color=red,label={[label distance=.05mm]30:$\hspace{-.1cm}\mbox{\tiny $x_{v_e}+y_e$}$}] at ({.8+4},{0}) (xv) {};   
   \node[ball,text width=.18cm,fill,color=black, scale=.75] at ({.8*cos(150)+4},{-.8*sin(150)}) (xn) {};   
   \fill[shade,thick] (8,0) circle (.8);  
   \node[text width=.18cm,color=black] at (8,0) (Fr) {$\displaystyle\psi_{\mathcal{R}}$};
   \node[ball,text width=.18cm,fill,color=red,label={[left=1cm]-30:$\hspace{-.1cm}\mbox{\tiny $x_{v'_e}+y_e$}$}] at (7.2,0) (xv2) {}; 
   \node[ball,text width=.18cm,fill,color=black, scale=.75] at ({.8*cos(30)+8},{.8*sin(30)}) (x2) {}; 
   \node[ball,text width=.18cm,fill,color=black, scale=.75] at ({.8+8},{0}) (xt) {};   
   \node[ball,text width=.18cm,fill,color=black, scale=.75] at ({.8*cos(-30)+8},{.8*sin(-30)}) (xn) {};   
   \draw[-,ultra thick,color=red] (xv) -- (xv2);
   \coordinate (lre) at ($(xv)!0.5!(xv2)$);
   \node[below=.15cm of lre, scale=.75] {$\displaystyle\left\{\nu_e-1\right\}$};
   \node[color=black] at (9.25,-0.1) (eq2) {$\displaystyle\;+$};
  \end{scope}
  \begin{scope}[shift={(8,0)}, scale={.9}, transform shape]
   \fill[shade,thick] (4,0) circle (.8);
   \node[ball,text width=.18cm,fill,color=red,label={[left=1.2cm]-30:$\mbox{\tiny $x_{v_e}+y_e$}$}] at ({.7*cos(-120)+4},{.7*sin(-120)}) (xv1) {};
   \node[ball,text width=.18cm,fill,color=red,label={[right=-.2cm]-30:$\mbox{\tiny $x_{v'_e}+y_e$}$}] at ({.7*cos(-60)+4},{.7*sin(-60)}) (xv2) {};
   \draw[-,ultra thick,color=red] (xv1.south) edge[bend right=100] (xv2.south);
   \coordinate (lre) at ($(xv1.south)!0.5!(xv2.south)$);
   \node[below=.15cm of lre, scale=.75] {$\displaystyle\left\{\nu_e-1\right\}$};
   \node[scale=1.125] at ($(4,0)+(.5,0)$) {$\displaystyle
	   				   \left.
   					    \begin{array}{l}
					     \phantom{ldots}\\
					     \phantom{ldots}\\
					     \phantom{ldots}
					    \end{array}
					   \right]$};
  \end{scope}
 \end{tikzpicture}
\end{equation}
with the operator $\hat{\mathcal{O}}_{e}$ being the differential operator appearing inside the square brackets of \eqref{eq:rrWFfin1}. Such a recursion relation, relates the contribution to wavefunction from a given graph $\mathcal{G}$ with internal states with order $\left\{\nu_e\right\}$ to the contribution to the wavefunction from the very same graph $\mathcal{G}$ but now with the order of the internal edges shifted by $-1$ one edge at a time.  Notice that, diagrammatically, a weight $\nu_e$ can be associated to the edge $e$ of $\mathcal{G}$, and thus the operator $\hat{\mathcal{O}}_{e}$ raises the weight of the edge $e$ by one. A straightforward manipulation of the recursion relation leads to following expression of the wavefunction $\psi_{\mathcal{G}}^{\mbox{\tiny $(\{\nu_e\})$}}$
\begin{equation}\eqlabel{eq:rrWFfin2}
 \psi_{\mathcal{G}}^{\mbox{\tiny $(\{\nu_e\})$}}\:=\:\sum_{e\in\mathcal{E}}\hat{\mathcal{O}}'_{e}\psi_{\mathcal{G}}^{\mbox{\tiny $(\nu_e-1,\{\nu_{\bar{e}}\})$}},
 \qquad
 \hat{\mathcal{O}}'_{e}\:\equiv\:
  \frac{2\left(\nu_e-\frac{3}{2}\right)}{\displaystyle\sum_{v\in\mathcal{V}} x_v}\left(\frac{\partial}{\partial x_{v_e}}+\frac{\partial}{\partial x_{v'_e}}\right)-
  \frac{\partial^2}{\partial x_{v_e} \partial x_{v'_e}}.
\end{equation}
Interestingly, for $\nu_e\,=\,3/2$ -- which corresponds to massless states for cosmologies with $\alpha\,=\,2/(d-1)$, including de Sitter in four dimensions, as well as to dS space-time ($\alpha\,=\,1$) with squared mass $m^2\,=\,(d^2-9)/4$ --, the operator $\hat{\mathcal{O}}'_{e}$ reduces to just the second derivative term.

A comment is now in order. Despite revealing the unsuspected connection among wavefunctions with different internal states, recursion relations gain power if they have an endpoint -- it is the seed the attention can be focused on -- or, in any case, basic terms in the recursion are known or computable. The order $\nu$ of a state can be either real or purely imaginary. If $\nu$ not only is real but it is also an integer or half-integer, the seed of the recursion relation can be taken to be $\nu\,=\,0$ an $\nu\,=\,1/2$ respectively. The latter corresponds to the massless scalar in flat space with time-dependent interaction, containing the conformally coupled scalar in FRW cosmologies. For de Sitter in four dimensions, the unitarity representations having $\nu\,\in\,\mathbb{R}_+$\footnote{Recall that because of the invariance of the propagator $G$ under the sign flip of the order $\nu$, we focused on $\nu\,\in\,\mathbb{R}_+$. Notice that however, the rescaling by of the propagator by $\left[(-y\eta_{v_e})(-y\eta_{v'_e})\right]^{1/2-\nu_e}$ breaks this symmetry. Luckily, the above formulas become valid also for $\nu\,\in\,\mathbb{R}_{-}$ -- up to the sign flip of $\nu$ -- if we simultaneously think about the propagator as rescaled by $\left[(-y\eta_{v_e})(-y\eta_{v'_e})\right]^{1/2+\nu_e}$, {\it i.e.} if we flip the sign of $\nu$ in the operator $\hat{\mathcal{W}}$, which maps the {\it integrand} into the actual wavefunction.} are such that, $\nu\,\in\,[0,\,3/2]$ and, consequently, the only states with $\nu\,\in\,\mathbb{Z}_+\,\mbox{ or }\,\mathbb{Z}_{\mbox{\tiny $+\frac{1}{2}$}}$ are given by $\nu\,=\,0,\,\frac{1}{2},\,1,\,\frac{3}{2}$. Notice further that which representations are unitary and which are not, and consequently the related values of $\nu$, changes with $d$ and $\alpha$, but the recursion relation is valid {\it irrespectively} of this: such an information is instead encoded into the operator $\hat{\mathcal{W}}$ mapping the {\it integrand} that the recursion relation is computing into the integrated wavefunction.

For arbitrary $\nu$, irrespectively of being real or purely imaginary, the recursion relations discussed do not have an endpoint. Further, for $\nu$ purely imaginary, the operator $\hat{\mathcal{O}}'_{e}$ would take the state out of the Hilbert space. In this case, one can take a perturbative approach, by considering the mass as a perturbative two point coupling around the point for which $\nu\,=\,0$. Importantly, we can even think of using this perturbative approach considering the full $\mu(\eta)$ as a perturbative coupling, {\it i.e.} introducing two-point corrections to the conformally coupled case: this would allow not to make any choice of the cosmology {\it at all}, while for the time being we have been restricting ourselves to a given class of $a(\eta)$'s. We postpone this discussion to future work, while in the rest of this paper we will focus on light states, with $\nu\,\in\,\mathbb{R}_+$.


\section{Edge-weighted graphs and the wavefunctions of the universe}\label{sec:EWG}

Let us restrict ourselves to the specific case $\nu_e\,=\,l_e+1/2$ ($l_e\,\in\,\mathbb{Z}_+$). Considering for all the $\nu_e$'s the value $\nu_e\,=\,1/2$ as the seed, then we can iterate \eqref{eq:rrWF3} to obtain
\begin{equation}\eqlabel{eq:rrWF4}
 \Delta G_{e}^{\mbox{\tiny $(\l_e)$}}\:=\:-i\left(\eta_{v_e}+\eta_{v'_e}\right)y_e^2\sum_{r_e\,=\,0}^{l_e-1}2\left(\l_e-r_e-\frac{1}{2}\right)\left(y_e^2\eta_{v_e}\eta_{v'_e}\right)^r G_{e}^{\mbox{\tiny $(l_e-r_e-1)$}} + 
					  \left(y_e^2\eta_{v_e}\eta_{v'_e}\right)^{l_e}\Delta G_{e}^{\mbox{\tiny $(0)$}}.
\end{equation}
Some comments are now in order. First, in the last term of \eqref{eq:rrWF4}, the total time translation operator acts on the propagator of a conformally coupled state: the time ordered terms get annihilated and, consequently, the non-time ordered part of $G_e$ contributes as $\Delta G_e\,=\,-2\,e^{iy_e(\eta_{v_e}+\eta_{v'_e})}$. Secondly, when $\Delta G_{e}^{\mbox{\tiny $(\l_e)$}}$ is inserted in \eqref{eq:rrWF2}, the factor of $\eta$ can be replaced by derivative acting on the external energies $x_{v_e}$ and $x_{v'_e}$, obtaining
\begin{equation}\eqlabel{eq:rrWFfin}
 \begin{split}
  \left(\sum_{v\in\mathcal{V}} x_v\right)\psi_{\mathcal{G}}^{\mbox{\tiny $(\{l_e\})$}}\:&=\:
   \sum_{e\in\mathcal{E}}\left(\frac{\partial}{\partial x_{v_e}}+\frac{\partial}{\partial x_{v'_e}}\right)\sum_{r_e=0}^{l_e-1}2\left(l_e-r_e-\frac{1}{2}\right)\left(-\frac{\partial^2}{\partial x_{v_e}\partial x_{v'_e}}\right)^{r_e}\psi_{\mathcal{G}}^{\mbox{\tiny $(\{l_{\bar{e}}\}; l_e-r_e-1)$}}\:+\\
  &+\:\sum_{e\in\mathcal{E}}\left(-\frac{\partial^2}{\partial x_{v_e}\partial x_{v'_e}}\right)^{l_e}\psi_{\mathcal{G}_{\mbox{\tiny L}}}^{\mbox{\tiny $(\{l_{\bar{e}}\})$}}\times \psi_{\mathcal{G}_{\mbox{\tiny R}}}^{\mbox{\tiny $(\{l_{\bar{e}}\})$}},
 \end{split}
\end{equation}
which can be schematically represented as
\begin{equation}\eqlabel{eq:rrWFpic}
 \begin{tikzpicture}[node distance=2cm, cross/.style={cross out, draw, inner sep=0pt, outer sep=0pt}, ball/.style = {circle, draw, align=center, anchor=north, inner sep=0}]
  \begin{scope}[shift={(2.125,0)}, scale={.9}, transform shape]
   \begin{scope}[shift={(-1,0)}, transform shape]
    \fill[shade,thick] (0,0) circle (.8);
    \node[text width=.18cm,color=black] at (0,0) (Fn) {$\displaystyle\psi_{\mbox{\tiny $\mathcal{G}$}}$};
    \node[below=.5cm of Fn] (lelhs) {$\displaystyle\left\{l_e\right\}$};
    \node[text width=.18cm,color=black] at (-2.8,-0.1) (sum) {$\displaystyle\left(\sum_{v\in\mathcal{V}}x_v\right)$};
    \node[ball,text width=.18cm,fill,color=black,label=left:$\mbox{\tiny $x_1$}$, scale=.75] at (-.8,0) (x1) {}; 
    \node[ball,text width=.18cm,fill,color=black,label=left:$\mbox{\tiny $x_2$}$, scale=.75] at ({.8*cos(150)},{.8*sin(150)}) (x2) {}; 
    \node[ball,text width=.18cm,fill,color=black,label={right:$\mbox{\tiny $x_{i-1}$}$}, scale=.75] at ({.8*cos(30)},{.8*sin(30)}) (xi1) {};
    \node[ball,text width=.18cm,fill,color=black,label={right:$\mbox{\tiny $x_i$}$}, scale=.75] at ({.8},{0}) (xi) {};   
    \node[ball,text width=.18cm,fill,color=black,label={right:$\mbox{\tiny $x_{i+1}$}$}, scale=.75] at ({.75*cos(-30)},{.75*sin(-30)}) (xi1b) {};  
    \node[ball,text width=.18cm,fill,color=black,label=left:$\mbox{\tiny $x_n$}$, scale=.75] at ({.8*cos(150)},{-.8*sin(150)}) (xn) {};   
   \end{scope}
   \node[color=black, scale=1.125] at (2.25,0) (eq) {$\displaystyle\quad=\:\sum_{e\in\mathcal{E}}\sum_{r=0}^{l_e-1}\hat{\mathcal{O}}_{r_e}^{\mbox{\tiny $(1)$}}
							\hspace{-.125cm}
							\left[
 							 \begin{array}{l}
							  \phantom{ldots}\\
							  \phantom{ldots}\\
							  \phantom{ldots}
							 \end{array}
 							\right.$};
   \fill[shade,thick] (4,0) circle (.8);  
   \node[text width=.18cm,color=black] at (4,0) (Fl) {$\displaystyle\psi_{\mathcal{L}}$};
   \node[ball,text width=.18cm,fill,color=black, scale=.75] at (3.2,0) (x1) {}; 
   \node[ball,text width=.18cm,fill,color=black, scale=.75] at ({.8*cos(150)+4},{.8*sin(150)}) (x2) {}; 
   \node[ball,text width=.18cm,fill,color=red,label={[label distance=.05mm]30:$\hspace{-.1cm}\mbox{\tiny $x_{v_e}+y_e$}$}] at ({.8+4},{0}) (xv) {};   
   \node[ball,text width=.18cm,fill,color=black, scale=.75] at ({.8*cos(150)+4},{-.8*sin(150)}) (xn) {};   
   \fill[shade,thick] (8,0) circle (.8);  
   \node[text width=.18cm,color=black] at (8,0) (Fr) {$\displaystyle\psi_{\mathcal{R}}$};
   \node[ball,text width=.18cm,fill,color=red,label={[left=1cm]-30:$\hspace{-.1cm}\mbox{\tiny $x_{v'_e}+y_e$}$}] at (7.2,0) (xv2) {}; 
   \node[ball,text width=.18cm,fill,color=black, scale=.75] at ({.8*cos(30)+8},{.8*sin(30)}) (x2) {}; 
   \node[ball,text width=.18cm,fill,color=black, scale=.75] at ({.8+8},{0}) (xt) {};   
   \node[ball,text width=.18cm,fill,color=black, scale=.75] at ({.8*cos(-30)+8},{.8*sin(-30)}) (xn) {};   
   \draw[-,ultra thick,color=red] (xv) -- (xv2);
   \coordinate (lre) at ($(xv)!0.5!(xv2)$);
   \node[below=.15cm of lre, scale=.75] {$\displaystyle\left\{l_e-r_e-1\right\}$};
   \node[color=black] at (9.25,-0.1) (eq2) {$\displaystyle\;+$};
  \end{scope}
  \begin{scope}[shift={(8,0)}, scale={.9}, transform shape]
   \fill[shade,thick] (4,0) circle (.8);
   \node[ball,text width=.18cm,fill,color=red,label={[left=1.2cm]-30:$\mbox{\tiny $x_{v_e}+y_e$}$}] at ({.7*cos(-120)+4},{.7*sin(-120)}) (xv1) {};
   \node[ball,text width=.18cm,fill,color=red,label={[right=-.2cm]-30:$\mbox{\tiny $x_{v'_e}+y_e$}$}] at ({.7*cos(-60)+4},{.7*sin(-60)}) (xv2) {};
   \draw[-,ultra thick,color=red] (xv1.south) edge[bend right=100] (xv2.south);
   \coordinate (lre) at ($(xv1.south)!0.5!(xv2.south)$);
   \node[below=.15cm of lre, scale=.75] {$\displaystyle\left\{l_e-r_e-1\right\}$};
   \node[scale=1.125] at ($(4,0)+(.5,0)$) {$\displaystyle
	   				   \left.
   					    \begin{array}{l}
					     \phantom{ldots}\\
					     \phantom{ldots}\\
					     \phantom{ldots}
					    \end{array}
					   \right]+$};
  \end{scope}
  \begin{scope}[shift={(2,-2)}, scale={.9}, transform shape]
   \node[color=black, scale=1.125] at (2.25,0) (eq) {$\displaystyle\quad+\:\sum_{e\in\mathcal{E}}\hat{\mathcal{O}}_{l_e}^{\mbox{\tiny $(2)$}}
							\hspace{-.125cm}
							\left[
 							 \begin{array}{l}
							  \phantom{ldots}\\
							  \phantom{ldots}\\
							  \phantom{ldots}
							 \end{array}
 							\right.$};
   \fill[shade,thick] (4,0) circle (.8);  
   \node[text width=.18cm,color=black] at (4,0) (Fl) {$\displaystyle\psi_{\mathcal{L}}$};
   \node[ball,text width=.18cm,fill,color=black, scale=.75] at (3.2,0) (x1) {}; 
   \node[ball,text width=.18cm,fill,color=black, scale=.75] at ({.8*cos(150)+4},{.8*sin(150)}) (x2) {}; 
   \node[ball,text width=.18cm,fill,color=red,label={[label distance=.05mm]30:$\hspace{-.1cm}\mbox{\tiny $x_{v_e}+y_e$}$}] at ({.8+4},{0}) (xv) {};   
   \node[ball,text width=.18cm,fill,color=black, scale=.75] at ({.8*cos(150)+4},{-.8*sin(150)}) (xn) {};   
   \fill[shade,thick] (8,0) circle (.8);  
   \node[text width=.18cm,color=black] at (8,0) (Fr) {$\displaystyle\psi_{\mathcal{R}}$};
   \node[ball,text width=.18cm,fill,color=red,label={[left=1cm]-30:$\hspace{-.1cm}\mbox{\tiny $x_{v'_e}+y_e$}$}] at (7.2,0) (xv2) {}; 
   \node[ball,text width=.18cm,fill,color=black, scale=.75] at ({.8*cos(30)+8},{.8*sin(30)}) (x2) {}; 
   \node[ball,text width=.18cm,fill,color=black, scale=.75] at ({.8+8},{0}) (xt) {};   
   \node[ball,text width=.18cm,fill,color=black, scale=.75] at ({.8*cos(-30)+8},{.8*sin(-30)}) (xn) {};   
   \draw[-,dashed,ultra thick,color=red] (xv) -- (xv2);
   \coordinate (lre) at ($(xv)!0.5!(xv2)$);
   \node[color=black] at (9.25,-0.1) (eq2) {$\displaystyle\;+$};
  \end{scope}
  \begin{scope}[shift={(8,-2)}, scale={.9}, transform shape]
   \fill[shade,thick] (4,0) circle (.8);
   \node[ball,text width=.18cm,fill,color=red,label={[left=1.2cm]-30:$\mbox{\tiny $x_{v_e}+y_e$}$}] at ({.7*cos(-120)+4},{.7*sin(-120)}) (xv1) {};
   \node[ball,text width=.18cm,fill,color=red,label={[right=-.2cm]-30:$\mbox{\tiny $x_{v'_e}+y_e$}$}] at ({.7*cos(-60)+4},{.7*sin(-60)}) (xv2) {};
   \draw[-,dashed,ultra thick,color=red] (xv1.south) edge[bend right=100] (xv2.south);
   \coordinate (lre) at ($(xv1.south)!0.5!(xv2.south)$);
   \node[scale=1.125] at ($(4,0)+(.5,0)$) {$\displaystyle
					   \left.
   					    \begin{array}{l}
					     \phantom{ldots}\\
					     \phantom{ldots}\\
					     \phantom{ldots}
					    \end{array}
					   \right]$};
  \end{scope}
 \end{tikzpicture}
\end{equation}
where the operators $\hat{\mathcal{O}}_{r_e}^{\mbox{\tiny $(1)$}}$ and $\hat{\mathcal{O}}_{l_e}^{\mbox{\tiny $(2)$}}$ are defined as 
\begin{equation}\eqlabel{eq:rrOps}
 \hat{\mathcal{O}}_{r_e}^{\mbox{\tiny $(1)$}}\,=\,2\left(l_e-r_e-\frac{1}{2}\right)\left(\frac{\partial}{\partial x_{v_e}}+\frac{\partial}{\partial x_{v'_e}}\right)
					  \left(-\frac{\partial^2}{\partial x_{v_e}\partial x_{v'_e}}\right)^{r_e},
 \qquad
 \hat{\mathcal{O}}_{l_e}^{\mbox{\tiny $(2)$}}\,=\,\left(-\frac{\partial^2}{\partial x_{v_e}\partial x_{v'_e}}\right)^{l_e}.
\end{equation}
The $\{l_e\}$ on the left-hand-side of \eqref{eq:rrWFpic} indicates that each edge $e$ has a weight $l_e\,\in\,\mathbb{Z}$, identifying the state that propagates on that edge; the solid red lines in the first line on the right-hand-side indicate that the corresponding edge $e$ has a lower weight $l_e-r_e-1$, while the dashed red lines in the second line indicate that the corresponding edge has been erased. Hence, the recursion relation in \eqref{eq:rrWFpic} -- and its functional expression in \eqref{eq:rrWFfin} -- states that the wavefunction $\psi_{\mbox{\tiny $\mathcal{G}$}}$, related to a graph $\mathcal{G}$ and having internal states labelled by the integers $l_e$ associated to the edges $e$ of $\mathcal{G}$, can be expressed in terms of wavefunctions related to the very same graph $\mathcal{G}$ but with lower masses $l_e-r_e-1$ as well as lower point and lower order wavefunctions.

For the concrete case of $l_e\,=\,1$ for any edge $e$ -- which corresponds to the case of all internal massless states in cosmologies with $\alpha\,=\,2/(d-1)$ as well as to states with squared mass $m^2\,=\,(d^2-9)/4$ in dS${}_{d+1}$--, then the order-raising operator $\hat{\mathcal{O}}'_{e}$ reduces just to the second derivative term in \eqref{eq:rrWFfin2} and the wavefunction can be expressed as
\begin{equation}\eqlabel{eq:WFnu1}
  \psi_{\mathcal{G}}^{\mbox{\tiny $(\{1\})$}}\:=\:n_e!\prod_{e\in\mathcal{E}}\left(-\frac{\partial^2}{\partial x_{v_e}\partial x_{v'_e}}\right)\psi_{\mathcal{G}}^{\mbox{\tiny $(\{0\})$}},
\end{equation}
and, consequently, the tree-level two-site graphs with weights $0$ and $1$ are related to each other via a two-dimensional wave equation with sources
\begin{equation}\eqlabel{eq:WF1we}
 \begin{tikzpicture}[overlay, node distance=2cm, cross/.style={cross out, draw, inner sep=0pt, outer sep=0pt}, ball/.style = {circle, draw, align=center, anchor=north, inner sep=0}]
 \coordinate [label=below:{\footnotesize $\displaystyle x_1$}] (v1) at (-3,-.125);
  \coordinate [label=below:{\footnotesize $\displaystyle x_2$}] (v2) at ($(v1)+(1cm,0)$);
  \draw[-,thick] (v1) -- node[above, scale=.75] {$1$} (v2); 
  \draw[fill,black] (v1) circle (2pt);
  \draw[fill,black] (v2) circle (2pt);  
  
  \node[right=.125cm of v2] (eq) {$\displaystyle=\,-\frac{\partial^2}{\partial x_1\partial x_2}$};
  \coordinate [label=below:{\footnotesize $\displaystyle x_1$}] (v1b) at ($(eq)+(1.25cm,0)$);
  \coordinate [label=below:{\footnotesize $\displaystyle x_2$}] (v2b) at ($(v1b)+(1cm,0)$);
  \draw[-,thick] (v1b) -- node[above, scale=.75] {$0$} (v2b); 
  \draw[fill,black] (v1b) circle (2pt);
  \draw[fill,black] (v2b) circle (2pt);
 \end{tikzpicture}
\end{equation}
\\

There is a further information that the recursion relation make manifest and which can be read off by just looking at the edge-weighted graphs: the order of the poles. Given a graph $\mathcal{G}$, while their locations $\mathfrak{p}(x,y)$ are associated to the subgraphs $\mathfrak{g}$ of $\mathcal{G}$ -- being the point where the sum of the energies which are external to $\mathfrak{g}$ vanishes --, their order $\mathfrak{o}(l)$ is fixed in terms of the weights $l_e$
\begin{equation}\eqlabel{eq:OrdPol}
 \mathfrak{p}(x,y)\:\equiv\:\sum_{v\in\mathfrak{g}}x_v+\sum_{e\in\mathcal{E}_{\mathfrak{g}}^{\mbox{\tiny ext}}}y_e
 \hspace{1.5cm}
 \mathfrak{o}(l)\:\equiv\:\sum_{e\in\mathcal{E}_{\mathfrak{g}}^{\mbox{\tiny int}}}2l_e+\sum_{e\in\mathcal{E}_{\mathfrak{g}}^{\mbox{\tiny ext}}}l_e+1
\end{equation}
where $\mathcal{E}_{\mathfrak{g}}^{\mbox{\tiny int}}$ and $\mathcal{E}_{\mathfrak{g}}^{\mbox{\tiny ext}}$ are the sets of edges which are respectively internal and external to $\mathfrak{g}$.


\subsection{Some examples}\label{subsec:ExEW}

It is useful to provide some explicit expressions of this new class of recursion relations. In the next two subsections we will discuss the two simplest examples in some detail: the two- three-site line graphs.


\subsubsection{Two-site line graph}\label{subsec:ExEw2sT}

The simplest example is given by the two-site line graph. In this case we can directly treat the case of a generic edge-weight $l$:
\begin{equation}\eqlabel{eq:WFl2pl}
 \begin{tikzpicture}[overlay, node distance=2cm, cross/.style={cross out, draw, inner sep=0pt, outer sep=0pt}, ball/.style = {circle, draw, align=center, anchor=north, inner sep=0}]
  \coordinate [label=below:{\footnotesize $\displaystyle x_1$}] (v1) at (-4,-.25);
  \coordinate [label=below:{\footnotesize $\displaystyle x_2$}] (v2) at ($(v1)+(1cm,0)$);
  \draw[-,thick] (v1) -- node[above, scale=.75] {$l$} (v2); 
  \draw[fill,black] (v1) circle (2pt);
  \draw[fill,black] (v2) circle (2pt);  
  
  \node[right=.125cm of v2] (eq) {$\displaystyle=\,\left[\frac{2(l-1)}{x_1+x_2}\left(\frac{\partial}{\partial x_1}+\frac{\partial}{\partial x_2}\right)-\frac{\partial^2}{\partial x_1\partial x_2}\right]$};
  \coordinate [label=below:{\footnotesize $\displaystyle x_1$}] (v1b) at ($(eq)+(3.25cm,0)$);
  \coordinate [label=below:{\footnotesize $\displaystyle x_2$}] (v2b) at ($(v1b)+(1cm,0)$);
  \draw[-,thick] (v1b) -- node[above, scale=.75] {$l-1$} (v2b); 
  \draw[fill,black] (v1b) circle (2pt);
  \draw[fill,black] (v2b) circle (2pt); 
 \end{tikzpicture}
\end{equation}
\\
As usual the singularities are given as the sum of the energies which are external to all the subgraphs. However, their are no longer simple poles, rather they are higher order poles, with the order given by \eqref{eq:OrdPol}:
\begin{equation}\eqlabel{eq:Sings}
 \begin{tikzpicture}[overlay, node distance=2cm, cross/.style={cross out, draw, inner sep=0pt, outer sep=0pt}, ball/.style = {circle, draw, align=center, anchor=north, inner sep=0}]
  \begin{scope}
   \coordinate [label=below:{\footnotesize $\displaystyle x_1$}] (v1) at (-5,0);
   \coordinate [label=below:{\footnotesize $\displaystyle x_2$}] (v2) at ($(v1)+(2,0)$);
   \draw[-,thick] (v1) -- node[above, scale=.75] {$l$} (v2); 
   \draw[-,thick] (v1) -- node[below, scale=.75] {$y$} (v2);
   \draw[fill,black] (v1) circle (2pt);
   \draw[fill,black] (v2) circle (2pt);
   \coordinate (ce) at ($(v1)!0.5!(v2)$);
   \draw[-,color=red!50!black] (ce) ellipse (1.25cm and .25cm);
   \node[below=.5cm of ce, scale=.875] (pol) {$\displaystyle x_1+x_2$};
   \node[below=.125cm of pol, scale=.875] (ord) {$\displaystyle\mathfrak{o}\,=\,2l+1$};
  \end{scope}
  \begin{scope}[shift={(4,0)}, transform shape]
   \coordinate [label=below:{\footnotesize $\displaystyle x_1$}] (v1) at (-5,0);
   \coordinate [label=below:{\footnotesize $\displaystyle x_2$}] (v2) at ($(v1)+(2,0)$);
   \draw[-,thick] (v1) -- node[above, scale=.75] {$l$} (v2); 
   \draw[-,thick] (v1) -- node[below, scale=.75] {$y$} (v2);
   \draw[fill,black] (v1) circle (2pt);
   \draw[fill,black] (v2) circle (2pt);
   \coordinate (ce) at ($(v1)!0.5!(v2)$);
   \draw[-,color=red!50!black] (v1) circle (4pt);
   \node[below=.5cm of ce, scale=.875] (pol) {$\displaystyle x_1+y$}; 
   \node[below=.125cm of pol, scale=.875] (ord) {$\displaystyle\mathfrak{o}\,=\,l+1$};     
  \end{scope}
  \begin{scope}[shift={(8,0)}, transform shape]
   \coordinate [label=below:{\footnotesize $\displaystyle x_1$}] (v1) at (-5,0);
   \coordinate [label=below:{\footnotesize $\displaystyle x_2$}] (v2) at ($(v1)+(2,0)$);
   \draw[-,thick] (v1) -- node[above, scale=.75] {$l$} (v2); 
   \draw[-,thick] (v1) -- node[below, scale=.75] {$y$} (v2);
   \draw[fill,black] (v1) circle (2pt);
   \draw[fill,black] (v2) circle (2pt);
   \coordinate (ce) at ($(v1)!0.5!(v2)$);
   \draw[-,color=red!50!black] (v2) circle (4pt);
   \node[below=.5cm of ce, scale=.875] (pol) {$\displaystyle y+x_2$};   
   \node[below=.125cm of pol, scale=.875] (ord) {$\displaystyle\mathfrak{o}\,=\,l+1$};   
  \end{scope}
 \end{tikzpicture} 
\end{equation}
\\

\noindent
In order to compute the wavefunction of the universe, we can iterate the recursion relation \eqref{eq:WFl2pl} until to reach the seed $l\,=\,0$ of the recursion, which is fixed by simple combinatorial rules. However, we can also make the following observation: from the order of the operator in \eqref{eq:WFl2pl}, it is straightforward to see that the contribution to the wavefunction with an internal $l$ state is a rational function of overall degree $\delta_{\psi}\,=\,-(2l+3)$. We can thus write the function associated to the two-site line graph as
\begin{equation}\eqlabel{eq:WFl2sl}
\begin{tikzpicture}[overlay, node distance=2cm, cross/.style={cross out, draw, inner sep=0pt, outer sep=0pt}, ball/.style = {circle, draw, align=center, anchor=north, inner sep=0}]
  \begin{scope}
   \coordinate [label=below:{\footnotesize $\displaystyle x_1$}] (v1) at (-5,0);
   \coordinate [label=below:{\footnotesize $\displaystyle x_2$}] (v2) at ($(v1)+(2,0)$);
   \draw[-,thick] (v1) -- node[above, scale=.75] {$l$} (v2); 
   \draw[-,thick] (v1) -- node[below, scale=.75] {$y$} (v2);
   \draw[fill,black] (v1) circle (2pt);
   \draw[fill,black] (v2) circle (2pt);
' 
    \node[right=.125cm of v2] (eq) {$\displaystyle=\,\sum_{r_1=0}^{l-1}\sum_{r_2=0}^{l-1}\frac{a_{r_1 r_2}^{\mbox{\tiny $(l)$}}}{(x_1+x_2)^{2l+1-r_1-r_2}(x_1+y)^{r_1+1}(y+x_2)^{r_2+1}}$};
  \end{scope}
 \end{tikzpicture}
\end{equation}
\\
The differential recursion relation \eqref{eq:WFl2pl} then translates into an algebraic one for the coefficients $a_{r_1 r_2}$:
\begin{equation}\eqlabel{eq:WFarr}
 \begin{split}
  &a_{r_1 r_2}^{\mbox{\tiny $(l)$}}\:=\:-(2l-1-r_1-r_2)[2(3l-2)-r_1-r_2]a_{r_1 r_2}^{\mbox{\tiny $(l-1)$}}-r_1[2(2l-1)-r_1-r_2]a_{r_1-1,r_2}^{\mbox{\tiny $(l-1)$}}-\\
  &\phantom{a_{r_1 r_2}^{\mbox{\tiny $(l)$}}\:=\:}-r_2[2(2l-1)-r_1-r_2]a_{r_1,r_2-1}^{\mbox{\tiny $(l-1)$}}-r_1r_2a_{r_1-1,r_2-1}^{\mbox{\tiny $(l-1)$}},
 \end{split}
\end{equation}
with $a_{r_1 r_2}^{\mbox{\tiny $(l)$}}\,=\,0$ for $r_j\,>\,l$ and $r_j\,<\,0$ ($j\,=\,1,\,2$). Notice that the expression \eqref{eq:WFl2sl} resemble a Laurent expansion of the wavefunction in $x_1+x_2$, making some of the physical content manifest: the coefficient of the term with highest order for $x_1+x_2$ is related to the flat-space scattering amplitude, and becomes proportional to it, with the proportionality coefficient given by $a_{00}^{\mbox{\tiny $(l)$}}$, on the sheet $x_1+x_2\,=\,0$. Such a coefficient, together with $a_{0l}^{\mbox{\tiny $(l)$}}$, $a_{l0}^{\mbox{\tiny $(l)$}}$ and $a_{ll}^{\mbox{\tiny $(l)$}}$, can be written in closed form
{\resizebox{\textwidth}{!}{
$\displaystyle
 a_{00}^{\mbox{\tiny $(l)$}}\:=\:\prod_{k=0}^{l-1}(-2)(2l-2k-1)(3l-3k-2),\quad
 a_{0l}^{\mbox{\tiny $(l)$}}\:=\:\prod_{k=0}^{l-1}\left[-(l-k)(3(l-k)-2)\right]\:=\:a_{l0}^{\mbox{\tiny $(l-1)$}},\quad
 a_{ll}^{\mbox{\tiny $(l)$}}\:=\:\prod_{k=1}^{l-1}[-(l-k)^2].
$
}}
Actually, any of the other terms can also be related to (derivative of) the high energy limit flat-space amplitude. This can be conveniently seen by introducing the variables $x_T\,\equiv\,x_1+x_2$, $x_L\,\equiv\,x_1+y$, $x_R\,\equiv\,y+x_2$

\begin{equation}\eqlabel{eq:WFl2sl2}
\begin{tikzpicture}[overlay, node distance=2cm, cross/.style={cross out, draw, inner sep=0pt, outer sep=0pt}, ball/.style = {circle, draw, align=center, anchor=north, inner sep=0}]
  \begin{scope}
   \coordinate [label=below:{\footnotesize $\displaystyle x_1$}] (v1) at (-6,0);
   \coordinate [label=below:{\footnotesize $\displaystyle x_2$}] (v2) at ($(v1)+(2,0)$);
   \draw[-,thick] (v1) -- node[above, scale=.75] {$l$} (v2); 
   \draw[-,thick] (v1) -- node[below, scale=.75] {$y$} (v2);
   \draw[fill,black] (v1) circle (2pt);
   \draw[fill,black] (v2) circle (2pt);
' 
    \node[right=.125cm of v2] (eq) {$\displaystyle=\,\sum_{r_1=0}^{l-1}\sum_{r_2=0}^{l-1}\frac{a_{r_1 r_2}^{\mbox{\tiny $(l)$}}}{r_1!r_2!}
		\frac{1}{x_T^{2l+1-r_1-r_1}}\left(\frac{\partial}{\partial x_L}\right)^{r_1}\psi_1(x_L)\left(\frac{\partial}{\partial x_R}\right)^{r_2}\psi_1(x_R)$};
  \end{scope}
 \end{tikzpicture}
\end{equation}
\\
where $\psi_1(x)\,=\,x^{-1}$ is nothing but the one-site wavefunction of the universe, {\it i.e.} the wavefunction for a contact interaction. As $x_T\,\longrightarrow\,0$, the order $(2l+1-u)$-th coefficient $\psi_{T}^{\mbox{\tiny $(2l+1-u)$}}$ (with $u\,\equiv\,r_1+r_2\,=\,$ fixed, $u\,\in\,[0,2l]$) can be written as
\begin{equation}\eqlabel{eq:WFl2sl3}
 \psi_{T}^{\mbox{\tiny $(2l+1-u)$}}\:=\:\sum_{r_1=0}^u\sum_{r_2=0}^u\frac{a_{r_1 r_2}^{\mbox{\tiny $(l)$}}}{r_1!r_2!}\left(\frac{\partial}{\partial x_L}\right)^{r_1}\left(\frac{\partial}{\partial x_R}\right)^{r_2}\mathcal{A}_2
\end{equation}
$\mathcal{A}_2$ being the (high energy limit of the) scattering amplitude. As $x_L\,\longrightarrow\,0$ we can also write all the coefficients in the Laurent expansion around such a point in terms of lower-point wavefunctions and, equivalently, in terms of the flat-space scattering amplitude:
\begin{equation}\eqlabel{eq:WFl2sl4}
 \begin{split}
  \psi_L^{\mbox{\tiny $(r_1)$}}\:&=\:(-1)^{2l-r_1}\sum_{r_2=0}^{l-1} a_{r_1 r_2}^{\mbox{\tiny $(l)$}}
   \left(\frac{\partial}{\partial x_{-}}\right)^{r_1}  \left(\frac{\partial}{\partial x_{+}}\right)^{r_2}
   \frac{1}{2y}\left[\psi_1(x_-)-\psi_1(x_+)\right]\:\equiv\\
  &\equiv\:(-1)^{2l-r_1+1}\sum_{r_2=0}^{l-1} a_{r_1 r_2}^{\mbox{\tiny $(l)$}}
   \left(\frac{\partial}{\partial x_{-}}\right)^{r_1}  \left(\frac{\partial}{\partial x_{+}}\right)^{r_2}
   \mathcal{A}_2  
 \end{split}
\end{equation}
where $x_{\mp}\,\equiv\,x_2\mp y$ (with $x_-\,\equiv\,\left.x_T\right|_{x_L=0}$). A similar formula can be obtained for the coefficients of the Laurent expansion as $x_R\,\longrightarrow\,0$. These formulas make manifest how the main physical information encoded is the (high energy limit of the) flat-space scattering amplitudes. In a sense, with the recursion relation \eqref{eq:WFl2pl} at hand (and, more generally, the recursion relations \eqref{eq:rrWFfin2} and \eqref{eq:rrWFpic} for higher point processes), this is not a big surprise: it relates the graph with the edge-weight $l$ to the one with edge-weight $0$ which was already proven to be reconstructible from the knowledge of the flat-space scattering amplitude and the requirement of Bunch-Davies condition (which translates into requiring the final answer to be function on sum of energies only) \cite{Benincasa:2018ssx}.


\subsubsection{Three-site line graph}\label{subsec:ExEw3sT}

Let us now consider the next-to-simplest case of the three-site line graph with edge-weights $l_{12}\,=\,0$ and $l_{23}\,=\,1$\footnote{The indices $ij$ in $l_{ij}$ indicates the labels of the sites that the edge the weight is associated to connects.}. Then, the recursion relation allows us to write such a graph as a differential operator acting on the same graphs but with all the edge-weights equal to zero:
\vspace{-.125cm}
\begin{equation}\eqlabel{eq:WF3sl}
 \begin{tikzpicture}[overlay, node distance=2cm, cross/.style={cross out, draw, inner sep=0pt, outer sep=0pt}, ball/.style = {circle, draw, align=center, anchor=north, inner sep=0}]
  \coordinate [label=below:{\footnotesize $\displaystyle x_1$}] (v1) at (-4,-.25);
  \coordinate [label=below:{\footnotesize $\displaystyle x_2$}] (v2) at ($(v1)+(1.5cm,0)$);
  \coordinate [label=below:{\footnotesize $\displaystyle x_3$}] (v3) at ($(v2)+(1.5cm,0)$);
  \draw[-,thick] (v1) -- node[above, scale=.75] {$0$} (v2); 
  \draw[-,thick] (v2) -- node[above, scale=.75] {$1$} (v3);   
  \draw[fill,black] (v1) circle (2pt);
  \draw[fill,black] (v2) circle (2pt); 
  \draw[fill,black] (v3) circle (2pt);   
  
  \node[right=.125cm of v3] (eq) {$\displaystyle=\,-\frac{\partial^2}{\partial x_2\partial x_3}$};
  \coordinate [label=below:{\footnotesize $\displaystyle x_1$}] (v1b) at ($(eq)+(1.25cm,0)$);
  \coordinate [label=below:{\footnotesize $\displaystyle x_2$}] (v2b) at ($(v1b)+(1.5cm,0)$);
  \coordinate [label=below:{\footnotesize $\displaystyle x_3$}] (v3b) at ($(v2b)+(1.5cm,0)$);
  \draw[-,thick] (v1b) -- node[above, scale=.75] {$0$} (v2b); 
  \draw[-,thick] (v2b) -- node[above, scale=.75] {$0$} (v3b);   
  \draw[fill,black] (v1b) circle (2pt);
  \draw[fill,black] (v2b) circle (2pt); 
  \draw[fill,black] (v3b) circle (2pt); 
 \end{tikzpicture}
\end{equation}
\\

\noindent
Indeed we know how to compute the $0$-edge-weight graph on the right-hand-side thanks to the combinatorial rules provided in \cite{Arkani-Hamed:2017fdk}, so we can just use them and apply the differential operator in \eqref{eq:WF3sl} -- which is the way to go if we were merely interested in the final answer. However, it is instructive to read off some of its features, such as the order of the poles 
$\displaystyle\mathfrak{o}\:\equiv\:\sum_{e\in\mathcal{E}_{\mathfrak{g}}^{\mbox{\tiny int}}}2l_e+\sum_{e\in\mathcal{E}_{\mathfrak{g}}^{\mbox{\tiny ext}}}l_e+1\mbox{ :}$

\begin{equation*}
 \begin{tikzpicture}[overlay, node distance=2cm, cross/.style={cross out, draw, inner sep=0pt, outer sep=0pt}, ball/.style = {circle, draw, align=center, anchor=north, inner sep=0}]
  \begin{scope}
   \coordinate [label=below:{\footnotesize $\displaystyle x_1$}] (v1) at (-6,-.25);
   \coordinate [label=below:{\footnotesize $\displaystyle x_2$}] (v2) at ($(v1)+(1.25cm,0)$);
   \coordinate [label=below:{\footnotesize $\displaystyle x_3$}] (v3) at ($(v2)+(1.25cm,0)$);
   \draw[-,thick] (v1) -- node[above, scale=.75] {$0$} (v2); 
   \draw[-,thick] (v1) -- node[below, scale=.75] {$y_{12}$} (v2);
   \draw[-,thick] (v2) -- node[above, scale=.75] {$1$} (v3);   
   \draw[-,thick] (v2) -- node[below, scale=.75] {$y_{23}$} (v3);   
   \draw[fill,black] (v1) circle (2pt);
   \draw[fill,black] (v2) circle (2pt); 
   \draw[fill,black] (v3) circle (2pt);   

   \draw[thick, color=red!50!black] (v2) ellipse (1.5cm and .25cm);
   \node[below=.5cm of v2, scale=.825] (xT) {$\displaystyle x_1+x_2+x_3,\quad \mathfrak{o}\,=\,3$};
  \end{scope}
  \begin{scope}[shift={(5,0)}, transform shape]
   \coordinate [label=below:{\footnotesize $\displaystyle x_1$}] (v1) at (-6,-.25);
   \coordinate [label=below:{\footnotesize $\displaystyle x_2$}] (v2) at ($(v1)+(1.25cm,0)$);
   \coordinate [label=below:{\footnotesize $\displaystyle x_3$}] (v3) at ($(v2)+(1.25cm,0)$);
   \draw[-,thick] (v1) -- node[above, scale=.75] {$0$} (v2); 
   \draw[-,thick] (v1) -- node[below, scale=.75] {$y_{12}$} (v2);
   \draw[-,thick] (v2) -- node[above, scale=.75] {$1$} (v3);   
   \draw[-,thick] (v2) -- node[below, scale=.75] {$y_{23}$} (v3);   
   \draw[fill,black] (v1) circle (2pt);
   \draw[fill,black] (v2) circle (2pt); 
   \draw[fill,black] (v3) circle (2pt);   

   \coordinate (ce) at ($(v1)!0.5!(v2)$);
   \draw[thick, color=red!50!black] (ce) ellipse (.75cm and .25cm);
   \node[below=.5cm of v2, scale=.825] (xT) {$\displaystyle x_1+x_2+y_{23},\quad \mathfrak{o}\,=\,2$};
  \end{scope}
  \begin{scope}[shift={(10,0)}, transform shape]
   \coordinate [label=below:{\footnotesize $\displaystyle x_1$}] (v1) at (-6,-.25);
   \coordinate [label=below:{\footnotesize $\displaystyle x_2$}] (v2) at ($(v1)+(1.25cm,0)$);
   \coordinate [label=below:{\footnotesize $\displaystyle x_3$}] (v3) at ($(v2)+(1.25cm,0)$);
   \draw[-,thick] (v1) -- node[above, scale=.75] {$0$} (v2); 
   \draw[-,thick] (v1) -- node[below, scale=.75] {$y_{12}$} (v2);
   \draw[-,thick] (v2) -- node[above, scale=.75] {$1$} (v3);   
   \draw[-,thick] (v2) -- node[below, scale=.75] {$y_{23}$} (v3);   
   \draw[fill,black] (v1) circle (2pt);
   \draw[fill,black] (v2) circle (2pt); 
   \draw[fill,black] (v3) circle (2pt);   

   \coordinate (ce) at ($(v2)!0.5!(v3)$);
   \draw[thick, color=red!50!black] (ce) ellipse (.75cm and .25cm);
   \node[below=.5cm of v2, scale=.825] (xT) {$\displaystyle x_1+x_2+y_{23},\quad \mathfrak{o}\,=\,3$};
  \end{scope}
  \begin{scope}[shift={(0,-1.75)}, transform shape]
   \coordinate [label=below:{\footnotesize $\displaystyle x_1$}] (v1) at (-6,-.25);
   \coordinate [label=below:{\footnotesize $\displaystyle x_2$}] (v2) at ($(v1)+(1.25cm,0)$);
   \coordinate [label=below:{\footnotesize $\displaystyle x_3$}] (v3) at ($(v2)+(1.25cm,0)$);
   \draw[-,thick] (v1) -- node[above, scale=.75] {$0$} (v2); 
   \draw[-,thick] (v1) -- node[below, scale=.75] {$y_{12}$} (v2);
   \draw[-,thick] (v2) -- node[above, scale=.75] {$1$} (v3);   
   \draw[-,thick] (v2) -- node[below, scale=.75] {$y_{23}$} (v3);   
   \draw[fill,black] (v1) circle (2pt);
   \draw[fill,black] (v2) circle (2pt); 
   \draw[fill,black] (v3) circle (2pt);   

   \draw[thick, color=red!50!black] (v1) circle (4pt);
   \node[below=.5cm of v2, scale=.825] (xT) {$\displaystyle x_1+y_{12},\quad \mathfrak{o}\,=\,1$};
  \end{scope}
  \begin{scope}[shift={(5,-1.75)}, transform shape]
   \coordinate [label=below:{\footnotesize $\displaystyle x_1$}] (v1) at (-6,-.25);
   \coordinate [label=below:{\footnotesize $\displaystyle x_2$}] (v2) at ($(v1)+(1.25cm,0)$);
   \coordinate [label=below:{\footnotesize $\displaystyle x_3$}] (v3) at ($(v2)+(1.25cm,0)$);
   \draw[-,thick] (v1) -- node[above, scale=.75] {$0$} (v2); 
   \draw[-,thick] (v1) -- node[below, scale=.75] {$y_{12}$} (v2);
   \draw[-,thick] (v2) -- node[above, scale=.75] {$1$} (v3);   
   \draw[-,thick] (v2) -- node[below, scale=.75] {$y_{23}$} (v3);   
   \draw[fill,black] (v1) circle (2pt);
   \draw[fill,black] (v2) circle (2pt); 
   \draw[fill,black] (v3) circle (2pt);   

   \draw[thick, color=red!50!black] (v2) circle (4pt);
   \node[below=.5cm of v2, scale=.825] (xT) {$\displaystyle y_{12}+x_2+y_{23},\quad \mathfrak{o}\,=\,2$};
  \end{scope}
  \begin{scope}[shift={(10,-1.75)}, transform shape]
   \coordinate [label=below:{\footnotesize $\displaystyle x_1$}] (v1) at (-6,-.25);
   \coordinate [label=below:{\footnotesize $\displaystyle x_2$}] (v2) at ($(v1)+(1.25cm,0)$);
   \coordinate [label=below:{\footnotesize $\displaystyle x_3$}] (v3) at ($(v2)+(1.25cm,0)$);
   \draw[-,thick] (v1) -- node[above, scale=.75] {$0$} (v2); 
   \draw[-,thick] (v1) -- node[below, scale=.75] {$y_{12}$} (v2);
   \draw[-,thick] (v2) -- node[above, scale=.75] {$1$} (v3);   
   \draw[-,thick] (v2) -- node[below, scale=.75] {$y_{23}$} (v3);   
   \draw[fill,black] (v1) circle (2pt);
   \draw[fill,black] (v2) circle (2pt); 
   \draw[fill,black] (v3) circle (2pt);   

   \draw[thick, color=red!50!black] (v3) circle (4pt);
   \node[below=.5cm of v2, scale=.825] (xT) {$\displaystyle y_{23}+x_3,\quad \mathfrak{o}\,=\,2$};
  \end{scope}
 \end{tikzpicture}
\end{equation*}
\\
\vspace{1.5cm}

\noindent
It can also determine the coefficient of the Laurent expansion of our edge-weighted graph as any of the singularity is approached, in terms of the residues of the wavefunction represented by the zero-edge-weighted graph. For example, as the total energy pole location is approached:

\begin{equation*}
 \begin{tikzpicture}[overlay, node distance=2cm, cross/.style={cross out, draw, inner sep=0pt, outer sep=0pt}, ball/.style = {circle, draw, align=center, anchor=north, inner sep=0}]
  \begin{scope}
   \coordinate [label=below:{\footnotesize $\displaystyle x_1$}] (v1) at (-6.5,-.25);
   \coordinate [label=below:{\footnotesize $\displaystyle x_2$}] (v2) at ($(v1)+(1.25cm,0)$);
   \coordinate [label=below:{\footnotesize $\displaystyle x_3$}] (v3) at ($(v2)+(1.25cm,0)$);
   \draw[-,thick] (v1) -- node[above, scale=.75] {$0$} (v2); 
   \draw[-,thick] (v1) -- node[below, scale=.75] {$y_{12}$} (v2);
   \draw[-,thick] (v2) -- node[above, scale=.75] {$1$} (v3);   
   \draw[-,thick] (v2) -- node[below, scale=.75] {$y_{23}$} (v3);   
   \draw[fill,black] (v1) circle (2pt);
   \draw[fill,black] (v2) circle (2pt); 
   \draw[fill,black] (v3) circle (2pt);   

   \draw[thick, color=red!50!black] (v2) ellipse (1.5cm and .25cm);   
   \node[right=.25cm of v3] (extT) {$\displaystyle\sim\frac{-2\mathcal{A}_3}{(x_1+x_2+x_3)^3}+\frac{1}{(x_1+x_2+x_3)^2}\left(\frac{\partial}{\partial y_{12}}+\frac{\partial}{\partial x_2}+\frac{\partial}{\partial y_{23}}\right)\mathcal{A}_3+\ldots$};
  \end{scope}
  \begin{scope}[shift={(0,-1.75)}, transform shape]
   \coordinate [label=below:{\footnotesize $\displaystyle x_1$}] (v1) at (-6.5,-.25);
   \coordinate [label=below:{\footnotesize $\displaystyle x_2$}] (v2) at ($(v1)+(1.25cm,0)$);
   \coordinate [label=below:{\footnotesize $\displaystyle x_3$}] (v3) at ($(v2)+(1.25cm,0)$);
   \draw[-,thick] (v1) -- node[above, scale=.75] {$0$} (v2); 
   \draw[-,thick] (v1) -- node[below, scale=.75] {$y_{12}$} (v2);
   \draw[-,thick] (v2) -- node[above, scale=.75] {$1$} (v3);   
   \draw[-,thick] (v2) -- node[below, scale=.75] {$y_{23}$} (v3);   
   \draw[fill,black] (v1) circle (2pt);
   \draw[fill,black] (v2) circle (2pt); 
   \draw[fill,black] (v3) circle (2pt);   

   \coordinate (ce) at ($(v1)!0.5!(v2)$);
   \draw[thick, color=red!50!black] (ce) ellipse (.75cm and .25cm);
   \node[right=.25cm of v3] (extL1) {$\displaystyle\sim\frac{1}{(x_1+x_2+y_{23})^2}\frac{\partial}{\partial x_3}\left[\mathcal{A}_2\times\frac{1}{2y_{23}}\left[\psi_1(x_3-y_{23})-\psi_1(x_3+y_{23})\right]\right]+\ldots$};
   \node[below=.25cm of extL1] (extL2) {$\displaystyle\hspace{-4.75cm}\equiv\frac{1}{(x_1+x_2+y_{23})^2}\frac{\partial}{\partial x_3}\left[-\mathcal{A}_2\times\mathcal{A}_2\right]+\ldots$};   
  \end{scope}
 \end{tikzpicture}
\end{equation*}
\\
\vspace{2.5cm}

\noindent
Notice that the operator in the last line acts just on the right-factor, so that the coefficient of $(x_1+x_2+y_{23})^{-2}$ really factorises as $\mathcal{A}_2\times\partial_{x_3}\mathcal{A}_2$.


\section{Perturbative mass}\label{sec:Pert}

The discussion so far has been restricted to a subclass of models identified by $\mu(\eta)^2\,=\,\mu_{\alpha}^2\eta^{-2}$ with $\mu_{\alpha}^2\,=\,-l(l+1)$ ($l\,\in\,\mathbb{Z}$), which includes states $m\,=\,0$ for cosmologies $a(\eta)\,\propto\,(-\eta)^{\alpha}$ for certain values of $\alpha\,=\,\alpha(l,d)$, as well as states with $m\,=\,m(l,d)$ (which can be non-zero) in $dS_{d+1}$. In these cases the recursion relations \eqref{eq:rrWfin1pic} and \eqref{eq:rrWFpic} have a natural seed, given by the massless scalar with time-dependent polynomial interactions in flat-space -- which contains the conformally coupled scalar in FRW cosmologies. In this section, we will extend such an analysis by considering the time-dependent mass perturbatively. This can be done in two ways: it is possible to consider $\mu^2(\eta)\,=\,\lambda_{2}(\eta)$, or $\mu(\eta)\,=\,\lambda_2(\eta)+\mu_{\alpha}\eta^{-2}$, with $\lambda_{2}$ being the dimensionless small expansion parameter. While in the first case, the free states are massless particles in flat space-time and the analysis applies to arbitrary $a(\eta)$, the second choice holds for cosmologies $a(\eta)\,=\,(-\eta)^{-\alpha}$, with $\lambda_2(\eta)\,\equiv\,m^2(-\eta)^{-2\alpha}$ and $\mu_{\alpha}\,=\,-l(l+1)$ for $\alpha\,=\,2l/(d-1)$: in this latter class of cases, the free states are labelled by $\nu\,=\,l+1/2$. Thus, performing a perturbative analysis in these two cases is equivalent to do perturbation theory around two different free propagation, which, for the above choices of cosmologies and parameters, corresponds to the scalar being conformally or minimally coupled respectively. However, as we just saw in the previous sections, the integrand defined by considering all the couplings in Fourier space satisfies recursion relations, which relate the $l\,\neq\,0$ states to the $l\,=\,0$ one, {\it i.e.} the massless free propagation. Hence, we will begin with analysing the case of a massless particles in flat-space with time-dependent couplings, including a two point coupling $\lambda_2(\eta)$, which will be also treated in its Fourier space
\begin{equation}\eqlabel{eq:2ptcFS}
 \lambda_2(\eta)\:=\:\int_{-\infty}^{+\infty}d\omega\,e^{i\omega\eta}\tilde{\lambda}_2(\omega).
\end{equation}
Thus, a generic contribution to the perturbative wavefunction of the universe can be represented via Feynman diagrams which now allow for two-point vertices, and its general form can be written as 
\begin{equation}\eqlabel{eq:PWF2ma}
 \tilde{\psi}_{\mathcal{G}_{\circ}}\:=\:\int_{-\infty}^{+\infty}\prod_{v\in\mathcal{V}}\left[dx_{v}\tilde{\lambda}_{k_v}(x_v-X_v)\right]
					\int_{-\infty}^{+\infty}\prod_{w\in\mathcal{V}_2}\left[d\omega_w\tilde{\lambda}_2(\omega_w)\right]\psi_{\mathcal{G}_{\circ}}(\{\omega_w;\,x_v,\,y_e\})
\end{equation}
where
\begin{equation}\eqlabel{eq:PWF2mb}
 \psi_{\mathcal{G}_{\circ}}(\{\omega_w,\varepsilon_v\})\:=\:\int_{-\infty}^0\prod_{v\in\mathcal{V}}\left[d\eta_v\,e^{ix_v\eta_v}\right]
							    \int_{-\infty}^0\prod_{w\in\mathcal{V}_2}\left[d\eta_w\,e^{i\omega_w\eta_w}\right]
							    \prod_{e\in\mathcal{E}}G_e(y_e;\,\eta_{v_e},\eta'_{v_e}),
\end{equation}
$\mathcal{V}_2$ being the set of two point vertices. Notice that all the two-point sites in formula \eqref{eq:PWF2mb} connect two edges of the graph $\mathcal{G}_{\circ}$: this means that such a formula considers just mass corrections to the internal propagators and not on the external states. Indeed nothing forbids to consider also (or only) mass corrections on the external states: the expression for the wavefunction integrand \eqref{eq:PWF2mb} is structurally the same with the extra exponential terms having argument $\hat{x}_w\,\equiv\,E_w+\omega_w$, $E$ being the energy of the state which is receiving the mass correction. Further, the arguments of the relevant two-point couplings in \eqref{eq:PWF2ma} gets shifted to $\hat{x}_w-E_w$, with the integration over $\hat{x}_w$. We will comment about this case separately, while, for the time being, we will focus on \eqref{eq:PWF2mb}.

Graphically, the two point couplings can be identified by a white site with valence $2$
\begin{figure}[H]
 \centering
 \begin{tikzpicture}[line join = round, line cap = round, ball/.style = {circle, draw, align=center, anchor=north, inner sep=0}, 
                     axis/.style={very thick, ->, >=stealth'}, pile/.style={thick, ->, >=stealth', shorten <=2pt, shorten>=2pt}, every node/.style={color=black}]
  \begin{scope}[shift={(-3,0)}, transform shape]
   \coordinate (w1) at (0,0);
   \coordinate (x2) at ($(w1)+(1.5,0)$);
   \draw[-,thick] (w1) -- (x2);
   \draw[fill=white] (w1) circle (2pt);
   \draw[fill,black] (x2) circle (2pt);
  \end{scope}
  \begin{scope}
   \coordinate (x1) at (0,0);
   \coordinate (x2) at ($(x1)+(1.5,0)$);
   \draw[-,thick] (x1) -- (x2);
   \draw[fill,black] (x1) circle (2pt);
   \draw[fill,black] (x2) circle (2pt);
  \end{scope}
  \begin{scope}[shift={(3,0)}, transform shape]
   \coordinate (x1) at (0,0);
   \coordinate (w1) at ($(x1)+(1,0)$);
   \coordinate (x2) at ($(x1)+(2,0)$);
   \draw[-,thick] (x1) -- (x2);
   \draw[fill,black] (x1) circle (2pt);
   \draw[fill=white] (w1) circle (2pt);
   \draw[fill,black] (x2) circle (2pt);
  \end{scope}
  \begin{scope}[shift={(6.5,0)}, transform shape]
   \coordinate (x1) at (0,0);
   \coordinate (w1) at ($(x1)+(1,0)$);
   \coordinate (w2) at ($(w1)+(1,0)$);
   \coordinate (x2) at ($(x1)+(3,0)$);
   \draw[-,thick] (x1) -- (x2);
   \draw[fill,black] (x1) circle (2pt);
   \draw[fill=white] (w1) circle (2pt);
   \draw[fill=white] (w2) circle (2pt);
   \draw[fill,black] (x2) circle (2pt);
  \end{scope}
 \end{tikzpicture}
\end{figure}

\noindent
with the first graph representing a mass correction to an external state, while the third and the fourth represent mass corrections to the two-site graph (appearing as second graph). Equivalently, the graphs above can be thought of as subgraphs so that they will represent mass correction to some internal (or external, in the case of the first graph) state in a more complicated graph. Importantly, two edges connected via a white site have the same $y$ associated to it because of spatial momentum conservation. This introduce a novel feature in the function form of the integrand: it is bounded to develop higher poles, which become explicit in some of its residues.

Notice that the formula \eqref{eq:PWF2mb} for the wavefunction integrand has the very same structure as the one without the two-point couplings: this means that it satisfies the very same recursion relation proven in \cite{Arkani-Hamed:2017fdk} and, consequently, the combinatorial rules on the graphs implementing it holds with no modification: one iteratively splits the graph in connected subgraphs associating to it the sum of the energies which are external to it, and sums over all the possibility in which such a decomposition can be done. Writing explicitly some example:
\begin{equation}\eqlabel{eq:2sgei}
 \begin{tikzpicture}[line join = round, line cap = round, ball/.style = {circle, draw, align=center, anchor=north, inner sep=0}, 
                     axis/.style={very thick, ->, >=stealth'}, pile/.style={thick, ->, >=stealth', shorten <=2pt, shorten>=2pt}, every node/.style={color=black}]
  \begin{scope}
   \coordinate[label=below:{\footnotesize $\tilde{x}_1$}] (x1) at (0,0);
   \coordinate[label=below:{\footnotesize $x_2$}] (x2) at ($(x1)+(1.5,0)$);
   \draw[-,thick] (x1) -- node[above] {\footnotesize $y$} (x2);
   \draw[fill=white] (x1) circle (2pt);
   \draw[fill,black] (x2) circle (2pt);
   \node[right=.5cm of x2] (eq) {$\displaystyle=\:\frac{1}{(\tilde{x}_1+x_2)(\tilde{x}_1+y)(y+x_2)}$};
  \end{scope}
  \begin{scope}[shift={(-4,-1.5)}, transform shape]
   \coordinate[label=below:{\footnotesize $x_1$}] (x1) at (0,0);
   \coordinate[label=below:{\footnotesize $\omega$}] (w1) at ($(x1)+(1,0)$);
   \coordinate[label=below:{\footnotesize $x_2$}] (x2) at ($(x1)+(2,0)$);
   \draw[-,thick] (x1) -- node[above] {\footnotesize $y$} (w1);
   \draw[-,thick] (w1) -- node[above] {\footnotesize $y$} (x2);
   \draw[fill,black] (x1) circle (2pt);
   \draw[fill=white] (w1) circle (2pt);
   \draw[fill,black] (x2) circle (2pt);
   \node[right=.5cm of x2] (eq) {$\displaystyle=\:\frac{x_1+2y+2\omega+x_2}{(x_1+\omega+x_2)(x_1+y)(\omega+2y)(y+x_2)(x_1+\omega+y)(y+\omega+x_2)}$};   
  \end{scope}
 \end{tikzpicture}
\end{equation}
From the second line in \eqref{eq:2sgei} it is easy to check that it develops a double pole: taking a residue iteratively in three out of the four variables, {\it e.g.} $\{x_1,\,x_2,\,y\}$ or $\{x_1,\,x_2,\,\omega\}$, one gets as a result $1/(2y)^2$ (or $1/(2\omega)^2$). This can be actually understood in full generality. Let us consider a generic graph $\mathcal{G}$ with black sites only. The iterated residue of the associated meromorphic function on the $x_i$'s is given by \cite{Arkani-Hamed:2017fdk}:
\begin{equation}\eqlabel{eq:ItRes}
 \mbox{Res}\left\{\psi_{\mathcal{G}},\{z_i(x)=0\}\right\}\,=\,\prod_{e\in\mathcal{E}}\frac{1}{2y_e}.
\end{equation}
If we now map the graph $\mathcal{G}$ into a graph $\mathcal{G}_{\circ}$ by substituting some of the black site connecting two edges only with white ones, then the meromorphic function associated to $\mathcal{G}_{\circ}$ is the very same one but with all the $y_e$'s related to edges which connect to each other via the white sites being now the same ones. Therefore, taking the iterated residues with respect the variables associated to the sites, one obtains \eqref{eq:ItRes} but with a number of the $y$'s collapsing onto each other, generating multiple poles. Looking at the example in the second line of \eqref{eq:2sgei}, the graph has just two edges connected via a white site, which allows us to predict that it has to develop a double pole $1/(2y)^2$.

As a final comment, the wavefunction with a massive internal state can be thought of series in the white site insertions on a given edge. Considering a two site graph, then it is given by
\begin{equation}\eqlabel{eq:WFser}
 \begin{tikzpicture}[line join = round, line cap = round, ball/.style = {circle, draw, align=center, anchor=north, inner sep=0}, 
                     axis/.style={very thick, ->, >=stealth'}, pile/.style={thick, ->, >=stealth', shorten <=2pt, shorten>=2pt}, every node/.style={color=black}]
   \coordinate[label=below:{\footnotesize $x_1$}] (x1) at (0,0);
   \coordinate[label=below:{\footnotesize $\omega_1$}] (w1) at ($(x1)+(1,0)$);
   \coordinate (t1) at ($(w1)+(.25,0)$);
   \coordinate (t2) at ($(t1)+(1,0)$);
   \coordinate[label=below:{\footnotesize $\omega_a$}] (w2) at ($(t2)+(.25,0)$);
   \coordinate[label=below:{\footnotesize $x_2$}] (x2) at ($(w2)+(1,0)$);
   \draw[-,thick] (x1) -- (w1) -- (t1);
   \draw[-,dotted] (t1) -- (t2);
   \draw[-,thick] (t2) -- (w2) -- (x2);
   \draw[fill,black] (x1) circle (2pt);
   \draw[fill=white] (w1) circle (2pt);
   \draw[fill=white] (w2) circle (2pt);
   \draw[fill,black] (x2) circle (2pt);
   \node[left=.5cm of x1] (eq) {$\displaystyle\psi_2^{\mbox{\tiny $(\lambda_2)$}}\:=\:\sum_{a=0}^{+\infty}\int_{-\infty}^{+\infty}\prod_{r=1}^{a}\left[d\omega_r\tilde{\lambda}_2(\omega_r)\right]$};
 \end{tikzpicture}
\end{equation}
where, for cosmologies $a(\eta)\,=\,(-\eta)^{-\alpha}$, the $\omega$-dependent coupling is $\tilde{\lambda}_2(\omega_r)\,=\,\lambda_2\,i^{2\alpha}\omega_r^{2\alpha-1}\vartheta(\omega_r)$. In other words, it is given by re-summing over all the line graphs integrated over the variables $\omega_r$ ($r\,=\,1,\ldots,a$) attached to their internal (white) sites. The study of a possible closed form for the re-summed two-site graph \eqref{eq:WFser} is postponed to future work\footnote{Despite it might look like the standard textbook discussion on resummation of the self-energy corrections on a two-point function, the structure we would need to re-sum does not have a geometric series structure.}. However, for the time being, there is a comment which can be made: at least for cosmologies with $\alpha\,\in\,\mathbb{Z}_{\frac{1}{2}+}$, the structure of the integrand immediately implies that the integration over the $\omega_r$ returns polylogarithms with rational coefficients. Let us write explicitly the simplest example in $dS$ ({\it i.e.} $\alpha\,=\,1$):
\begin{equation}\eqlabel{eq:c1int}
 \begin{tikzpicture}[line join = round, line cap = round, ball/.style = {circle, draw, align=center, anchor=north, inner sep=0}, 
                     axis/.style={very thick, ->, >=stealth'}, pile/.style={thick, ->, >=stealth', shorten <=2pt, shorten>=2pt}, every node/.style={color=black}]
  \begin{scope}
   \coordinate[label=below:{\footnotesize $x_1$}] (x1) at (0,0);
   \coordinate[label=below:{\footnotesize $\omega$}] (w1) at ($(x1)+(1,0)$);
   \coordinate[label=below:{\footnotesize $x_2$}] (x2) at ($(x1)+(2,0)$);
   \draw[-,thick] (x1) -- node[above] {\footnotesize $y$} (w1);
   \draw[-,thick] (w1) -- node[above] {\footnotesize $y$} (x2);
   \draw[fill,black] (x1) circle (2pt);
   \draw[fill=white] (w1) circle (2pt);
   \draw[fill,black] (x2) circle (2pt);
   \node[left=.125cm of x1, scale=.9] (int) {$\displaystyle\int_{0}^{+\infty}\hspace{-.325cm}d\omega\,\omega$};
   \node[right=.125cm of x2, scale=.9] (eq) {$\displaystyle=\:\frac{(x_1+x_2)\log{(x_1+x_2)}+2y\log{(2y)}-(x_1+y)\log{(x_1+y)}-(y+x_2)\log{(y+x_2)}}{(y^2-x_1^2)(y^2-x_2^2)}$};
  \end{scope}
 \end{tikzpicture}
\end{equation}
Notice that the integrated expression above seems to have poles in $y-x_i$, which are not really expected for the Bunch-Davies wavefunction. In fact, if we compute the residues of such poles they are indeed zero!

Let us close this section commenting on the perturbative treatment with $\nu\,=\,l+1/2$ ($l\,\in\,\mathbb{Z}_+$) states as free states. As for the case just discussed, the structure \eqref{eq:PWF2mb} of the wavefunction integrand stays unchanged, with the propagators now being the propagators for the $l$-states. This means that the recursion relations \eqref{eq:rrWfin1pic} and \eqref{eq:rrWFpic} hold. Hence, in this case we have edge-weighted white/black-site graphs, which, because of the recursion relation proved in this paper, can be rewritten as a differential operator acting on the related $l\,=\,0$ edge-weighted ones ({\it i.e.} the ones discussed above) upon iteration:
\begin{equation}\eqlabel{eq:bwrr}
 \begin{tikzpicture}[line join = round, line cap = round, ball/.style = {circle, draw, align=center, anchor=north, inner sep=0}, 
                     axis/.style={very thick, ->, >=stealth'}, pile/.style={thick, ->, >=stealth', shorten <=2pt, shorten>=2pt}, every node/.style={color=black}]
  \begin{scope}  
   \coordinate[label=below:{\footnotesize $x_1$}] (x1) at (0,0);
   \coordinate[label=below:{\footnotesize $\omega_1$}] (w1) at ($(x1)+(1,0)$);
   \coordinate (t1) at ($(w1)+(.25,0)$);
   \coordinate (t2) at ($(t1)+(1,0)$);
   \coordinate[label=below:{\footnotesize $\omega_a$}] (w2) at ($(t2)+(.25,0)$);
   \coordinate[label=below:{\footnotesize $x_2$}] (x2) at ($(w2)+(1,0)$);
   \draw[-,thick] (x1) -- node[above] {\footnotesize $l_{11}$} (w1) -- (t1);
   \draw[-,dotted] (t1) -- node[above] {\footnotesize $l_e$} (t2);
   \draw[-,thick] (t2) -- (w2) -- node[above] {\footnotesize $l_{a2}$} (x2);
   \draw[fill,black] (x1) circle (2pt);
   \draw[fill=white] (w1) circle (2pt);
   \draw[fill=white] (w2) circle (2pt);
   \draw[fill,black] (x2) circle (2pt);
   \node[right=.125cm of x2, scale=.9] (eq) {$\displaystyle=\:\sum_{e\in\mathcal{E}}
			           \left[\frac{2(l_e-1)}{\displaystyle\sum_{v\in\mathcal{V}}x_v}\left(\frac{\partial}{\partial x_{v_e}}+\frac{\partial}{\partial x_{v'_e}}\right)-\frac{\partial^2}{\partial x_{v_e}\partial x_{v'_e}}\right]$};
  \end{scope}
  \begin{scope}[shift={(10.75,0)}, transform shape] 
   \coordinate[label=below:{\footnotesize $x_1$}] (x1) at (0,0);
   \coordinate[label=below:{\footnotesize $\omega_1$}] (w1) at ($(x1)+(1,0)$);
   \coordinate (t1) at ($(w1)+(.25,0)$);
   \coordinate (t2) at ($(t1)+(1,0)$);
   \coordinate[label=below:{\footnotesize $\omega_a$}] (w2) at ($(t2)+(.25,0)$);
   \coordinate[label=below:{\footnotesize $x_2$}] (x2) at ($(w2)+(1,0)$);
   \draw[-,thick] (x1) -- node[above] {\footnotesize $l_{11}$} (w1) -- (t1);
   \draw[-,dotted] (t1) -- node[above] {\footnotesize $l_e-1$} (t2);
   \draw[-,thick] (t2) -- (w2) -- node[above] {\footnotesize $l_{a2}$} (x2);
   \draw[fill,black] (x1) circle (2pt);
   \draw[fill=white] (w1) circle (2pt);
   \draw[fill=white] (w2) circle (2pt);
   \draw[fill,black] (x2) circle (2pt);
  \end{scope}
 \end{tikzpicture}
\end{equation}
where $x_v\,\equiv\,\omega_v$ for the internal sites. For example, the $l\,=\,1$ edge-weighted graph with one white site can be written as:
\begin{equation}\eqlabel{eq:bwrrl1}
 \begin{tikzpicture}[line join = round, line cap = round, ball/.style = {circle, draw, align=center, anchor=north, inner sep=0}, 
                     axis/.style={very thick, ->, >=stealth'}, pile/.style={thick, ->, >=stealth', shorten <=2pt, shorten>=2pt}, every node/.style={color=black}]
  \begin{scope}
   \coordinate[label=below:{\footnotesize $x_1$}] (x1) at (0,0);
   \coordinate[label=below:{\footnotesize $\omega$}] (w1) at ($(x1)+(1,0)$);
   \coordinate[label=below:{\footnotesize $x_2$}] (x2) at ($(x1)+(2,0)$);
   \draw[-,thick] (x1) -- node[above] {\footnotesize $1$} (w1);
   \draw[-,thick] (w1) -- node[above] {\footnotesize $1$} (x2);
   \draw[fill,black] (x1) circle (2pt);
   \draw[fill=white] (w1) circle (2pt);
   \draw[fill,black] (x2) circle (2pt);
   \node[right=.125cm of x2] (eq1) {$\displaystyle =$};
  \end{scope}
  \begin{scope}[shift={(6.25,0)}, transform shape]
   \coordinate[label=below:{\footnotesize $x_1$}] (x1) at (0,0);
   \coordinate[label=below:{\footnotesize $\omega$}] (w1) at ($(x1)+(1,0)$);
   \coordinate[label=below:{\footnotesize $x_2$}] (x2) at ($(x1)+(2,0)$);
   \draw[-,thick] (x1) -- node[above] {\footnotesize $0$} (w1);
   \draw[-,thick] (w1) -- node[above] {\footnotesize $0$} (x2);
   \draw[fill,black] (x1) circle (2pt);
   \draw[fill=white] (w1) circle (2pt);
   \draw[fill,black] (x2) circle (2pt);
   \node[left=.125cm of x1, scale=.9] (eq1) {$\displaystyle\left(-\frac{\partial^2}{\partial x_1\partial\omega}\right)\left(-\frac{\partial^2}{\partial\omega\partial x_2}\right)$};
  \end{scope}
 \end{tikzpicture}
\end{equation}
which in the $dS$ case integrates to
\begin{equation}\eqlabel{eq:c1l1int}
 \begin{tikzpicture}[line join = round, line cap = round, ball/.style = {circle, draw, align=center, anchor=north, inner sep=0}, 
                     axis/.style={very thick, ->, >=stealth'}, pile/.style={thick, ->, >=stealth', shorten <=2pt, shorten>=2pt}, every node/.style={color=black}]
  \begin{scope}
   \coordinate[label=below:{\footnotesize $x_1$}] (x1) at (0,0);
   \coordinate[label=below:{\footnotesize $\omega$}] (w1) at ($(x1)+(1,0)$);
   \coordinate[label=below:{\footnotesize $x_2$}] (x2) at ($(x1)+(2,0)$);
   \draw[-,thick] (x1) -- node[above] {\footnotesize $1$} (w1);
   \draw[-,thick] (w1) -- node[above] {\footnotesize $1$} (x2);
   \draw[fill,black] (x1) circle (2pt);
   \draw[fill=white] (w1) circle (2pt);
   \draw[fill,black] (x2) circle (2pt);
   \node[left=.125cm of x1, scale=.9] (int) {$\displaystyle\int_{0}^{+\infty}\hspace{-.325cm}d\omega\,\omega$};
   \node[right=.125cm of x2, scale=.9] (eq) {$\displaystyle=\:$};
  \end{scope}
  \begin{scope}[shift={(4,0)}, transform shape]
   \coordinate[label=below:{\footnotesize $x_1$}] (x1) at (0,0);
   \coordinate[label=below:{\footnotesize $\omega$}] (w1) at ($(x1)+(1,0)$);
   \coordinate[label=below:{\footnotesize $x_2$}] (x2) at ($(x1)+(2,0)$);
   \draw[-,thick] (x1) -- node[above] {\footnotesize $0$} (w1);
   \draw[-,thick] (w1) -- node[above] {\footnotesize $0$} (x2);
   \draw[fill,black] (x1) circle (2pt);
   \draw[fill=white] (w1) circle (2pt);
   \draw[fill,black] (x2) circle (2pt);
   \node[left=.125cm of x1, scale=.9] (op) {$\displaystyle\frac{\partial^2}{\partial x_1\partial x_2}$};
   \node[right=.1cm of x2] (val) {$\displaystyle\Bigg|_{\omega\,=\,0}\:=\:\frac{4\left[(x_1+x_2)^3+x_1y x_2\right]}{2y(x_1+x_2)^3(x_1+y)^3(y+x_2)^3}.$};
  \end{scope}
 \end{tikzpicture}
\end{equation}
 
Summarising, the combinatorial structures encountered earlier extends in the case of perturbative mass, both around the massless flat-space scalars (which contains the conformally coupled one) as well as the scalars with time-dependent mass $\mu(\eta)\,=\,-l(l+1)\eta^{-2}$ (containing the minimally coupled scalars). It would be astonishing if the peculiar structure this perturbative expansion would allow us to re-sum it. As already mentioned, we leave this exploration for future work.


\section{Cosmological polytopes and edge-weighted graphs}\label{sec:CPl1}

Graphs with $l\,=\,0$ return the wavefunction of the universe for massless scalars, which are naturally associated with the so-called {\it cosmological polytopes}. In this section we can show how this is also true for the case $l\,=\,1$, {\it i.e.} for internal massless scalars in cosmologies $a(\eta)\,\propto\,(-\eta)^{-\alpha}$. 


\subsection{Cosmological polytopes and the wavefunction of the universe: a concise review}\label{subsec:CPrev}

Given the space of $n_e$ triangles $\left\{\Delta_i\right\}$ identified via their midpoints $(\mathbf{x}_i,\,\mathbf{y}_i,\,\mathbf{x}'_i)$, cosmological polytopes are defined as the convex hulls of the  $3\,n_e$ vertices of such triangles intersected in the midpoints $(\mathbf{x}_i,\,\mathbf{x}'_i)$ of at most two out of their three sides (see Figure \ref{fig:CP}).

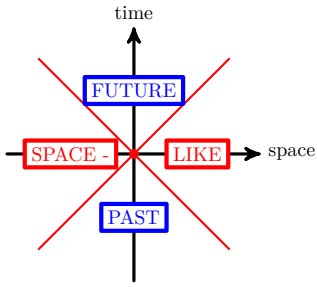
\begin{figure}[h]
 \centering
 \begin{tikzpicture}[line join = round, line cap = round, ball/.style = {circle, draw, align=center, anchor=north, inner sep=0}, 
                     axis/.style={very thick, ->, >=stealth'}, pile/.style={thick, ->, >=stealth', shorten <=2pt, shorten>=2pt}, every node/.style={color=black}, scale={1.25}]
   \begin{scope}[shift={(0,2.5)}, scale={.5}]
   \coordinate (A) at (0,0);
   \coordinate (B) at (-1.75,-2.25);
   \coordinate (C) at (+1.75,-2.25);
   \coordinate [label=left:{\footnotesize $\displaystyle {\bf x}_i$}] (m1) at ($(A)!0.5!(B)$);
   \coordinate [label=right:{\footnotesize $\displaystyle \;{\bf x'}_i$}] (m2) at ($(A)!0.5!(C)$);
   \coordinate [label=below:{\footnotesize $\displaystyle {\bf y}_i$}] (m3) at ($(B)!0.5!(C)$);
   \tikzset{point/.style={insert path={ node[scale=2.5*sqrt(\pgflinewidth)]{.} }}} 

   \draw[color=blue,fill=blue] (m1) circle (2pt);
   \draw[color=blue,fill=blue] (m2) circle (2pt);
   \draw[color=red,fill=red] (m3) circle (2pt);

   \draw[-, very thick, color=blue] (B) -- (A);
   \draw[-, very thick, color=blue] (A) -- (C);  
   \draw[-, very thick, color=red] (B) -- (C);    
  \end{scope}
  \begin{scope}[shift={(2.5,2.5)}, scale={.5}]
   \coordinate (A) at (0,0);
   \coordinate (B) at (-1.75,-2.25);
   \coordinate (C) at (+1.75,-2.25);
   \coordinate [label=left:{\footnotesize $\displaystyle {\bf x}_j$}] (m1) at ($(A)!0.5!(B)$);
   \coordinate [label=right:{\footnotesize $\displaystyle \;{\bf x'}_j$}] (m2) at ($(A)!0.5!(C)$);
   \coordinate [label=below:{\footnotesize $\displaystyle {\bf y}_j$}] (m3) at ($(B)!0.5!(C)$);
   \tikzset{point/.style={insert path={ node[scale=2.5*sqrt(\pgflinewidth)]{.} }}} 

   \draw[color=blue,fill=blue] (m1) circle (2pt);
   \draw[color=blue,fill=blue] (m2) circle (2pt);
   \draw[color=red,fill=red] (m3) circle (2pt);

   \draw[-, very thick, color=blue] (B) -- (A);
   \draw[-, very thick, color=blue] (A) -- (C);  
   \draw[-, very thick, color=red] (B) -- (C);    
  \end{scope}
  \begin{scope}[scale={.4}, shift={(-7,2)}, transform shape]
   \pgfmathsetmacro{\factor}{1/sqrt(2)};  
   \coordinate  (B2) at (1.5,-3,-1.5*\factor);
   \coordinate  (A1) at (-1.5,-3,-1.5*\factor);
   \coordinate  (B1) at (1.5,-3.75,1.5*\factor);
   \coordinate  (A2) at (-1.5,-3.75,1.5*\factor);  
   \coordinate  (C1) at (0.75,-.65,.75*\factor);
   \coordinate  (C2) at (0.4,-6.05,.75*\factor);
   \coordinate (Int) at (intersection of A2--B2 and B1--C1);
   \coordinate (Int2) at (intersection of A1--B1 and A2--B2);

   \tikzstyle{interrupt}=[
    postaction={
        decorate,
        decoration={markings,
                    mark= at position 0.5 
                          with
                          {
                            \node[rectangle, color=white, fill=white, below=-.1 of Int] {};
                          }}}
   ]
   
   \draw[interrupt,thick,color=red] (B1) -- (C1);
   \draw[-,very thick,color=blue] (A1) -- (B1);
   \draw[-,very thick,color=blue] (A2) -- (B2);
   \draw[-,very thick,color=blue] (A1) -- (C1);
   \draw[-, dashed, very thick, color=red] (A2) -- (C2);
   \draw[-, dashed, thick, color=blue] (B2) -- (C2);

   \coordinate[label=below:{\Large ${\bf x'}_i$}] (x2) at ($(A1)!0.5!(B1)$);
   \draw[fill,color=blue] (x2) circle (2.5pt);   
   \coordinate[label=left:{\Large ${\bf x}_i$}] (x1) at ($(C1)!0.5!(A1)$);
   \draw[fill,color=blue] (x1) circle (2.5pt);
   \coordinate[label=right:{\Large ${\bf x}_j$}] (x3) at ($(B2)!0.5!(C2)$);
   \draw[fill,color=blue] (x3) circle (2.5pt);

   \node[right=2.25cm of x2] (cht) {$\displaystyle\xrightarrow{\substack{\mbox{convex} \\ \mbox{hull}}}$};
  \end{scope}
  \begin{scope}[scale={.4}, shift={(-.5,2)}, transform shape]
   \pgfmathsetmacro{\factor}{1/sqrt(2)};  
   \coordinate  (B2) at (1.5,-3,-1.5*\factor);
   \coordinate  (A1) at (-1.5,-3,-1.5*\factor);
   \coordinate  (B1) at (1.5,-3.75,1.5*\factor);
   \coordinate  (A2) at (-1.5,-3.75,1.5*\factor);  
   \coordinate  (C1) at (0.75,-.65,.75*\factor);
   \coordinate  (C2) at (0.4,-6.05,.75*\factor);

   \draw[-,dashed,fill=blue!30, opacity=.7] (A1) -- (B2) -- (C1) -- cycle;
   \draw[-,thick,fill=blue!20, opacity=.7] (A1) -- (A2) -- (C1) -- cycle;
   \draw[-,thick,fill=blue!20, opacity=.7] (B1) -- (B2) -- (C1) -- cycle;
   \draw[-,thick,fill=blue!35, opacity=.7] (A2) -- (B1) -- (C1) -- cycle;

   \draw[-,dashed,fill=red!30, opacity=.3] (A1) -- (B2) -- (C2) -- cycle;
   \draw[-,dashed, thick, fill=red!50, opacity=.5] (B2) -- (B1) -- (C2) -- cycle;
   \draw[-,dashed,fill=red!40, opacity=.3] (A1) -- (A2) -- (C2) -- cycle;
   \draw[-,dashed, thick, fill=red!45, opacity=.5] (A2) -- (B1) -- (C2) -- cycle;
  \end{scope}
  \begin{scope}[scale={.5}, shift={(4.5,.75)}, transform shape]
   \pgfmathsetmacro{\factor}{1/sqrt(2)};
   \coordinate  (c1b) at (0.75,0,-.75*\factor);
   \coordinate  (b1b) at (-.75,0,-.75*\factor);
   \coordinate  (a2b) at (0.75,-.65,.75*\factor);
   
   \coordinate  (c2b) at (1.5,-3,-1.5*\factor);
   \coordinate  (b2b) at (-1.5,-3,-1.5*\factor);
   \coordinate  (a1b) at (1.5,-3.75,1.5*\factor); 

  \coordinate (Int1) at (intersection of b2b--c2b and b1b--a1b);
   \coordinate (Int2) at (intersection of b2b--c2b and c1b--a1b);
   \coordinate (Int3) at (intersection of b2b--a2b and b1b--a1b);
   \coordinate (Int4) at (intersection of a2b--c2b and c1b--a1b);
   \tikzstyle{interrupt}=[
    postaction={
        decorate,
        decoration={markings,
                    mark= at position 0.5 
                          with
                          {
                            \node[rectangle, color=white, fill=white] at (Int1) {};
                            \node[rectangle, color=white, fill=white] at (Int2) {};                            
                          }}}
   ]

   \node at (c1b) (c1c) {};
   \node at (b1b) (b1c) {};
   \node at (a2b) (a2c) {};
   \node at (c2b) (c2c) {};
   \node at (b2b) (b2c) {};
   \node at (a1b) (a1c) {};

   \draw[interrupt,thick,color=red] (b2b) -- (c2b);
   \draw[-,very thick,color=red] (b1b) -- (c1b);
   \draw[-,very thick,color=blue] (b1b) -- (a1b);
   \draw[-,very thick,color=blue] (a1b) -- (c1b);   
   \draw[-,very thick,color=blue] (b2b) -- (a2b);
   \draw[-,very thick,color=blue] (a2b) -- (c2b);

   \node[ball,text width=.15cm,fill,color=blue, above=-.06cm of Int3, label=left:{\large ${\bf x}_i$}] (Inta) {};
   \node[ball,text width=.15cm,fill,color=blue, above=-.06cm of Int4, label=right:{\large ${\bf x'}_i$}] (Intb) {};

   \node[right=.875cm of Intb, scale=.75] (chl) {$\displaystyle\xrightarrow{\substack{\mbox{convex} \\ \mbox{hull}}}$};
  \end{scope}
  \begin{scope}[scale={.5}, shift={(9,.75)}, transform shape]
   \pgfmathsetmacro{\factor}{1/sqrt(2)};
   \coordinate  (c1b) at (0.75,0,-.75*\factor);
   \coordinate  (b1b) at (-.75,0,-.75*\factor);
   \coordinate  (a2b) at (0.75,-.65,.75*\factor);
   
   \coordinate  (c2b) at (1.5,-3,-1.5*\factor);
   \coordinate  (b2b) at (-1.5,-3,-1.5*\factor);
   \coordinate  (a1b) at (1.5,-3.75,1.5*\factor);

   \draw[-,dashed,fill=green!50,opacity=.6] (c1b) -- (b1b) -- (b2b) -- (c2b) -- cycle;
   \draw[draw=none,fill=red!60, opacity=.45] (c2b) -- (b2b) -- (a1b) -- cycle;
   \draw[-,fill=blue!,opacity=.3] (c1b) -- (b1b) -- (a2b) -- cycle; 
   \draw[-,fill=green!50,opacity=.4] (b1b) -- (a2b) -- (a1b) -- (b2b) -- cycle;
   \draw[-,fill=green!45!black,opacity=.2] (c1b) -- (a2b) -- (a1b) -- (c2b) -- cycle;  
  \end{scope}
 \end{tikzpicture}
 \caption{Cosmological polytopes constructed from the space of $n_e\,=\,2$ triangles. The (red) blue sides of the triangles (first line in the picture) are (non)-intersectable. Out of these two triangles, there are two ways of constructing more complicated objects: they can be intersected on the midpoint of one of the two (blue) intersectable sides ($\mathbf{x}'_i\,=\,\mathbf{x}'_j$), or in both ($\mathbf{x}_i\,=\,\mathbf{x}_j$; $\mathbf{x}'_i\,=\,\mathbf{x}'_j$), originating the polytopes on the bottom left and right respectively.}
 \label{fig:CP}
\end{figure}
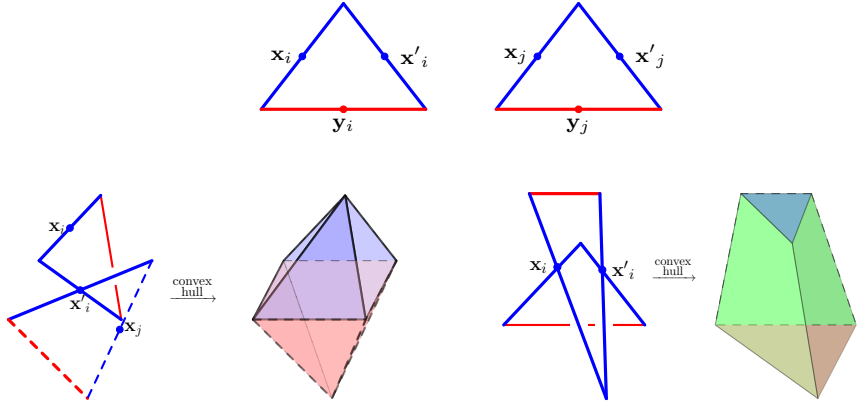

\begin{wrapfigure}{l}{4.5cm}
 \begin{tikzpicture}[line join = round, line cap = round, ball/.style = {circle, draw, align=center, anchor=north, inner sep=0}, 
                     axis/.style={very thick, ->, >=stealth'}, pile/.style={thick, ->, >=stealth', shorten <=2pt, shorten>=2pt}, every node/.style={color=black}, scale={1.125}]
  \begin{scope}[scale={2.5}]
   \draw[axis] (-0.6,0) -- (+0.6,0) node(xline)[right, scale=.75]{space};
   \draw[axis] (0,-0.6) -- (0,+0.6) node(yline)[above, scale=.75]{time};
   \fill[red] (0,0) circle (.65pt);    
   \draw[-, thick, color=red] (-0.45,+0.45) -- (+0.45,-0.45);
   \draw[-, thick, color=red] (-0.45,-0.45) -- (+0.45,+0.45);
   \node[draw, ultra thick, align=center, color=blue, fill=white, scale=.75] at (0,+0.3) {FUTURE};
   \node[draw, ultra thick, align=center, color=blue, fill=white, scale=.75] at (0,-0.3) {PAST};
   \node[draw, ultra thick, align=center, color=red, fill=white, scale=.75] at (-0.3,0) {SPACE -};
   \node[draw, ultra thick, align=center, color=red, fill=white, scale=.75] at (+0.3,0) {LIKE}; 
  \end{scope}
 \end{tikzpicture}
\end{wrapfigure}

Interestingly, this construction has the space-time causal structure imprinted: the two intersectable edges of a triangle correspond to the two space-time regions with a definite causal relation ({\it past} and {\it future}), while the non-intersectable one represents the region with no causal relation ({\it space-like}). Or, turning the table around, the causal structure of the space-time provides a rationale to having triangles as fundamental objects as well as to the prescription of considering the class of polytopes generated by intersecting at most two out of the three sides of the triangles. While the $n_e$ non-intersected triangles live in $\mathbb{P}^{3n_e-1}$, a cosmological polytope generated by intersecting them by imposing $r$ constraints live in $\mathbb{P}^{3n_e-r-1}$.

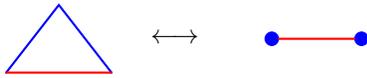
\begin{wrapfigure}{l}{5cm}
 \begin{tikzpicture}[shift={(1,0)}, line join = round, line cap = round, ball/.style = {circle, draw, align=center, anchor=north, inner sep=0}, scale={.4}]
  \coordinate (A) at (0,0);
  \coordinate (B) at (-1.75,-2.25);
  \coordinate (C) at (+1.75,-2.25);
  \coordinate  (m2) at ($(A)!0.5!(C)$);

  \draw[-, thick, color=blue] (B) -- (A); 
  \draw[-, thick, color=blue] (A) -- (C); 
  \draw[-, thick, color=red] (B) -- (C);

  \node[right=.75cm of m2.east] (lra) {$\longleftrightarrow$};
  \node[ball,text width=.18cm,fill,color=blue,right=.75cm of lra.east] (x1) {};
  \node[ball,text width=.18cm,fill,color=blue,right=1cm of x1.east] (x2) {};
  \draw[-,thick,color=red] (x1.east) -- (x2.west);
 \end{tikzpicture}
\end{wrapfigure}
There is a $1-1$ correspondence between cosmological polytopes and the $l\,=\,0$ graphs: any triangle $\triangle_i$, characterised via its midpoints $(\mathbf{x}_i,\,\mathbf{y}_i,\,\mathbf{x}'_i)$, is associated to a two-site graph, with each site corresponding to the intersectable sides of $\triangle_i$ and its only edge with the non-intersectable one. Thus, a cosmological polytope $\mathcal{P}_{\mathcal{G}}$ generated by intersecting $n_e$ triangles, is associated to a graph $\mathcal{G}$ with $n_e$ edges constructed from a collection of two-site graphs by identifying some of their vertices:
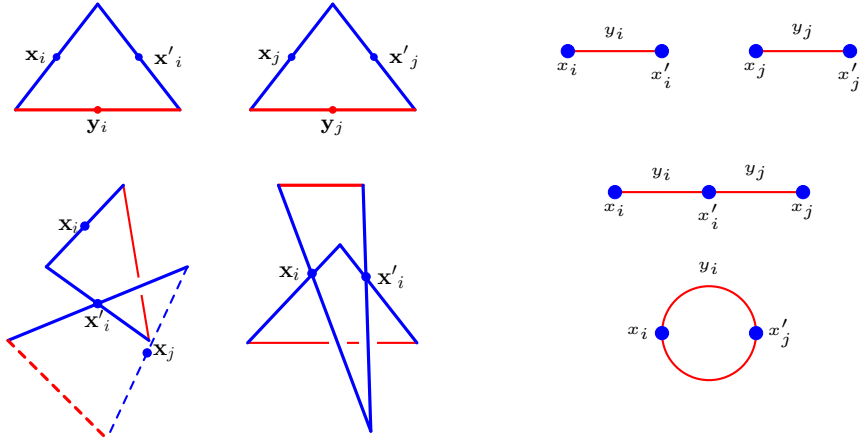
\begin{equation*}
 \begin{tikzpicture}[line join = round, line cap = round, ball/.style = {circle, draw, align=center, anchor=north, inner sep=0}, 
                     axis/.style={very thick, ->, >=stealth'}, pile/.style={thick, ->, >=stealth', shorten <=2pt, shorten>=2pt}, every node/.style={color=black}, scale={1.25}]
  \begin{scope}[shift={(5,2)}, transform shape]
    \coordinate[label=below:{\tiny $x_i$}] (v1) at (0,0);
    \coordinate[label=below:{\tiny $x'_i$}] (v2) at ($(v1)+(1,0)$);
    \coordinate[label=below:{\tiny $x_j$}] (v3) at ($(v2)+(1,0)$);
    \coordinate[label=below:{\tiny $x'_j$}] (v4) at ($(v3)+(1,0)$);
    \coordinate[label=above:{\tiny $y_i$}] (yi) at ($(v1)!0.5!(v2)$);
    \coordinate[label=above:{\tiny $y_j$}] (yj) at ($(v3)!0.5!(v4)$);
    \draw[thick, color=red] (v1) -- (v2);
    \draw[thick, color=red] (v3) -- (v4);
    \draw[fill=blue, color=blue] (v1) circle (2pt);
    \draw[fill=blue, color=blue] (v2) circle (2pt);
    \draw[fill=blue, color=blue] (v3) circle (2pt);
    \draw[fill=blue, color=blue] (v4) circle (2pt);
    \coordinate (t1) at ($(v2)!0.5!(v3)$);
    \coordinate[label=below:{\tiny $x'_i$}] (s2) at ($(t1)+(0,-1.5)$);
    \coordinate[label=below:{\tiny $x_i$}] (s1) at ($(s2)-(1,0)$);
    \coordinate[label=below:{\tiny $x_j$}] (s3) at ($(s2)+(1,0)$);
    \coordinate[label=above:{\tiny $y_i$}] (yyi) at ($(s1)!0.5!(s2)$);
    \coordinate[label=above:{\tiny $y_j$}] (yyj) at ($(s2)!0.5!(s3)$);
    \draw[thick, color=red] (s1) -- (s2) -- (s3);
    \draw[fill=blue, color=blue] (s1) circle (2pt);
    \draw[fill=blue, color=blue] (s2) circle (2pt);
    \draw[fill=blue, color=blue] (s3) circle (2pt);
    \coordinate[label=left:{\tiny $x_i$}] (n1) at ($(s1)!0.5!(s2)+(0,-1.5)$);
    \coordinate[label=right:{\tiny $x'_j$}] (n2) at ($(s2)!0.5!(s3)+(0,-1.5)$);
    \coordinate (nc) at ($(n1)!0.5!(n2)$);
    \coordinate[label=above:{\tiny $y_i$}] (yyyi) at ($(nc)+(0,.5cm)$);
    \draw[thick, color=red] (nc) circle (.5cm);
    \draw[fill=blue, color=blue] (n1) circle (2pt);
    \draw[fill=blue, color=blue] (n2) circle (2pt);
  \end{scope}
  \begin{scope}[shift={(0,2.5)}, scale={.5}]
   \coordinate (A) at (0,0);
   \coordinate (B) at (-1.75,-2.25);
   \coordinate (C) at (+1.75,-2.25);
   \coordinate [label=left:{\footnotesize $\displaystyle {\bf x}_i$}] (m1) at ($(A)!0.5!(B)$);
   \coordinate [label=right:{\footnotesize $\displaystyle \;{\bf x'}_i$}] (m2) at ($(A)!0.5!(C)$);
   \coordinate [label=below:{\footnotesize $\displaystyle {\bf y}_i$}] (m3) at ($(B)!0.5!(C)$);
   \tikzset{point/.style={insert path={ node[scale=2.5*sqrt(\pgflinewidth)]{.} }}} 

   \draw[color=blue,fill=blue] (m1) circle (2pt);
   \draw[color=blue,fill=blue] (m2) circle (2pt);
   \draw[color=red,fill=red] (m3) circle (2pt);

   \draw[-, very thick, color=blue] (B) -- (A);
   \draw[-, very thick, color=blue] (A) -- (C);  
   \draw[-, very thick, color=red] (B) -- (C);    
  \end{scope}
  \begin{scope}[shift={(2.5,2.5)}, scale={.5}]
   \coordinate (A) at (0,0);
   \coordinate (B) at (-1.75,-2.25);
   \coordinate (C) at (+1.75,-2.25);
   \coordinate [label=left:{\footnotesize $\displaystyle {\bf x}_j$}] (m1) at ($(A)!0.5!(B)$);
   \coordinate [label=right:{\footnotesize $\displaystyle \;{\bf x'}_j$}] (m2) at ($(A)!0.5!(C)$);
   \coordinate [label=below:{\footnotesize $\displaystyle {\bf y}_j$}] (m3) at ($(B)!0.5!(C)$);
   \tikzset{point/.style={insert path={ node[scale=2.5*sqrt(\pgflinewidth)]{.} }}} 

   \draw[color=blue,fill=blue] (m1) circle (2pt);
   \draw[color=blue,fill=blue] (m2) circle (2pt);
   \draw[color=red,fill=red] (m3) circle (2pt);

   \draw[-, very thick, color=blue] (B) -- (A);
   \draw[-, very thick, color=blue] (A) -- (C);  
   \draw[-, very thick, color=red] (B) -- (C);    
  \end{scope}
  \begin{scope}[scale={.5}, shift={(0,2)}, transform shape]
   \pgfmathsetmacro{\factor}{1/sqrt(2)};  
   \coordinate  (B2) at (1.5,-3,-1.5*\factor);
   \coordinate  (A1) at (-1.5,-3,-1.5*\factor);
   \coordinate  (B1) at (1.5,-3.75,1.5*\factor);
   \coordinate  (A2) at (-1.5,-3.75,1.5*\factor);  
   \coordinate  (C1) at (0.75,-.65,.75*\factor);
   \coordinate  (C2) at (0.4,-6.05,.75*\factor);
   \coordinate (Int) at (intersection of A2--B2 and B1--C1);
   \coordinate (Int2) at (intersection of A1--B1 and A2--B2);

   \tikzstyle{interrupt}=[
    postaction={
        decorate,
        decoration={markings,
                    mark= at position 0.5 
                          with
                          {
                            \node[rectangle, color=white, fill=white, below=-.1 of Int] {};
                          }}}
   ]
   
   \draw[interrupt,thick,color=red] (B1) -- (C1);
   \draw[-,very thick,color=blue] (A1) -- (B1);
   \draw[-,very thick,color=blue] (A2) -- (B2);
   \draw[-,very thick,color=blue] (A1) -- (C1);
   \draw[-, dashed, very thick, color=red] (A2) -- (C2);
   \draw[-, dashed, thick, color=blue] (B2) -- (C2);

   \coordinate[label=below:{\Large ${\bf x'}_i$}] (x2) at ($(A1)!0.5!(B1)$);
   \draw[fill,color=blue] (x2) circle (2.5pt);   
   \coordinate[label=left:{\Large ${\bf x}_i$}] (x1) at ($(C1)!0.5!(A1)$);
   \draw[fill,color=blue] (x1) circle (2.5pt);
   \coordinate[label=right:{\Large ${\bf x}_j$}] (x3) at ($(B2)!0.5!(C2)$);
   \draw[fill,color=blue] (x3) circle (2.5pt);
  \end{scope}
  \begin{scope}[scale={.6}, shift={(3.75,.75)}, transform shape]
   \pgfmathsetmacro{\factor}{1/sqrt(2)};
   \coordinate  (c1b) at (0.75,0,-.75*\factor);
   \coordinate  (b1b) at (-.75,0,-.75*\factor);
   \coordinate  (a2b) at (0.75,-.65,.75*\factor);
   
   \coordinate  (c2b) at (1.5,-3,-1.5*\factor);
   \coordinate  (b2b) at (-1.5,-3,-1.5*\factor);
   \coordinate  (a1b) at (1.5,-3.75,1.5*\factor); 

   \coordinate (Int1) at (intersection of b2b--c2b and b1b--a1b);
   \coordinate (Int2) at (intersection of b2b--c2b and c1b--a1b);
   \coordinate (Int3) at (intersection of b2b--a2b and b1b--a1b);
   \coordinate (Int4) at (intersection of a2b--c2b and c1b--a1b);
   \tikzstyle{interrupt}=[
    postaction={
        decorate,
        decoration={markings,
                    mark= at position 0.5 
                          with
                          {
                            \node[rectangle, color=white, fill=white] at (Int1) {};
                            \node[rectangle, color=white, fill=white] at (Int2) {};                            
                          }}}
   ]

   \node at (c1b) (c1c) {};
   \node at (b1b) (b1c) {};
   \node at (a2b) (a2c) {};
   \node at (c2b) (c2c) {};
   \node at (b2b) (b2c) {};
   \node at (a1b) (a1c) {};

   \draw[interrupt,thick,color=red] (b2b) -- (c2b);
   \draw[-,very thick,color=red] (b1b) -- (c1b);
   \draw[-,very thick,color=blue] (b1b) -- (a1b);
   \draw[-,very thick,color=blue] (a1b) -- (c1b);   
   \draw[-,very thick,color=blue] (b2b) -- (a2b);
   \draw[-,very thick,color=blue] (a2b) -- (c2b);

   \node[ball,text width=.15cm,fill,color=blue, above=-.06cm of Int3, label=left:{\large ${\bf x}_i$}] (Inta) {};
   \node[ball,text width=.15cm,fill,color=blue, above=-.06cm of Int4, label=right:{\large ${\bf x'}_i$}] (Intb) {};

  \end{scope}
 \end{tikzpicture}
\end{equation*}

Each vertex and edge are then labelled respectively by $x_v$ and $y_e$ which are related to the midpoints $\mathbf{x}_i$ and $\mathbf{y}_i$. These labels can be identified with the energies of the sites $v$ and edge $e$. Vice versa, starting with a graph $\mathcal{G}$, it is possible to define the space of the energies associates to its sites and edges $\mathcal{Y}\,=\,(x's,y's)\,\in\,\mathbb{P}^{n_v+n_e-1}$, $n_v$ and $n_e$ being the number of vertices and edges of $\mathcal{G}$, with a basis formed by the vectors $\mathbf{x}_v\,\equiv\,x_v\mathbf{X}_v$ and $\mathbf{y}_e\,\equiv\,y_e\mathbf{Y}_e$ associated to vertices and edges. Each edge of $\mathcal{G}$ is then associated with the set of vertices 
$\{\mathbf{x}_i-\mathbf{y}_i+\mathbf{x}'_i,\, \mathbf{x}_i+\mathbf{y}_i-\mathbf{x}'_i,\, -\mathbf{x}_i+\mathbf{y}_i+\mathbf{x}'_i,\}$. The cosmological polytope $\mathcal{P}_{\mathcal{G}}$ associated to the graph $\mathcal{G}$ is thus the convex hull defined by these vertices. Finally, any cosmological polytope $\mathcal{P}_{\mathcal{G}}\,\in\,\mathbb{P}^{n_v+n_e-1}$ has an associated canonical differential top form $\omega(\mathcal{Y};\,\mathcal{P}_{\mathcal{G}})$, $\mathcal{Y}$ being a generic point of $\mathcal{P}_{\mathcal{G}}$, which is uniquely fixed by the requirement that it has logarithmic singularities on (and only on) all the faces of $\mathcal{P}_{\mathcal{G}}$. Defining the coefficient $\Omega(\mathcal{Y};\,\mathcal{P}_{\mathcal{G}})$ of the canonical form $\omega(\mathcal{Y};\,\mathcal{P}_{\mathcal{G}})$ by stripping the universal top-form measure $\langle\mathcal{Y},d^N\mathcal{Y}\rangle$ out ($N\,\equiv\,n_v+n_e$), it returns (the integrand of) the wavefunction of the universe $\Psi_{\mathcal{G}}(x,y)$ associated to the graph $\mathcal{G}$ \cite{Arkani-Hamed:2017fdk}
\begin{equation}\eqlabel{eq:CFWF}
 \omega(\mathcal{Y};\,\mathcal{P}_{\mathcal{G}})\:\equiv\:\Omega(\mathcal{Y};\,\mathcal{P}_{\mathcal{G}})\langle\mathcal{Y},d^N\mathcal{Y}\rangle\:=\:\frac{\Psi_{\mathcal{G}}(x_v,y_e)}{\mbox{Vol}\{GL(1)\}}\prod_{\substack{v\in\mathcal{V} \\ e\in\mathcal{E}}}dx_vdy_e.
\end{equation}
When any singularity is reached, the canonical form $\omega(\mathcal{Y};\,\mathcal{P}_{\mathcal{G}})$ reduces to a lower-dimensional one which characterises the lower dimensional polytope $\mathcal{P}_{\mathfrak{g}}$ corresponding to the face of $\mathcal{P}_{\mathcal{G}}$ identified by a subgraph $\mathfrak{g}$ of $\mathcal{G}$.

The definition of the cosmological polytopes as convex hull of the vertices of $n_e$ intersecting triangles allows a direct combinatorial characterisation of their faces as well: given a certain cosmological polytope $\mathcal{P}_{\mathcal{G}}$, any of its facets is defined as the collection $\mathcal{V}_{\mathcal{F}}$ of vertices $\mathbf{V}_a^I$ ($a\,=\,1,\ldots\,3n_e$) of $\mathcal{P}_{\mathcal{G}}$ such that $\mathcal{W}_I\mathbf{V}^I\,=\,0$, $\mathcal{W}_I\,\equiv\,\tilde{x}_v\tilde{\mathbf{X}}_{vI}+\tilde{y}_e\mathbf{\tilde{Y}}_{eI}$ being an hyperplane in $\mathbb{P}^{n_v+n_e-1}$ with $\mathbf{\tilde{X}}_v\cdot\mathbf{x}_{v'}\,=\,\delta_{vv'}$, $\mathbf{\tilde{Y}}_e\cdot\mathbf{y}_{e'}\,=\,\delta_{ee'}$ and $\mathbf{\tilde{X}}_v\cdot\mathbf{y}_{e'}\,=\,0$, compatibly with the constraints on the midpoints of the sides of the generating triangles -- those vertices of $\mathcal{P}_{\mathcal{G}}$ which are not on the facet identified by a certain hyperplane $\mathcal{W}_I$, satisfy instead the inequality $\mathcal{W}\cdot\mathbf{V}_a\,\ge\,0$. Each of these hyperplanes is in a $1-1$ correspondence with a subgraph $\mathfrak{g}$ of the graph $\mathcal{G}$ associated to $\mathcal{P}_{\mathcal{G}}$, and it is given by $\mathcal{W}\,=\,\sum_{v\in\mathfrak{g}}\tilde{x}_v\mathbf{\tilde{X}}_v+\sum_{e\in\mathcal{E}_{\mathfrak{g}}^{\mbox{\tiny ext}}}\tilde{y}_e\mathbf{\tilde{Y}}_e$, with $\mathcal{E}_{\mathfrak{g}}^{\mbox{\tiny ext}}$ being the set of edges external to $\mathfrak{g}$. It is possible to graphically keep track of the vertices belonging to a certain hyperplane $\mathcal{W}$ introducing a marking on the associated graph $\mathcal{G}$ indicating the vertices which do not belong to $\mathcal{W}$
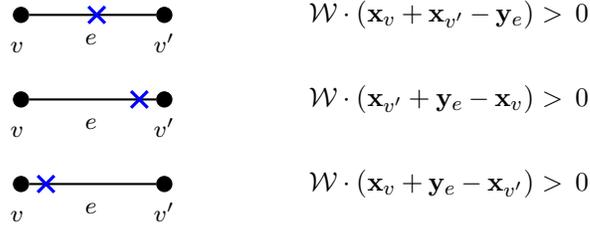
\begin{equation*}
 \begin{tikzpicture}[ball/.style = {circle, draw, align=center, anchor=north, inner sep=0}, cross/.style={cross out, draw, minimum size=2*(#1-\pgflinewidth), inner sep=0pt, outer sep=0pt}, scale={1.125}, transform shape]
  \begin{scope}
   \node[ball,text width=.18cm,fill,color=black,label=below:{\footnotesize $v\phantom{'}$}] at (0,0) (v1) {};
   \node[ball,text width=.18cm,fill,color=black,label=below:{\footnotesize $v'$},right=1.5cm of v1.east] (v2) {};  
   \draw[-,thick,color=black] (v1.east) edge node [text width=.18cm,below=.1cm,midway] {\footnotesize $e$} (v2.west);
   \node[very thick, cross=4pt, rotate=0, color=blue, right=.7cm of v1.east]{};
   \coordinate (x) at ($(v1)!0.5!(v2)$);
   \node[right=1.5cm of v2, scale=.9] (lb1) {$\mathcal{W}\cdot({\bf x}_v+{\bf x}_{v'}-{\bf y}_e)>\,0$};  
  \end{scope}
  \begin{scope}[shift={(0,-1)}]
   \node[ball,text width=.18cm,fill,color=black,label=below:{\footnotesize $v\phantom{'}$}] at (0,0) (v1) {};
   \node[ball,text width=.18cm,fill,color=black,label=below:{\footnotesize $v'$},right=1.5cm of v1.east] (v2) {};  
   \draw[-,thick,color=black] (v1.east) edge node [text width=.18cm,below=.1cm,midway] {\footnotesize $e$} (v2.west);
   \node[very thick, cross=4pt, rotate=0, color=blue, left=.1cm of v2.west]{};
   \coordinate (x) at ($(v1)!0.5!(v2)$);
   \node[right=1.5cm of v2, scale=.9] (lb1) {$\mathcal{W}\cdot({\bf x}_{v'}+{\bf y}_e-{\bf x}_v)>\,0$};  
  \end{scope}
  \begin{scope}[shift={(0,-2)}]
   \node[ball,text width=.18cm,fill,color=black,label=below:{\footnotesize $v\phantom{'}$}] at (0,0) (v1) {};
   \node[ball,text width=.18cm,fill,color=black,label=below:{\footnotesize $v'$},right=1.5cm of v1.east] (v2) {};  
   \draw[-,thick,color=black] (v1.east) edge node [text width=.18cm,below=.1cm,midway] {\footnotesize $e$} (v2.west);
   \node[very thick, cross=4pt, rotate=0, color=blue, right=.1cm of v1.east]{};
   \coordinate (x) at ($(v1)!0.5!(v2)$);
   \node[right=1.5cm of v2, scale=.9] (lb1) {$\mathcal{W}\cdot({\bf x}_v+{\bf y}_e-{\bf x}_{v'})>\,0$};  
  \end{scope}
 \end{tikzpicture}
\end{equation*}
If $\mathcal{G}\,=\,\mathfrak{g}$, then all the edges are marked in their middle and the hyperplane identifying this facet is given by $\mathcal{W}\,=\,\sum_{v}\tilde{x}_v{\bf \tilde{X}}_v$, {\it i.e.} it is the scattering facet and it is identified by the collection of $2n_e$ vertices $\{\mathbf{x}_v+\mathbf{y}_e-\mathbf{x'}_v,\,-\mathbf{x}_v+\mathbf{y}_e+\mathbf{x'}_v\}$. For a more general facet, the graph gets marked in the middle of the edges of $\mathcal{G}$ which are internal to $\mathfrak{g}$ and in the extreme close to $\mathfrak{g}$ for the edges of $\mathcal{G}$ which are external to $\mathfrak{g}$:
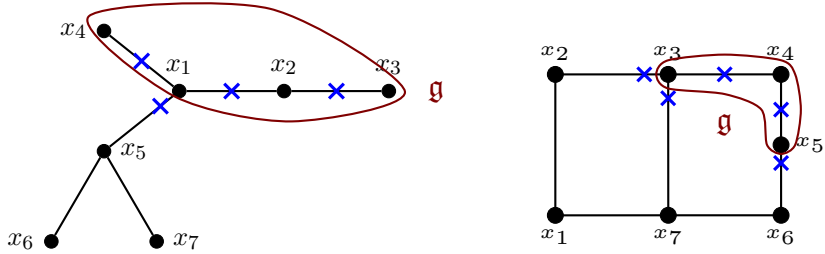
\begin{equation*}
  \begin{tikzpicture}[ball/.style = {circle, draw, align=center, anchor=north, inner sep=0}, cross/.style={cross out, draw, minimum size=2*(#1-\pgflinewidth), inner sep=0pt, outer sep=0pt}, scale=1, transform shape]
   \begin{scope}
    \node[ball,text width=.18cm,fill,color=black,label=above:{$x_1$}] at (0,0) (x1) {};    
    \node[ball,text width=.18cm,fill,color=black,right=1.2cm of x1.east, label=above:{$x_2$}] (x2) {};    
    \node[ball,text width=.18cm,fill,color=black,right=1.2cm of x2.east, label=above:{$x_3$}] (x3) {};
    \node[ball,text width=.18cm,fill,color=black, label=left:{$x_4$}] at (-1,.8) (x4) {};    
    \node[ball,text width=.18cm,fill,color=black, label=right:{$x_5$}] at (-1,-.8) (x5) {};    
    \node[ball,text width=.18cm,fill,color=black, label=left:{$x_6$}] at (-1.7,-2) (x6) {};    
    \node[ball,text width=.18cm,fill,color=black, label=right:{$x_7$}] at (-.3,-2) (x7) {};

    \node[above=.35cm of x5.north] (ref2) {};
    \coordinate (Int2) at (intersection of x5--x1 and ref2--x2);  

    \coordinate (t1) at (x3.east);
    \coordinate (t2) at (x4.west);
    \coordinate (t3) at (x1.south west);
    \coordinate (t4) at (x2.south);

    \draw[-,thick,color=black] (x1) -- (x2) -- (x3); 
    \draw[-,thick,color=black] (x1) -- (x4);
    \draw[-,thick,color=black] (x5) -- (x1);
    \draw[-,thick,color=black] (x5) -- (x7);   
    \draw[-,thick,color=black] (x5) -- (x6); 
    \draw[red!50!black, thick] plot [smooth cycle] coordinates {(3,-.1) (1.2,1) (-1.2,.9) (t3) (1.5,-.5)};
    \node[color=red!50!black,right=.3cm of x3.east] {\large $\mathfrak{g}$}; 

    \coordinate (m1) at ($(x1)!0.5!(x4)$);
    \coordinate (m2) at ($(x1)!0.5!(x2)$);
    \coordinate (m3) at ($(x2)!0.5!(x3)$);
    \coordinate (e1) at ($(x1)!0.25!(x5)$);
    \node[very thick, cross=4pt, rotate=0, color=blue] at (m1) {};
    \node[very thick, cross=4pt, rotate=0, color=blue] at (m2) {};
    \node[very thick, cross=4pt, rotate=0, color=blue] at (m3) {};  
    \node[very thick, cross=4pt, rotate=0, color=blue] at (e1) {};  
   \end{scope}
   \begin{scope}[shift={(5,-1.75)}, scale={1.5}, transform shape]
   \coordinate[label=below:{\tiny $x_1$}] (v1) at (0,0);
   \coordinate[label=above:{\tiny $x_2$}] (v2) at ($(v1)+(0,1.25)$);
   \coordinate[label=above:{\tiny $x_3$}] (v3) at ($(v2)+(1,0)$);
   \coordinate[label=above:{\tiny $x_4$}] (v4) at ($(v3)+(1,0)$);
   \coordinate[label=right:{\tiny $x_5$}] (v5) at ($(v4)-(0,.625)$);   
   \coordinate[label=below:{\tiny $x_6$}] (v6) at ($(v5)-(0,.625)$);
   \coordinate[label=below:{\tiny $x_7$}] (v7) at ($(v6)-(1,0)$);
   \draw[thick] (v1) -- (v2) -- (v3) -- (v4) -- (v5) -- (v6) -- (v7) -- cycle;
   \draw[thick] (v3) -- (v7);
   \draw[fill=black] (v1) circle (2pt);
   \draw[fill=black] (v2) circle (2pt);
   \draw[fill=black] (v3) circle (2pt);
   \draw[fill=black] (v4) circle (2pt);
   \draw[fill=black] (v5) circle (2pt);
   \draw[fill=black] (v6) circle (2pt);
   \draw[fill=black] (v7) circle (2pt);   
   \coordinate (v12) at ($(v1)!0.5!(v2)$);   
   \coordinate (v23) at ($(v2)!0.5!(v3)$);
   \coordinate (v34) at ($(v3)!0.5!(v4)$);
   \coordinate (v45) at ($(v4)!0.5!(v5)$);   
   \coordinate (v56) at ($(v5)!0.5!(v6)$);   
   \coordinate (v67) at ($(v6)!0.5!(v7)$);
   \coordinate (v71) at ($(v7)!0.5!(v1)$);   
   \coordinate (v37) at ($(v3)!0.5!(v7)$);   
   \node[very thick, cross=4pt, rotate=0, color=blue, scale=.625] at (v34) {};
   \node[very thick, cross=4pt, rotate=0, color=blue, scale=.625] at (v45) {};
   \node[very thick, cross=4pt, rotate=0, color=blue, scale=.625, left=.15cm of v3] (v3l) {};
   \node[very thick, cross=4pt, rotate=0, color=blue, scale=.625, below=.15cm of v3] (v3b) {};   
   \node[very thick, cross=4pt, rotate=0, color=blue, scale=.625, below=.1cm of v5] (v5b){};
   \coordinate (a) at ($(v3l)!0.5!(v3)$);
   \coordinate (b) at ($(v3)+(0,.125)$);
   \coordinate (c) at ($(v34)+(0,.175)$);
   \coordinate (d) at ($(v4)+(0,.125)$);
   \coordinate (e) at ($(v4)+(.125,0)$);
   \coordinate (f) at ($(v45)+(.175,0)$);
   \coordinate (g) at ($(v5)+(.125,0)$);
   \coordinate (h) at ($(v5b)!0.5!(v5)$);
   \coordinate (i) at ($(v5)-(.125,0)$);
   \coordinate (j) at ($(v45)-(.175,0)$);
   \coordinate (k) at ($(v34)-(0,.175)$);
   \coordinate (l) at ($(v3)-(0,.125)$);
   \draw [thick, red!50!black] plot [smooth cycle] coordinates {(a) (b) (c) (d) (e) (f) (g) (h) (i) (j) (k) (l)};
   \node[below=.05cm of k, color=red!50!black] {\footnotesize $\displaystyle\mathfrak{g}$};   
  \end{scope}
 \end{tikzpicture}  
\end{equation*}
and the facets are identified by the collection of: the vertices $\{\mathbf{x}_v+\mathbf{y}_e-\mathbf{x'}_v,\,-\mathbf{x}_v+\mathbf{y}_e+\mathbf{x'}_v\}$ for each edge $e$ marked in the middle, $\{\mathbf{x}_v+\mathbf{y}_e-\mathbf{x'}_v,\,\mathbf{x}_v-\mathbf{y}_e+\mathbf{x'}_v$ for each edge marked close to the vertex $\mathbf{x'}_v$, and all the three vertices for those edges which are unmarked.


\subsection{Cosmological polytopes and $l\,=\,1$ states}\label{subsec:CPl1}

As reminded above, a cosmological polytope -- but in truth any positive geometry -- is uniquely characterised by its canonical form, with the properties of having logarithmic singularities on (and only on) its boundary and with the residue of any of its singularities being a lower-dimensional canonical form.

The recursion relation for the edge-weighted graphs crucially involves differential operators. This immediately implies that for an edge-weighted graph $\mathcal{G}_{\{l_e\}}$ with any $l_e\,\neq\,0$, the related wavefunction of the universe is characterised by poles with order higher than one. This fact seems to exclude the possibility of a description of any wavefunction of the universe associated to edge-weighted graphs with non-zero edge weights via any positive geometry. In this section we instead want to argue that it is indeed possible to extract it from combinatorial and geometrical objects in the same class of the cosmological polytopes.


\subsubsection{Cosmological polytopes, derivatives and high order poles}\label{subsubsec:CPdhp}

Let us begin with considering the space of two triangles with vertices $\{\mathbf{x}_i-\mathbf{y}_i+\mathbf{x}'_i,\,\mathbf{x}_i+\mathbf{y}_i-\mathbf{x}'_i,\,-\mathbf{x}_i+\mathbf{y}_i+\mathbf{x}'_i,\}$ ($i\,=\,1,\,2$). The general prescription for generating more complex objects is to intersect them on the midpoints of at most two of their sides, which we previous referred to them as {\it intersectable sides}. However, this prescription can be extended with the inclusion of the vertices of the triangles, {\it i.e.} we can allow to intersect the triangles in the midpoints of their intersectable edges and in their vertices as well. An idea which goes along these lines was discussed in \cite{Benincasa:2018ssx}, where the triangles were allowed to intersect also on the vertex shared by their intersectable sides -- which was dubbed for this reason as {\it intersectable vertex}. 

In our present context, we allow the two triangles to intersect in one of their midpoints and in the vertices opposite to them, {\it e.g.} imposing the constraints $\mathbf{x}'_1\,=\,\mathbf{x}'_2$ and $\mathbf{x}_1+\mathbf{y}_1-\mathbf{x}'_1\,=\,\mathbf{x}_2+\mathbf{y}_2-\mathbf{x}'_2$:

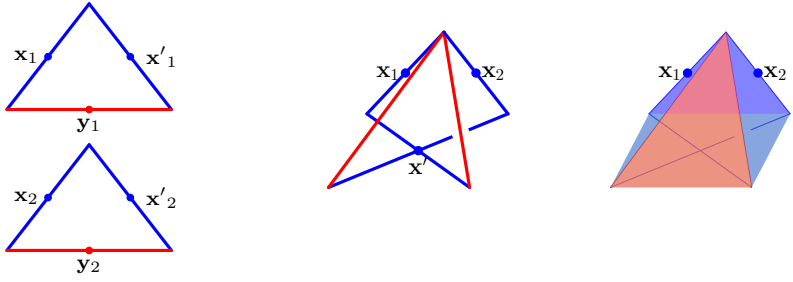
\begin{equation*}
 \begin{tikzpicture}[line join = round, line cap = round, ball/.style = {circle, draw, align=center, anchor=north, inner sep=0}, 
                     axis/.style={very thick, ->, >=stealth'}, pile/.style={thick, ->, >=stealth', shorten <=2pt, shorten>=2pt}, every node/.style={color=black}, scale={1.25}]
  \begin{scope}[scale={.5}]
   \coordinate (A) at (0,0);
   \coordinate (B) at (-1.75,-2.25);
   \coordinate (C) at (+1.75,-2.25);
   \coordinate [label=left:{\footnotesize $\displaystyle {\bf x}_1$}] (m1) at ($(A)!0.5!(B)$);
   \coordinate [label=right:{\footnotesize $\displaystyle \;{\bf x'}_1$}] (m2) at ($(A)!0.5!(C)$);
   \coordinate [label=below:{\footnotesize $\displaystyle {\bf y}_1$}] (m3) at ($(B)!0.5!(C)$);
   \tikzset{point/.style={insert path={ node[scale=2.5*sqrt(\pgflinewidth)]{.} }}} 

   \draw[color=blue,fill=blue] (m1) circle (2pt);
   \draw[color=blue,fill=blue] (m2) circle (2pt);
   \draw[color=red,fill=red] (m3) circle (2pt);

   \draw[-, very thick, color=blue] (B) -- (A);
   \draw[-, very thick, color=blue] (A) -- (C);  
   \draw[-, very thick, color=red] (B) -- (C);    
  \end{scope}
  \begin{scope}[shift={(0,-1.5)}, scale={.5}]
   \coordinate (A) at (0,0);
   \coordinate (B) at (-1.75,-2.25);
   \coordinate (C) at (+1.75,-2.25);
   \coordinate [label=left:{\footnotesize $\displaystyle {\bf x}_2$}] (m1) at ($(A)!0.5!(B)$);
   \coordinate [label=right:{\footnotesize $\displaystyle \;{\bf x'}_2$}] (m2) at ($(A)!0.5!(C)$);
   \coordinate [label=below:{\footnotesize $\displaystyle {\bf y}_2$}] (m3) at ($(B)!0.5!(C)$);
   \tikzset{point/.style={insert path={ node[scale=2.5*sqrt(\pgflinewidth)]{.} }}} 

   \draw[color=blue,fill=blue] (m1) circle (2pt);
   \draw[color=blue,fill=blue] (m2) circle (2pt);
   \draw[color=red,fill=red] (m3) circle (2pt);

   \draw[-, very thick, color=blue] (B) -- (A);
   \draw[-, very thick, color=blue] (A) -- (C);  
   \draw[-, very thick, color=red] (B) -- (C);    
  \end{scope}
  \begin{scope}[scale={.5}, shift={(7,.25)}, transform shape]
   \pgfmathsetmacro{\factor}{1/sqrt(2)};  
   \coordinate  (B2) at (1.5,-3,-1.5*\factor);
   \coordinate  (A1) at (-1.5,-3,-1.5*\factor);
   \coordinate  (B1) at (1.5,-3.75,1.5*\factor);
   \coordinate  (A2) at (-1.5,-3.75,1.5*\factor);  
   \coordinate  (C1) at (0.75,-.65,.75*\factor);
   \coordinate  (C2) at (0.4,-6.05,.75*\factor);
   \coordinate (Int) at (intersection of A2--B2 and B1--C1);
   \coordinate (Int2) at (intersection of A1--B1 and A2--B2);

   \tikzstyle{interrupt}=[
    postaction={
        decorate,
        decoration={markings,
                    mark= at position 0.5 
                          with
                          {
                            \node[rectangle, color=white, fill=white, below=-.1 of Int] {};
                          }}}
   ]
  
   \draw[interrupt,very thick,color=blue] (A1) -- (B1); 
   \draw[interrupt,very thick,color=blue] (A2) -- (B2);
   \draw[-,very thick,color=red] (B1) -- (C1);
   \draw[-,very thick,color=blue] (A1) -- (C1);
   \draw[-, very thick, color=red] (A2) -- (C1);
   \draw[-, very thick, color=blue] (B2) -- (C1);

   \coordinate[label=below:{\Large ${\bf x'}$}] (x2) at ($(A1)!0.5!(B1)$);
   \draw[fill,color=blue] (x2) circle (2.5pt);   
   \coordinate[label=left:{\Large ${\bf x}_1$}] (x1) at ($(C1)!0.5!(A1)$);
   \draw[fill,color=blue] (x1) circle (2.5pt);
   \coordinate[label=right:{\Large ${\bf x}_2$}] (x3) at ($(B2)!0.5!(C1)$);
   \draw[fill,color=blue] (x3) circle (2.5pt);
  \end{scope}
  \begin{scope}[scale={.5}, shift={(13,.25)}, transform shape]
   \pgfmathsetmacro{\factor}{1/sqrt(2)};  
   \coordinate  (B2) at (1.5,-3,-1.5*\factor);
   \coordinate  (A1) at (-1.5,-3,-1.5*\factor);
   \coordinate  (B1) at (1.5,-3.75,1.5*\factor);
   \coordinate  (A2) at (-1.5,-3.75,1.5*\factor);  
   \coordinate  (C1) at (0.75,-.65,.75*\factor);
   \coordinate  (C2) at (0.4,-6.05,.75*\factor);
   \coordinate (Int) at (intersection of A2--B2 and B1--C1);
   \coordinate (Int2) at (intersection of A1--B1 and A2--B2);

   \tikzstyle{interrupt}=[
    postaction={
        decorate,
        decoration={markings,
                    mark= at position 0.5 
                          with
                          {
                            \node[rectangle, color=white, fill=white, below=-.1 of Int] {};
                          }}}
   ]
  
   \draw[interrupt,color=blue] (A1) -- (B1); 
   \draw[interrupt,color=blue] (A2) -- (B2);
   \draw[-,color=red] (B1) -- (C1);
   \draw[-,color=blue] (A1) -- (C1);
   \draw[-, color=red] (A2) -- (C1);
   \draw[-, color=blue] (B2) -- (C1);

   \draw[draw=none,fill=green!80,opacity=.3] (A2) -- (B1) -- (B2) -- (A1) -- cycle;
   \draw[draw=none,fill=blue!60, opacity=.45] (C1) -- (B2) -- (A1) -- cycle;
   \draw[draw=none,fill=red!60, opacity=.7] (C1) -- (A2) -- (B1) -- cycle;
   \draw[draw=none,fill=blue!70, opacity=.5] (C1) -- (A1) -- (A2) -- cycle;
   \draw[draw=none,fill=blue!70, opacity=.5] (C1) -- (B2) -- (B1) -- cycle;  

   \coordinate[label=left:{\Large ${\bf x}_1$}] (x1) at ($(C1)!0.5!(A1)$);
   \draw[fill,color=blue] (x1) circle (2.5pt);
   \coordinate[label=right:{\Large ${\bf x}_2$}] (x3) at ($(B2)!0.5!(C1)$);
   \draw[fill,color=blue] (x3) circle (2.5pt);
  \end{scope}
 \end{tikzpicture}
\end{equation*}
The polytope generated in this way is a square pyramid in $\mathbb{P}^3$ with vertices
\begin{equation}\eqlabel{eq:VerPir}
 \{\mathbf{x}_1-\mathbf{y}_1+\mathbf{x}',\quad \mathbf{x}_1+\mathbf{y}_1-\mathbf{x}',\quad  -\mathbf{x}_1+\mathbf{y}_1+\mathbf{x}',\quad 2\mathbf{x}'-\mathbf{h}_2,\quad \mathbf{h}_2\},
\end{equation}
where $\mathbf{h}_2\,\equiv\,\mathbf{x}_2-\mathbf{y}_2+\mathbf{x}'\,=\,-2\mathbf{y}_2+\mathbf{x}_1+\mathbf{y}_1+\mathbf{x}'$, together with $\mathbf{x}_1$, $\mathbf{y}_1$ and $\mathbf{x}'$, parametrises the (ungauged) degrees of freedom.

One comment is now in order. This construction is equivalent to just intersect a triangle $\{\mathbf{x}_1-\mathbf{y}_1+\mathbf{x}',\, \mathbf{x}_1+\mathbf{y}_1-\mathbf{x}',\,  -\mathbf{x}_1+\mathbf{y}_1+\mathbf{x}'_1\}$ with a segment $\{2\mathbf{x}'_2-\mathbf{h}_2,\quad \mathbf{h}_2\}$, with the latter which can be thought of as a projection of a triangle $\{\mathbf{x}_2-\mathbf{h}_2+\mathbf{x}'_2,\, \mathbf{x}_2+\mathbf{h}_2-\mathbf{x}'_2,\,  -\mathbf{x}_2+\mathbf{y}_2+\mathbf{x}'_2\}$ through the cone with origin $\mathbf{O}\,\equiv\,\mathbf{x}_2-\mathbf{x}'_2$:

\begin{equation*}
 \begin{tikzpicture}[line join = round, line cap = round, ball/.style = {circle, draw, align=center, anchor=north, inner sep=0},  
                     axis/.style={very thick, ->, >=stealth'}, pile/.style={thick, ->, >=stealth', shorten <=2pt, shorten>=2pt}, every node/.style={color=black}, scale={1.25}]
  \begin{scope}[scale={.5}]
   \coordinate (A) at (0,0);
   \coordinate (B) at (-1.75,-2.25);
   \coordinate (C) at (+1.75,-2.25);
   \coordinate [label=left:{\footnotesize $\displaystyle {\bf x}_1$}] (m1) at ($(A)!0.5!(B)$);
   \coordinate [label=right:{\footnotesize $\displaystyle \;{\bf x'}_1$}] (m2) at ($(A)!0.5!(C)$);
   \coordinate [label=below:{\footnotesize $\displaystyle {\bf y}_1$}] (m3) at ($(B)!0.5!(C)$);
   \tikzset{point/.style={insert path={ node[scale=2.5*sqrt(\pgflinewidth)]{.} }}} 

   \draw[color=blue,fill=blue] (m1) circle (2pt);
   \draw[color=blue,fill=blue] (m2) circle (2pt);
   \draw[color=red,fill=red] (m3) circle (2pt);

   \draw[-, very thick, color=blue] (B) -- (A);
   \draw[-, very thick, color=blue] (A) -- (C);  
   \draw[-, very thick, color=red] (B) -- (C);    
  \end{scope}
  \begin{scope}[shift={(0,-1.5)}, scale={.5}]
   \coordinate (A) at (0,0);
   \coordinate (B) at (-1.75,-2.25);
   \coordinate (C) at (+1.75,-2.25);
   \coordinate [label=below:{\footnotesize $\displaystyle 2{\bf x'}_2-{\bf h}_2$}] (m1) at (B);
   \coordinate [label=below:{\footnotesize $\displaystyle \;{\bf h}_2$}] (m2) at (C);
   \coordinate [label=above:{\footnotesize $\displaystyle {\bf x'}_2$}] (m3) at ($(B)!0.5!(C)$);
   \tikzset{point/.style={insert path={ node[scale=2.5*sqrt(\pgflinewidth)]{.} }}} 

   \draw[color=blue,fill=red] (m3) circle (2pt);

   \draw[-, very thick, color=blue] (B) -- (C);    
   \draw[color=red, fill=red] (C) circle (2pt); 
   \draw[color=blue, fill=blue] (B) circle (2pt); 
  \end{scope}
  \begin{scope}[scale={.5}, shift={(7,.25)}, transform shape]
   \pgfmathsetmacro{\factor}{1/sqrt(2)};  
   \coordinate  (B2) at (1.5,-3,-1.5*\factor);
   \coordinate  (A1) at (-1.5,-3,-1.5*\factor);
   \coordinate  (B1) at (1.5,-3.75,1.5*\factor);
   \coordinate  (A2) at (-1.5,-3.75,1.5*\factor);  
   \coordinate  (C1) at (0.75,-.65,.75*\factor);
   \coordinate  (C2) at (0.4,-6.05,.75*\factor);
   \coordinate (Int) at (intersection of A2--B2 and B1--C1);
   \coordinate (Int2) at (intersection of A1--B1 and A2--B2);

   \tikzstyle{interrupt}=[
    postaction={
        decorate,
        decoration={markings,
                    mark= at position 0.5 
                          with
                          {
                            \node[rectangle, color=white, fill=white, below=-.1 of Int] {};
                          }}}
   ]
  
   \draw[interrupt,very thick,color=blue] (A1) -- (B1); 
   \draw[interrupt,very thick,color=blue] (A2) -- (B2);
   \draw[-,very thick,color=red] (B1) -- (C1);
   \draw[-,very thick,color=blue] (A1) -- (C1);

   \coordinate[label=below:{\Large ${\bf x'}$}] (x2) at ($(A1)!0.5!(B1)$);
   \draw[fill,color=blue] (x2) circle (2.5pt);   
   \coordinate[label=left:{\Large ${\bf x}_1$}] (x1) at ($(C1)!0.5!(A1)$);
   \draw[fill,color=blue] (x1) circle (2.5pt);
   \draw[fill,color=red] (A2) circle (2.5pt);
   \draw[fill,color=blue] (B2) circle (2.5pt);

   \coordinate [label=above:{$\displaystyle 2{\bf x'}_2-{\bf h}_2$}] (B2b) at (B2);
   \coordinate [label=below:{$\displaystyle \;{\bf h}_2$}] (A2b) at (A2);
  \end{scope}
  \begin{scope}[scale={.5}, shift={(13,.25)}, transform shape]
   \pgfmathsetmacro{\factor}{1/sqrt(2)};  
   \coordinate  (B2) at (1.5,-3,-1.5*\factor);
   \coordinate  (A1) at (-1.5,-3,-1.5*\factor);
   \coordinate  (B1) at (1.5,-3.75,1.5*\factor);
   \coordinate  (A2) at (-1.5,-3.75,1.5*\factor);  
   \coordinate  (C1) at (0.75,-.65,.75*\factor);
   \coordinate  (C2) at (0.4,-6.05,.75*\factor);
   \coordinate (Int) at (intersection of A2--B2 and B1--C1);
   \coordinate (Int2) at (intersection of A1--B1 and A2--B2);

   \tikzstyle{interrupt}=[
    postaction={
        decorate,
        decoration={markings,
                    mark= at position 0.5 
                          with
                          {
                            \node[rectangle, color=white, fill=white, below=-.1 of Int] {};
                          }}}
   ]
  
   \draw[interrupt,color=blue] (A1) -- (B1); 
   \draw[interrupt,color=blue] (A2) -- (B2);
   \draw[-,color=red] (B1) -- (C1);
   \draw[-,color=blue] (A1) -- (C1);

   \draw[draw=none,fill=green!80,opacity=.3] (A2) -- (B1) -- (B2) -- (A1) -- cycle;
   \draw[draw=none,fill=blue!60, opacity=.45] (C1) -- (B2) -- (A1) -- cycle;
   \draw[draw=none,fill=red!60, opacity=.7] (C1) -- (A2) -- (B1) -- cycle;
   \draw[draw=none,fill=blue!70, opacity=.5] (C1) -- (A1) -- (A2) -- cycle;
   \draw[draw=none,fill=blue!70, opacity=.5] (C1) -- (B2) -- (B1) -- cycle;  

   \coordinate[label=left:{\Large ${\bf x}_1$}] (x1) at ($(C1)!0.5!(A1)$);
   \draw[fill,color=blue] (x1) circle (2.5pt);

   \coordinate [label=right:{$\displaystyle 2{\bf x'}_2-{\bf h}_2$}] (B2b) at (B2);
   \coordinate [label=below:{$\displaystyle \;{\bf h}_2$}] (A2b) at (A2);
  \end{scope}
 \end{tikzpicture}
\end{equation*}
This point of view makes the connection to a graph straightforward: while, as reviewed in Section \ref{subsec:CPrev}, a triangle is in a $1-1$ correspondence with a two-site line graph, the segment is with a one-site one-loop graph -- obtaining the segment as a projection of a triangle through a cone with origin $\mathbf{O}\,\equiv\,\mathbf{x}_2-\mathbf{x}'_2$ corresponds from the graph point of view to take a two-site line graph and merging its two sites (this is nothing but the procedure to obtain a $(L+1)$-loop wavefunctions from an $L$-loop one defined in \cite{Benincasa:2018ssx}). Hence, the square pyramid with vertices $\{\mathbf{x}_1-\mathbf{y}_1+\mathbf{x}',\, \mathbf{x}_1+\mathbf{y}_1-\mathbf{x}',\,  -\mathbf{x}_1+\mathbf{y}_1+\mathbf{x}',\, 2\mathbf{x}'-\mathbf{h}_2,\, \mathbf{h}_2\}$ obtained by intersecting a triangle and a segment as just described is in a $1-1$ correspondence with the graph obtaining by merging a two-site line graph and a one-site one-loop graph:

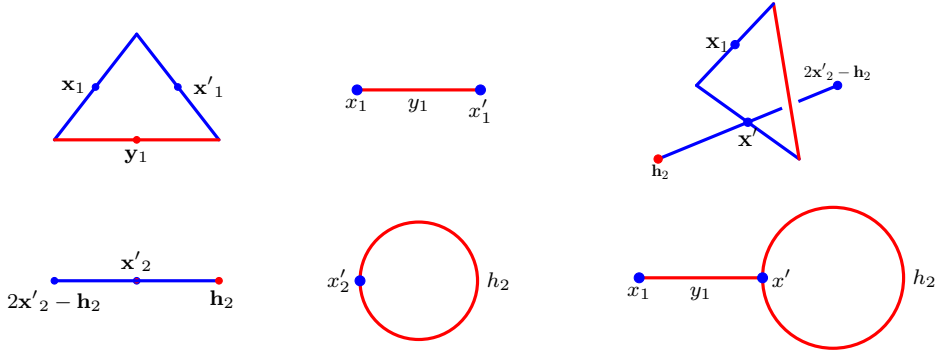
\begin{equation*}
 \begin{tikzpicture}[line join = round, line cap = round, ball/.style = {circle, draw, align=center, anchor=north, inner sep=0}, 
                     axis/.style={very thick, ->, >=stealth'}, pile/.style={thick, ->, >=stealth', shorten <=2pt, shorten>=2pt}, every node/.style={color=black}, scale={1.25}]
  \begin{scope}[scale={.5}]
   \coordinate (A) at (0,0);
   \coordinate (B) at (-1.75,-2.25);
   \coordinate (C) at (+1.75,-2.25);
   \coordinate [label=left:{\footnotesize $\displaystyle {\bf x}_1$}] (m1) at ($(A)!0.5!(B)$);
   \coordinate [label=right:{\footnotesize $\displaystyle \;{\bf x'}_1$}] (m2) at ($(A)!0.5!(C)$);
   \coordinate [label=below:{\footnotesize $\displaystyle {\bf y}_1$}] (m3) at ($(B)!0.5!(C)$);
   \tikzset{point/.style={insert path={ node[scale=2.5*sqrt(\pgflinewidth)]{.} }}} 

   \draw[color=blue,fill=blue] (m1) circle (2pt);
   \draw[color=blue,fill=blue] (m2) circle (2pt);
   \draw[color=red,fill=red] (m3) circle (2pt);

   \draw[-, very thick, color=blue] (B) -- (A);
   \draw[-, very thick, color=blue] (A) -- (C);  
   \draw[-, very thick, color=red] (B) -- (C);    
  \end{scope}
  \begin{scope}[shift={(0,-1.5)}, scale={.5}]
   \coordinate (A) at (0,0);
   \coordinate (B) at (-1.75,-2.25);
   \coordinate (C) at (+1.75,-2.25);
   \coordinate [label=below:{\footnotesize $\displaystyle 2{\bf x'}_2-{\bf h}_2$}] (m1) at (B);
   \coordinate [label=below:{\footnotesize $\displaystyle \;{\bf h}_2$}] (m2) at (C);
   \coordinate [label=above:{\footnotesize $\displaystyle {\bf x'}_2$}] (m3) at ($(B)!0.5!(C)$);
   \tikzset{point/.style={insert path={ node[scale=2.5*sqrt(\pgflinewidth)]{.} }}} 

   \draw[color=blue,fill=red] (m3) circle (2pt);

   \draw[-, very thick, color=blue] (B) -- (C);    
   \draw[color=red, fill=red] (C) circle (2pt); 
   \draw[color=blue, fill=blue] (B) circle (2pt); 
  \end{scope}
  \begin{scope}[shift={(3,.25)}, scale={.75}]
   \coordinate (A) at (0,0);
   \coordinate (B) at (-1.75,-2.25);
   \coordinate (C) at (+1.75,-2.25);
   \coordinate [label=below:{\footnotesize $\displaystyle x_1$}] (m1) at ($(A)!0.5!(B)$);
   \coordinate [label=below:{\footnotesize $\displaystyle x'_1$}] (m2) at ($(A)!0.5!(C)$);
   \coordinate [label=below:{\footnotesize $\displaystyle y_1$}] (m3) at ($(m1)!0.5!(m2)$);
   \tikzset{point/.style={insert path={ node[scale=2.5*sqrt(\pgflinewidth)]{.} }}} 

   \draw[-, very thick, color=red] (m1) -- (m2);    
   \draw[color=blue,fill=blue] (m1) circle (2pt);
   \draw[color=blue,fill=blue] (m2) circle (2pt);
  \end{scope}
  \begin{scope}[shift={(3,-1.5)}, scale={.5}]
   \coordinate (A) at (0,0);
   \coordinate (B) at (-1.75,-2.25);
   \coordinate (C) at (+1.75,-2.25);
    \coordinate (m3) at ($(B)!0.5!(C)$); 
    \coordinate [label=left:{\footnotesize $\displaystyle x'_2$}] (x) at ($(m3)+(-1.25cm,0)$);
    \coordinate [label=right:{\footnotesize $\displaystyle h_2$}] (h) at ($(m3)+(1.25cm,0)$);    
   \tikzset{point/.style={insert path={ node[scale=2.5*sqrt(\pgflinewidth)]{.} }}} 

   \draw[very thick, color=red] (m3) circle (1.25cm); 
   \draw[fill,color=blue] (x) circle (3pt);
  \end{scope}
  \begin{scope}[scale={.5}, shift={(13,1.5)}, transform shape]
   \pgfmathsetmacro{\factor}{1/sqrt(2)};  
   \coordinate  (B2) at (1.5,-3,-1.5*\factor);
   \coordinate  (A1) at (-1.5,-3,-1.5*\factor);
   \coordinate  (B1) at (1.5,-3.75,1.5*\factor);
   \coordinate  (A2) at (-1.5,-3.75,1.5*\factor);  
   \coordinate  (C1) at (0.75,-.65,.75*\factor);
   \coordinate  (C2) at (0.4,-6.05,.75*\factor);
   \coordinate (Int) at (intersection of A2--B2 and B1--C1);
   \coordinate (Int2) at (intersection of A1--B1 and A2--B2);

   \tikzstyle{interrupt}=[
    postaction={
        decorate,
        decoration={markings,
                    mark= at position 0.5 
                          with
                          {
                            \node[rectangle, color=white, fill=white, below=-.1 of Int] {};
                          }}}
   ]
  
   \draw[interrupt,very thick,color=blue] (A1) -- (B1); 
   \draw[interrupt,very thick,color=blue] (A2) -- (B2);
   \draw[-,very thick,color=red] (B1) -- (C1);
   \draw[-,very thick,color=blue] (A1) -- (C1);

   \coordinate[label=below:{\Large ${\bf x'}$}] (x2) at ($(A1)!0.5!(B1)$);
   \draw[fill,color=blue] (x2) circle (2.5pt);   
   \coordinate[label=left:{\Large ${\bf x}_1$}] (x1) at ($(C1)!0.5!(A1)$);
   \draw[fill,color=blue] (x1) circle (2.5pt);
   \draw[fill,color=red] (A2) circle (2.5pt);
   \draw[fill,color=blue] (B2) circle (2.5pt);

   \coordinate [label=above:{$\displaystyle 2{\bf x'}_2-{\bf h}_2$}] (B2b) at (B2);
   \coordinate [label=below:{$\displaystyle \;{\bf h}_2$}] (A2b) at (A2);
  \end{scope}
  \begin{scope}[shift={(6,.-1.75)}, scale={.75}]
   \coordinate (A) at (0,0);
   \coordinate (B) at (-1.75,-2.25);
   \coordinate (C) at (+1.75,-2.25);
   \coordinate [label=below:{\footnotesize $\displaystyle x_1$}] (m1) at ($(A)!0.5!(B)$);
   \coordinate [label=right:{\footnotesize $\displaystyle x'$}] (m2) at ($(A)!0.5!(C)$);
   \coordinate [label=below:{\footnotesize $\displaystyle y_1$}] (m3) at ($(m1)!0.5!(m2)$);
   \coordinate [label=right:{\footnotesize $\displaystyle h_2$}] (m4) at ($(m2)+(2,0)$);
   \tikzset{point/.style={insert path={ node[scale=2.5*sqrt(\pgflinewidth)]{.} }}} 

   \draw[-, very thick, color=red] (m1) -- (m2);    
   \draw[very thick, color=red] ($(m2)!0.5!(m4)$) circle (1cm);

   \draw[color=blue,fill=blue] (m1) circle (2pt);
   \draw[color=blue,fill=blue] (m2) circle (2pt);

  \end{scope}
 \end{tikzpicture}
\end{equation*}

The canonical form of the square pyramid returns the wavefunction of the universe for the two-site tadpole-like graph above. If $\{\mathbf{V}_a^I\}$ ($a\,=\,1,\,\ldots,5$, with the labels identifying the vertices in the order as they appear in \eqref{eq:VerPir}) is the set of the five vertices of the square pyramid, then the coefficient of its canonical form can be easily computed via a contour integral \cite{Arkani-Hamed:2017tmz, Arkani-Hamed:2017fdk}
\begin{equation}\eqlabel{eq:CCFpyd}
 \begin{split}
  \Omega(\mathcal{Y};\,\mathcal{P}_{\mathcal{G}})\:&=\:\frac{1}{(2\pi i)3!}\int_{\mathbb{R}^4}\prod_{j=1}^5\frac{dc_j}{c_j-i\varepsilon_j}\,\delta^{\mbox{\tiny $(4)$}}\left(\mathcal{Y}-\sum_{j=1}^5 c_j\mathbf{V}^{\mbox{\tiny $(j)$}}\right)\:=\\
  &=\:\frac{x_1+y_1+2x_2+2h_2}{(x_1+x_2+2h_2)(x_1+x_2)(x_1+y_1)(y_1+x_2+h_2)(y_1+x_2)}.
 \end{split}
\end{equation}
From this expression it is immediate to see that it reduces to (minus) the first derivative with respect to $x_2$ of the canonical form of the triangle, {\it i.e.} of the wavefunction of the universe related to a(n $l=0$) two-site graph! Explicitly
\begin{equation}\eqlabel{eq:CCFh20}
 \lim_{h_2\longrightarrow 0} \Omega(\mathcal{Y};\,\mathcal{P}_{\mathcal{G}})\:=\:\frac{x_1+y_1+2x_2}{(x_1+x_2)^2(x_1+y_1)(y_1+x_2)^2}\:\equiv\:-\frac{\partial}{\partial x_2}\frac{1}{(x_1+x_2)(x_1+y)(y+x_2)}. 
\end{equation}
This can be even more straightforwardly seen by considering one of the two triangulations of the square pyramid, specifically $\{1234\}+\{1235\}$
\begin{equation*}
 \begin{tikzpicture}[line join = round, line cap = round, ball/.style = {circle, draw, align=center, anchor=north, inner sep=0}, 
                     axis/.style={very thick, ->, >=stealth'}, pile/.style={thick, ->, >=stealth', shorten <=2pt, shorten>=2pt}, every node/.style={color=black}, scale={1.25}]
  \begin{scope}[scale={.5}, transform shape]
   \pgfmathsetmacro{\factor}{1/sqrt(2)};  
   \coordinate[label=right:{$\mathbf{4}$}] (B2) at (1.5,-3,-1.5*\factor);
   \coordinate[label=left:{$\mathbf{1}$}] (A1) at (-1.5,-3,-1.5*\factor);
   \coordinate[label=right:{$\mathbf{3}$}] (B1) at (1.5,-3.75,1.5*\factor);
   \coordinate[label=left:{$\mathbf{5}$}] (A2) at (-1.5,-3.75,1.5*\factor);  
   \coordinate[label=above:{$\mathbf{2}$}] (C1) at (0.75,-.65,.75*\factor);
   \coordinate  (C2) at (0.4,-6.05,.75*\factor);
   \coordinate (Int) at (intersection of A2--B2 and B1--C1);
   \coordinate (Int2) at (intersection of A1--B1 and A2--B2);

   \tikzstyle{interrupt}=[
    postaction={
        decorate,
        decoration={markings,
                    mark= at position 0.5 
                          with
                          {
                            \node[rectangle, color=white, fill=white, below=-.1 of Int] {};
                          }}}
   ]
  
   \draw[draw=none,fill=green!80,opacity=.3] (A2) -- (B1) -- (B2) -- (A1) -- cycle;
   \draw[draw=none,fill=blue!60, opacity=.45] (C1) -- (B2) -- (A1) -- cycle;
   \draw[draw=none,fill=red!60, opacity=.7] (C1) -- (A2) -- (B1) -- cycle;
   \draw[draw=none,fill=blue!70, opacity=.5] (C1) -- (A1) -- (A2) -- cycle;
   \draw[draw=none,fill=blue!70, opacity=.5] (C1) -- (B2) -- (B1) -- cycle;  

   \node[right=1.75cm of B2, scale=1.5] (eq) {$\displaystyle =$};
  \end{scope}
  \begin{scope}[scale={.5}, shift={(7,0)}, transform shape]
   \pgfmathsetmacro{\factor}{1/sqrt(2)};  
   \coordinate[label=right:{$\mathbf{4}$}] (B2) at (1.5,-3,-1.5*\factor);
   \coordinate[label=left:{$\mathbf{1}$}] (A1) at (-1.5,-3,-1.5*\factor);
   \coordinate[label=right:{$\mathbf{3}$}] (B1) at (1.5,-3.75,1.5*\factor);
   \coordinate[label=left:{$\mathbf{5}$}] (A2) at (-1.5,-3.75,1.5*\factor);  
   \coordinate[label=above:{$\mathbf{2}$}] (C1) at (0.75,-.65,.75*\factor);
   \coordinate  (C2) at (0.4,-6.05,.75*\factor);
   \coordinate (Int) at (intersection of A2--B2 and B1--C1);
   \coordinate (Int2) at (intersection of A1--B1 and A2--B2);

   \tikzstyle{interrupt}=[
    postaction={
        decorate,
        decoration={markings,
                    mark= at position 0.5 
                          with
                          {
                            \node[rectangle, color=white, fill=white, below=-.1 of Int] {};
                          }}}
   ]
  
   \draw[draw=none,fill=green!80,opacity=.6] (A1) -- (B1) -- (C1) -- cycle;
   \draw[draw=none,fill=green!80,opacity=.3] (A1) -- (B1) -- (A2) -- cycle;
   \draw[draw=none,fill=blue!60, opacity=.15] (C1) -- (B2) -- (A1) -- cycle;
   \draw[draw=none,fill=red!60, opacity=.7] (C1) -- (A2) -- (B1) -- cycle;
   \draw[draw=none,fill=blue!70, opacity=.5] (C1) -- (A1) -- (A2) -- cycle;
   \draw[draw=none,fill=blue!70, opacity=.15] (C1) -- (B2) -- (B1) -- cycle;  

   \node[right=1.75cm of B2, scale=1.5] (pl) {$\displaystyle +$};
  \end{scope}
  \begin{scope}[scale={.5}, shift={(14,0)}, transform shape]
   \pgfmathsetmacro{\factor}{1/sqrt(2)};  
   \coordinate[label=right:{$\mathbf{4}$}] (B2) at (1.5,-3,-1.5*\factor);
   \coordinate[label=left:{$\mathbf{1}$}] (A1) at (-1.5,-3,-1.5*\factor);
   \coordinate[label=right:{$\mathbf{3}$}] (B1) at (1.5,-3.75,1.5*\factor);
   \coordinate[label=left:{$\mathbf{5}$}] (A2) at (-1.5,-3.75,1.5*\factor);  
   \coordinate[label=above:{$\mathbf{2}$}] (C1) at (0.75,-.65,.75*\factor);
   \coordinate  (C2) at (0.4,-6.05,.75*\factor);
   \coordinate (Int) at (intersection of A2--B2 and B1--C1);
   \coordinate (Int2) at (intersection of A1--B1 and A2--B2);

   \tikzstyle{interrupt}=[
    postaction={
        decorate,
        decoration={markings,
                    mark= at position 0.5 
                          with
                          {
                            \node[rectangle, color=white, fill=white, below=-.1 of Int] {};
                          }}}
   ]
  
   \draw[draw=none,fill=green!80,opacity=.1] (A2) -- (A1) -- (B1) -- cycle;
   \draw[draw=none,fill=green!80,opacity=.4] (B2) -- (A1) -- (B1) -- cycle;   
   \draw[draw=none,fill=blue!60, opacity=.5] (C1) -- (B2) -- (A1) -- cycle;
   \draw[draw=none,fill=green!80,opacity=.7] (C1) -- (A1) -- (B1) -- cycle;   
   \draw[draw=none,fill=red!60, opacity=.3] (C1) -- (A2) -- (B1) -- cycle;
   \draw[draw=none,fill=blue!70, opacity=.2] (C1) -- (A1) -- (A2) -- cycle;
   \draw[draw=none,fill=blue!70, opacity=.5] (C1) -- (B2) -- (B1) -- cycle;  
  \end{scope}
 \end{tikzpicture}
\end{equation*}
\begin{equation}\eqlabel{eq:CCfpydT}
 \begin{split}
  \Omega(\mathcal{Y};\,\mathcal{P}_{\mathcal{G}})\:&=\:\frac{\langle1234\rangle^3}{\langle\mathcal{Y}123\rangle\langle\mathcal{Y}234\rangle\langle\mathcal{Y}341\rangle\langle\mathcal{Y}412\rangle}+
						       \frac{\langle3215\rangle^3}{\langle\mathcal{Y}321\rangle\langle\mathcal{Y}215\rangle\langle\mathcal{Y}153\rangle\langle\mathcal{Y}532\rangle}\:=\\
						   &=\:\frac{1}{2h_2(x_1+x_2)(x_1+y_1)(y_1+x_2)}-\frac{1}{2h_2(x_1+x_2+2h_2)(x_1+y_1)(y_1+x_2+2h_2)}\:=\\
  						   &=\:-\frac{\Omega_{\mathcal{T}}(x_1,y_1,x_2+2h_2)-\Omega_{\mathcal{T}}(x_1,y_1,x_2)}{2h_2}
 \end{split}
\end{equation}
where the last line just serves to make explicit how, in the limit $h_2\,\longrightarrow\,0$, this triangulation of the square-pyramid matches the textbook definition of the derivative of the canonical form coefficient $\Omega_{\mathcal{T}}$ of the triangle $\mathcal{T}$.

Thus, one can start from a set of $n_e$ triangles and $n_h$ segments, and intersect them in their midpoints -- or, equivalently, one can consider a set of $n_e+n_h$ triangles and intersect $n_e$ of them on the midpoints of at most two of their intersectable edges, and each of the other $n_h$ ones on one of its midpoints {\it and} the vertex opposite to it\footnote{The difference between the two ways of looking at this procedure boils down simply to a different choice of basis in $\mathbb{R}^{3n_e+n_h}$: when we intersect triangles and segments in their midpoints the basis is chosen once for all by the choice of the parametrisation of the vertices, which for the segments is conveniently chosen by thinking them as a projection of a triangle through a certain cone; when we instead consider the intersection of just triangles in their intersectable midpoints or in one of the midpoints and the vertex opposite to it, the basis depends on the way the constraints $\{{\bf x'}_i\,=\,{\bf x'}_j,\: {\bf x}_i+{\bf y}_i-{\bf x'}_i\,=\,{\bf x}_j+{\bf y}_j-{\bf x'}_j\}$ are parametrised -- the choice in \eqref{eq:VerPir}, which matches the triangle-segment construction, is $\mathbf{h}_j\,=\,{\bf x}_j-{\bf y}_j+{\bf x'}_j$, but one could analogously take $\mathbf{h}_j\,=\,{\bf x}_j-{\bf y}_j$, corresponding just to a shift $h_j\,\longrightarrow\,h_j+x'_j$. This change of basis indeed reflects in the way that the graph related to the polytope is labelled: with the choice just mentioned, the site with a tadpoles labelled by $x_j-h_j$ (rather than by $x_j$).}.

This procedure beautifully provides a combinatorial, geometric and graph theoretical understanding of derivative operators: allowing the triangles to intersect in one of their midpoints and in the vertices opposite to them -- or, equivalently, introducing a segment as a second building block and allowing it to intersect with a triangle in their midpoints -- generates a polytope whose canonical form coefficient is nothing but the Newton's difference quotient, which, from a graph theoretical point of view corresponds to glue a one-loop one-site graph to the original graph on one of its vertices. Then, taking the limit $h_j\,\longrightarrow\,0$ for all $j\,=\,1,\ldots,\,n_h$, the canonical form constructed in this way reduces -- up to a sign $(-1)^{n_h}$ -- to the action of an $n_h$-order derivative operator onto the canonical form of the cosmological polytope constructed our of the $n_e$ triangles in the usual way.


\subsubsection{The wavefunction of the universe for $l=1$ states}

Let us now consider the space of $n_e$ triangles with vertices $\{\mathbf{x}_i-\mathbf{y}_i+\mathbf{x'}_i,\,\mathbf{x}_i+\mathbf{y}_i-\mathbf{x'}_i,\,-\mathbf{x}_i+\mathbf{y}_i+\mathbf{x'}_i\}$ ($i\,=\,1,\ldots,n_e$), and of $2n_e$ segments with vertices $\{2\mathbf{\tilde{x}}_j-\mathbf{h}_j,\,\mathbf{h}_j\}$ ($j\,=\,1,\,\ldots,2n_e$). From it, we can define the space of $n_e$ polytopes $\mathcal{P}_{\mathfrak{t}}^{\mbox{\tiny $(i)$}}$ ($i\,=\,1,\ldots,n_e$) by intersecting each of the triangles with two segments, one for each midpoint of its intersectable edges. Consequently, from the discussion in the previous section, a polytope $\mathcal{P}_{\mathfrak{t}}^{\mbox{\tiny $(i)$}}$  is identified as the convex hull of the following vertices
\begin{equation}\eqlabel{eq:PB2}
 \left\{
  \mathbf{x}_i-\mathbf{y}_i+\mathbf{x'}_i,\quad\mathbf{x}_i+\mathbf{y}_i-\mathbf{x'}_i,\quad-\mathbf{x}_i+\mathbf{y}_i+\mathbf{x'}_i,\quad2\mathbf{x}_i-\mathbf{h}_i,\quad\mathbf{h}_i,\quad2\mathbf{x'}_i-\mathbf{h'}_i,\quad\mathbf{h'}_i
 \right\}.	
\end{equation}
One feature of $\mathcal{P}_{\mathfrak{t}}^{\mbox{\tiny $(i)$}}$ is that the pair of vertices of each intersectable side of the triangle belongs to the same plane of the vertices of one segment identifying a square face with midpoint $\mathbf{x}_i$ ($\mathbf{x'}_i$) -- this is just a consequence of the constraints defining each $\mathcal{P}_{\mathfrak{t}}^{\mbox{\tiny $(i)$}}$.

Thus, given a set of $n_e$ $\mathcal{P}_{\mathfrak{t}}^{\mbox{\tiny $(i)$}}$'s, we can generate more complicated polytopes $\mathcal{P}_{\mathcal{G}_{\mathfrak{t}}}$ by intersecting them in the midpoints of their square facets. The resulting polytope $\mathcal{P}_{\mathcal{G}_{\mathfrak{t}}}$ lives in $\mathbb{P}^{5n_e-r-1}$, $r$ being the number of constraints on the midpoints. For example, given two polytopes $\mathcal{P}_{\mathfrak{t}}^{\mbox{\tiny $(i)$}}$ ($i\,=\,1,2$) identified by the vertices \eqref{eq:PB2}, we can glue them in one of such midpoints via the constraint $\mathbf{x'}_1\,=\,\mathbf{x'}_2\,(\equiv\,\mathbf{x'})$. The polytope $\mathcal{P}_{\mathcal{G}_{\mathfrak{t}}}$ lives in $\mathbb{P}^8$ and is then the convex hull of the vertices
\begin{equation*}
 \begin{split}
  &\left\{
    \mathbf{x}_1-\mathbf{y}_1+\mathbf{x'},\quad\mathbf{x}_1+\mathbf{y}_1-\mathbf{x'},\quad-\mathbf{x}_1+\mathbf{y}_1+\mathbf{x'},\quad2\mathbf{x}_1-\mathbf{h}_1,\quad\mathbf{h}_1,\quad2\mathbf{x'}-\mathbf{h'}_1,\quad\mathbf{h'}_1,
   \right.\\
  &\left.
   \hspace{.125cm}\mathbf{x}_2-\mathbf{y}_2+\mathbf{x'},\quad\mathbf{x}_2+\mathbf{y}_2-\mathbf{x'},\quad-\mathbf{x}_2+\mathbf{y}_2+\mathbf{x'},\quad2\mathbf{x}_2-\mathbf{h}_2,\quad\mathbf{h}_2,\quad2\mathbf{x'}-\mathbf{h'}_2,\quad\mathbf{h'}_2
  \right\}.
 \end{split}
\end{equation*}
Importantly, any collection of such intersecting $\mathcal{P}_{\mathfrak{t}}^{\mbox{\tiny $(i)$}}$'s has a graph associated to it. Recall that any $\mathcal{P}_{\mathfrak{t}}^{\mbox{\tiny $(i)$}}$ is in turn defined as the intersection of one triangle and two segments. Recall further that the triangle can be represented by a two-site line graph, while the segment is represented by a one-site one-loop graph. In both cases, the sites of the graphs represent the sides of the triangle/segments which they can be intersected on. Hence, being the intersection of one triangle and two segments, the polytope $\mathcal{P}_{\mathfrak{t}}^{\mbox{\tiny $(i)$}}$ is represented by the graph $\mathfrak{t}$ obtained by gluing two one-loop one-site graphs with a two-site line graph, one at each site of the latter
\begin{equation}
 \begin{tikzpicture}[line join = round, line cap = round, ball/.style = {circle, draw, align=center, anchor=north, inner sep=0}, 
                     axis/.style={very thick, ->, >=stealth'}, pile/.style={thick, ->, >=stealth', shorten <=2pt, shorten>=2pt}, every node/.style={color=black}, scale={1.25}]
  \begin{scope}[scale={.75}]
   \coordinate (A) at (0,0);
   \coordinate (B) at (-1.75,-2.25);
   \coordinate (C) at (+1.75,-2.25);
   \coordinate [label=left:{\footnotesize $\displaystyle x_1$}] (m1) at ($(A)!0.5!(B)$);
   \coordinate [label=right:{\footnotesize $\displaystyle x'$}] (m2) at ($(A)!0.5!(C)$);
   \coordinate [label=below:{\footnotesize $\displaystyle y_1$}] (m3) at ($(m1)!0.5!(m2)$);
   \coordinate [label=right:{\footnotesize $\displaystyle h_2$}] (m4) at ($(m2)+(.5,0)$);
   \coordinate [label=left:{\footnotesize $\displaystyle h_1$}] (m5) at ($(m1)-(.5,0)$);   
   \tikzset{point/.style={insert path={ node[scale=2.5*sqrt(\pgflinewidth)]{.} }}} 

   \draw[-, very thick, color=red] (m1) -- (m2);    
   \draw[very thick, color=red] ($(m2)!0.5!(m4)$) circle (.25cm);
   \draw[very thick, color=red] ($(m1)!0.5!(m5)$) circle (.25cm);

   \draw[color=blue,fill=blue] (m1) circle (2pt);
   \draw[color=blue,fill=blue] (m2) circle (2pt);

   \node[left=2cm of m1] (eq) {$\displaystyle\xleftrightarrow{\hspace*{2cm}}$};
   \node[left=2cm of eq] (Pt) {$\displaystyle\mathcal{P}_{\mathfrak{t}}$};
  \end{scope}
 \end{tikzpicture}
\end{equation}
Thus, a collection of $\mathcal{P}_{\mathfrak{t}}^{\mbox{\tiny $(i)$}}$'s is represented by a collection of graphs $\mathfrak{t}_i$'s depicted above, and a polytope $\mathcal{P}_{\mathcal{G}_{\mathfrak{t}}}$ is then represented by the intersection of the $\mathfrak{t}_i$'s in their sites. For example\footnote{It is worth to stress that the disposition of the one-loop one-site subgraph (nested, internal, external, etc.) is meaningless: they are drawn as they are for pictorial convenience.}:

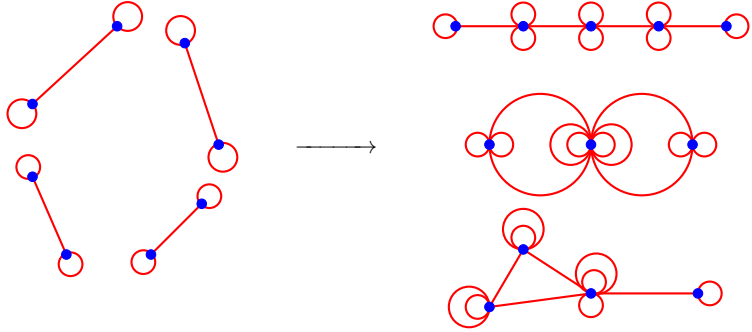
\begin{equation*}
 \begin{tikzpicture}[ball/.style = {circle, draw, align=center, anchor=north, inner sep=0}, scale={.9}, transform shape]
  \coordinate (x1) at (-.5,0) {};
  \coordinate (x2) at (-1.75,-1.15) {};
  \coordinate (x3) at (-1.75,-2.225) {};
  \coordinate (x4) at (.5,-.25) {};
  \coordinate (x5) at (1,-1.75) {};
  \coordinate (x6) at (-1.25,-3.375) {};
  \coordinate (x7) at (.75,-2.625) {};
  \coordinate (x8) at (0,-3.375) {};

  \draw[-,thick,color=red] (x1) -- (x2);
  \draw[-,thick,color=red] (x4) -- (x5);
  \draw[-,thick,color=red] (x3) -- (x6);
  \draw[-,thick,color=red] (x7) -- (x8);

  \coordinate (c12r) at ($(x1)!-0.125!(x2)$);
  \draw[thick, color=red] (c12r) circle (.2125cm);
  \coordinate (c12l) at ($(x2)!-0.125!(x1)$);
  \draw[thick, color=red] (c12l) circle (.2125cm);

  \coordinate (c45r) at ($(x4)!-0.125!(x5)$);
  \draw[thick, color=red] (c45r) circle (.2125cm);
  \coordinate (c45l) at ($(x5)!-0.125!(x4)$);
  \draw[thick, color=red] (c45l) circle (.2125cm);

  \coordinate (c36r) at ($(x3)!-0.125!(x6)$);
  \draw[thick, color=red] (c36r) circle (.175cm);
  \coordinate (c36l) at ($(x6)!-0.125!(x3)$);
  \draw[thick, color=red] (c36l) circle (.175cm);

  \coordinate (c78r) at ($(x7)!-0.15!(x8)$);
  \draw[thick, color=red] (c78r) circle (.175cm);
  \coordinate (c78l) at ($(x8)!-0.15!(x7)$);
  \draw[thick, color=red] (c78l) circle (.175cm);

  \draw[fill, color=blue] (x1) circle (2pt);
  \draw[fill, color=blue] (x2) circle (2pt);
  \draw[fill, color=blue] (x3) circle (2pt);
  \draw[fill, color=blue] (x4) circle (2pt);
  \draw[fill, color=blue] (x5) circle (2pt);
  \draw[fill, color=blue] (x6) circle (2pt);
  \draw[fill, color=blue] (x7) circle (2pt);
  \draw[fill, color=blue] (x8) circle (2pt);

  \node[right=1cm of x5.east] (arr) {$\displaystyle \xrightarrow{\hspace{1cm}}$};

  \coordinate (xa) at ($(x1.east)+(5cm, 0)$);
  \coordinate (xb) at ($(xa.east)+(1cm,0)$);
  \coordinate (xc) at ($(xb.east)+(1cm,0)$) ;
  \coordinate (xd) at ($(xc.east)+(1cm,0)$);
  \coordinate (xe) at ($(xd.east)+(1cm,0)$);
  \draw[-,thick,color=red] (xa.east) -- (xb.west);
  \draw[-,thick,color=red] (xb.east) -- (xc.west);
  \draw[-,thick,color=red] (xc.east) -- (xd.west);
  \draw[-,thick,color=red] (xd.east) -- (xe.west);

  \coordinate (cal) at ($(xa)!-0.15!(xb)$);
  \draw[thick, color=red] (cal) circle (.175cm);
  \coordinate (cbu) at ($(xb)+(0,.175cm)$);
  \draw[thick, color=red] (cbu) circle (.175cm);
  \coordinate (cbd) at ($(xb)-(0,.175cm)$);
  \draw[thick, color=red] (cbd) circle (.175cm);
  \coordinate (ccu) at ($(xc)+(0,.175cm)$);
  \draw[thick, color=red] (ccu) circle (.175cm);
  \coordinate (ccd) at ($(xc)-(0,.175cm)$);
  \draw[thick, color=red] (ccd) circle (.175cm);
  \coordinate (cdu) at ($(xd)+(0,.175cm)$);
  \draw[thick, color=red] (cdu) circle (.175cm);
  \coordinate (cdd) at ($(xd)-(0,.175cm)$);
  \draw[thick, color=red] (cdd) circle (.175cm);
  \coordinate (cer) at ($(xe)!-0.15!(xd)$);
  \draw[thick, color=red] (cer) circle (.175cm);

  \draw[fill, color=blue] (xa) circle (2pt);
  \draw[fill, color=blue] (xb) circle (2pt);
  \draw[fill, color=blue] (xc) circle (2pt);
  \draw[fill, color=blue] (xd) circle (2pt);
  \draw[fill, color=blue] (xe) circle (2pt);

  \coordinate (xg) at ($(xc)-(0,1.75cm)$);
  \coordinate (xf) at ($(xg)-(1.5cm,0)$);
  \coordinate (xh) at ($(xg)+(1.5cm,0)$);

  \draw[thick, color=red] ($(xf)!0.5!(xg)$) circle (.75);
  \draw[thick, color=red] ($(xg)!0.5!(xh)$) circle (.75);

  \coordinate (xfl) at ($(xf.west)+(-.175cm,0)$);
  \draw[thick, color=red] (xfl) circle (.175cm);
  \coordinate (xfr) at ($(xf.east)+(.175cm,0)$);
  \draw[thick, color=red] (xfr) circle (.175cm);

  \coordinate (xgla) at ($(xg.west)+(-.175cm,0)$);
  \draw[thick, color=red] (xgla) circle (.175cm);
  \coordinate (xglb) at ($(xg.west)+(-.3cm,0)$);
  \draw[thick, color=red] (xglb) circle (.3cm);
  \coordinate (xgra) at ($(xg.east)+(.175cm,0)$);
  \draw[thick, color=red] (xgra) circle (.175cm);
  \coordinate (xgrb) at ($(xg.east)+(.3cm,0)$);
  \draw[thick, color=red] (xgrb) circle (.3cm);

  \coordinate (xhl) at ($(xh.west)+(-.175cm,0)$);
  \draw[thick, color=red] (xhl) circle (.175cm);
  \coordinate (xhr) at ($(xh.east)+(.175cm,0)$);
  \draw[thick, color=red] (xhr) circle (.175cm);

  \draw[fill,blue] (xf) circle (2pt);
  \draw[fill,blue] (xg) circle (2pt);
  \draw[fill,blue] (xh) circle (2pt);

  \coordinate (xi) at ($(xb)-(0,3.3cm)$);
  \coordinate (xj) at ($(xf)-(0,2.4cm)$);
  \coordinate (xk) at ($(xg)-(0,2.2cm)$);
  \coordinate (xl) at ($(xk.east)+(1.58cm,0)$);

  \draw[-,thick,color=red] (xi) -- (xj) -- (xk) -- (xi);
  \draw[-,thick,color=red] (xk) -- (xl);

  \coordinate (ciu) at ($(xi)+(0,+.175cm)$);
  \draw[thick,color=red] (ciu) circle (.175cm);
  \coordinate (cid) at ($(xi)+(0,.3cm)$);
  \draw[thick,color=red] (cid) circle (.3cm);

  \coordinate (cju) at ($(xj)+(-.175cm,0)$);
  \draw[thick,color=red] (cju) circle (.175cm);
  \coordinate (cjd) at ($(xj)+(-.3cm,0)$);
  \draw[thick,color=red] (cjd) circle (.3cm);

  \coordinate (cka) at ($(xk)+(.045cm,+.169cm)$);
  \draw[thick,color=red] (cka) circle (.175cm);
  \coordinate (ckb) at ($(xk)+(.077cm,.29cm)$);
  \draw[thick,color=red] (ckb) circle (.3cm);
  \coordinate (ckd) at ($(xk)+(0,-.175cm)$);
  \draw[thick,color=red] (ckd) circle (.175cm);

  \coordinate (cl) at ($(xl)+(.175cm,0)$);
  \draw[thick,color=red] (cl) circle (.175cm);

  \draw[fill,blue] (xi) circle (2pt);
  \draw[fill,blue] (xj) circle (2pt);
  \draw[fill,blue] (xk) circle (2pt);
  \draw[fill,blue] (xl) circle (2pt);  

 \end{tikzpicture}
\end{equation*}
As shown in the previous subsection, the presence of one-loop one-site subgraphs in the graph $\mathcal{G}_{\mathfrak{t}}$ implies that the associated canonical form is nothing but a Newton's difference quotient of the canonical form of the related graph $\mathcal{G}$ {\it without} one-loop one-site subgraphs with respect to the variables associated to the sites of $\mathcal{G}_{\mathfrak{t}}$ where the one-loop one-site subgraphs are glued.

Thus, given a graph $\mathcal{G}_{\mathfrak{t}}$, there is a polytope $\mathcal{P}_{\mathcal{G}_{\mathfrak{t}}}$ living in\footnote{Here we write the dimension of the projective space where the polytope lives in terms of the data of the graph $\{n_v,\,n_e,\,n_h\}$, which are respectively the numbers of vertices, edges and tadpoles, as well as using the relation among these parameters, {\it i.e.} $n_v\,=\,n_e+1-L$ and $n_h\,=\,2n_e$ , $L$ being the number of internal loops.} $\mathbb{P}^{n_v+n_e+n_h-1}\,\equiv\,\mathbb{P}^{4n_e-L}$ associated to it, whose canonical form coefficient $\Omega(\mathcal{Y};\,\mathcal{P}_{\mathcal{G}_{\mathfrak{t}}})$ is the Newton' difference quotient of the wavefunction of the universe for $l\,=\,0$ with respect to a subset of variables which can be equivalently identified with {\it midpoints} of special hyperplanes in $\mathcal{P}_{\mathcal{G}_{\mathfrak{t}}})$ and with the sites of $\mathcal{G}_{\mathfrak{t}}$ with one-loop one-site graphs. In the limit $h_j\,\longrightarrow\,0$ ($\forall\:j\,=\,1,\ldots,2n_e$), the canonical form for $\mathcal{P}_{\mathcal{G}_{\mathfrak{t}}}$ returns the wavefunction of the universe for $l\,=\,1$ states, reproducing \eqref{eq:WFnu1}

\begin{equation}\eqlabel{eq:CFWFl1}
 \lim_{\{h_j\}\longrightarrow 0}\Omega(\mathcal{Y};\,\mathcal{P}_{\mathcal{G}_{\mathfrak{t}}})\:=\:(-1)^{n_e}\Psi_{\mathcal{G}_{\mathfrak{t}}}(x_v,y_e)
\end{equation}


\subsection{Facets of the polytopes $\mathcal{P}_{\mathcal{G}_{\mathfrak{t}}}$}\label{subsec:FacP}

The polytopes $\mathcal{P}_{\mathcal{G}_{\mathfrak{t}}}$ are nothing but specific sub-class of the standard cosmological polytopes. Consequently, the structure of their faces can still be analysed via the marking described in Section \ref{subsec:CPrev}. There is just one subtlety which is encoded in the presence of one-loop one-site subgraphs in the associated graph $\mathcal{G}_{\mathfrak{t}}$. In order to discuss it, we can just consider a one-loop one-site graph, {\it i.e.} the segment polytope
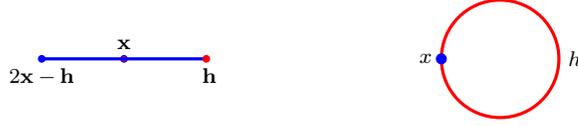
\begin{equation*}
 \begin{tikzpicture}[line join = round, line cap = round, ball/.style = {circle, draw, align=center, anchor=north, inner sep=0}, 
                     axis/.style={very thick, ->, >=stealth'}, pile/.style={thick, ->, >=stealth', shorten <=2pt, shorten>=2pt}, every node/.style={color=black}, scale={1.25}]
  \begin{scope}[scale={.5}]
   \coordinate (A) at (0,0);
   \coordinate (B) at (-1.75,-2.25);
   \coordinate (C) at (+1.75,-2.25);
   \coordinate [label=below:{\footnotesize $\displaystyle 2{\bf x}-{\bf h}$}] (m1) at (B);
   \coordinate [label=below:{\footnotesize $\displaystyle \;{\bf h}$}] (m2) at (C);
   \coordinate [label=above:{\footnotesize $\displaystyle {\bf x}$}] (m3) at ($(B)!0.5!(C)$);
   \tikzset{point/.style={insert path={ node[scale=2.5*sqrt(\pgflinewidth)]{.} }}} 

   \draw[color=blue,fill=red] (m3) circle (2pt);

   \draw[-, very thick, color=blue] (B) -- (C);    
   \draw[color=red, fill=red] (C) circle (2pt); 
   \draw[color=blue, fill=blue] (B) circle (2pt); 
  \end{scope}
  \begin{scope}[shift={(4,0)}, scale={.5}]
   \coordinate (A) at (0,0);
   \coordinate (B) at (-1.75,-2.25);
   \coordinate (C) at (+1.75,-2.25);
    \coordinate (m3) at ($(B)!0.5!(C)$); 
    \coordinate [label=left:{\footnotesize $\displaystyle x$}] (x) at ($(m3)+(-1.25cm,0)$);
    \coordinate [label=right:{\footnotesize $\displaystyle h$}] (h) at ($(m3)+(1.25cm,0)$);    
   \tikzset{point/.style={insert path={ node[scale=2.5*sqrt(\pgflinewidth)]{.} }}} 

   \draw[very thick, color=red] (m3) circle (1.25cm); 
   \draw[fill,color=blue] (x) circle (3pt);
  \end{scope}
 \end{tikzpicture}
\end{equation*}
It has two boundaries, identified by the two vertices $\{\mathbf{V}_1,\,\mathbf{V}_2\}\,=\{\mathbf{2x}-\mathbf{h},\,\mathbf{h}\}$ and which belongs to the straight lines $x+2h\,=\,0$ and $x\,=\,0$ respectively. Now, the associated graph can be thought of being obtained from a two-site line graph by merging its two sites into one. Such a constraint implies that the two vertices which are kept track of via a marking on the two extremes of the edge of the two-site line graph are made coincident. However, for consistency with the previous notation we will keep marking both. Therefore, the two facets of the segment polytope can be indicated as 
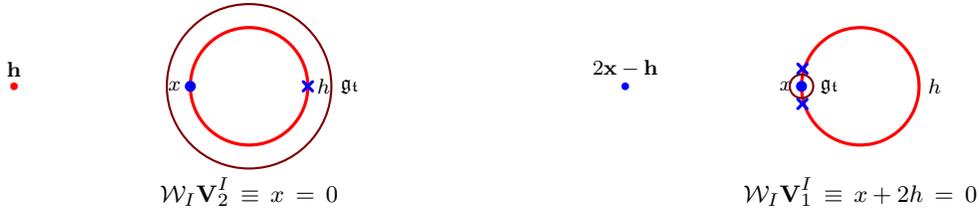
\begin{equation*}
 \begin{tikzpicture}[line join = round, line cap = round, ball/.style = {circle, draw, align=center, anchor=north, inner sep=0}, cross/.style={cross out, draw, minimum size=2*(#1-\pgflinewidth), inner sep=0pt, outer sep=0pt},
                     axis/.style={very thick, ->, >=stealth'}, pile/.style={thick, ->, >=stealth', shorten <=2pt, shorten>=2pt}, every node/.style={color=black}, scale={1.25}]
  \begin{scope}[scale={.5}]
   \coordinate (A) at (0,0);
   \coordinate (B) at (-1.75,-2.25);
   \coordinate (C) at (+1.75,-2.25);
   \coordinate [label=above:{\footnotesize $\displaystyle {\bf h}$}] (m3) at ($(B)!0.5!(C)$);
   \tikzset{point/.style={insert path={ node[scale=2.5*sqrt(\pgflinewidth)]{.} }}} 

   \draw[fill,red] (m3) circle (2pt);i
  \end{scope}
  \begin{scope}[shift={(2.5,0)}, scale={.5}]
   \coordinate (A) at (0,0);
   \coordinate (B) at (-1.75,-2.25);
   \coordinate (C) at (+1.75,-2.25);
    \coordinate (m3) at ($(B)!0.5!(C)$); 
    \coordinate [label=left:{\footnotesize $\displaystyle x$}] (x) at ($(m3)+(-1.25cm,0)$);
    \coordinate [label=right:{\footnotesize $\displaystyle h$}] (h) at ($(m3)+(1.25cm,0)$);    
   \tikzset{point/.style={insert path={ node[scale=2.5*sqrt(\pgflinewidth)]{.} }}} 

   \draw[very thick, color=red] (m3) circle (1.25cm); 
   \draw[fill,color=blue] (x) circle (3pt);
   \draw[thick,color=red!50!black] (m3) circle (1.75cm);
   \coordinate  [label=right:{\footnotesize $\displaystyle\mathfrak{g}_{\mathfrak{t}}$}] (gt) at ($(m3)+(1.75cm,0)$);
   \node[very thick, cross=4pt, rotate=0, color=blue, scale=.625] (X2) at (h) {};   
  
   \coordinate [label={\small $\displaystyle\mathcal{W}_I\mathbf{V}_2^I\,\equiv\,x\,=\,0$}] (hyp1) at ($(m3)-(0,2.75cm)$);
  \end{scope}
  \begin{scope}[shift={(6.5,0)}, scale={.5}]
   \coordinate (A) at (0,0);
   \coordinate (B) at (-1.75,-2.25);
   \coordinate (C) at (+1.75,-2.25);
   \coordinate [label=above:{\footnotesize $\displaystyle 2{\bf x} - {\bf h}$}] (m3) at ($(B)!0.5!(C)$);
   \tikzset{point/.style={insert path={ node[scale=2.5*sqrt(\pgflinewidth)]{.} }}} 

   \draw[fill,blue] (m3) circle (2pt);i
  \end{scope}
  \begin{scope}[shift={(9,0)}, scale={.5}]
   \coordinate (A) at (0,0);
   \coordinate (B) at (-1.75,-2.25);
   \coordinate (C) at (+1.75,-2.25);
    \coordinate (m3) at ($(B)!0.5!(C)$); 
    \coordinate [label=left:{\footnotesize $\displaystyle x$}] (x) at ($(m3)+(-1.25cm,0)$);
    \coordinate [label=right:{\footnotesize $\displaystyle h$}] (h) at ($(m3)+(1.25cm,0)$);    
   \tikzset{point/.style={insert path={ node[scale=2.5*sqrt(\pgflinewidth)]{.} }}} 

   \draw[very thick, color=red] (m3) circle (1.25cm); 
   \draw[fill,color=blue] (x) circle (3pt);
   \draw[thick,color=red!50!black] (x) circle (.25cm);
   \coordinate [label=left:{\footnotesize $\displaystyle\mathfrak{g}_{\mathfrak{t}}$}] (gt) at ($(x)+(1cm,0)$);
   \coordinate (va) at ($(x)+(.02,.375)$);
   \coordinate (vb) at ($(x)+(.02,-.375)$);
   \node[very thick, cross=4pt, rotate=0, color=blue, scale=.625] (X1a) at (va) {};      
   \node[very thick, cross=4pt, rotate=0, color=blue, scale=.625] (X1b) at (vb) {};      

    \coordinate [label={\small $\displaystyle\mathcal{W}_I\mathbf{V}_1^I\,\equiv\,x+2h\,=\,0$}] (hyp1) at ($(m3)-(0,2.75cm)$);
  \end{scope}
 \end{tikzpicture}
\end{equation*}
where the two vertices indicated by a marking close to the only site indicate the very same vertex $\mathbf{h}$. So, given a generic graph $\mathcal{G}_{\mathfrak{t}}$, the vertices on the facets of $\mathcal{P}_{\mathcal{G}_{\mathfrak{t}}}$ can be kept track of following the very same rule as for the standard cosmological polytopes keeping in mind that the markings associated to the two ends of the edge of the one-loop one-site subgraphs identify the very same vertex. With this in mind we can analyse the facets of $\mathcal{P}_{\mathcal{G}_{\mathfrak{t}}}$. 

A first observation is that all the facets corresponding to subgraphs containing the lowest codimension graph without tadpoles lives in $\mathbb{P}^{4n_e-L-1}$ and has $4n_e$ vertices.
 
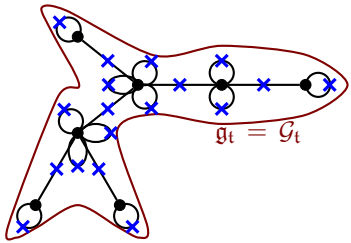
\begin{wrapfigure}{l}{4.5cm}
  \begin{tikzpicture}[ball/.style = {circle, draw, align=center, anchor=north, inner sep=0}, cross/.style={cross out, draw, minimum size=2*(#1-\pgflinewidth), inner sep=0pt, outer sep=0pt}, scale=.8, transform shape]
   \begin{scope}
    \node[ball,text width=.18cm,fill,color=black] at (0,0) (x1) {};    
    \node[ball,text width=.18cm,fill,color=black,right=1.2cm of x1.east] (x2) {};    
    \node[ball,text width=.18cm,fill,color=black,right=1.2cm of x2.east] (x3) {};
    \node[ball,text width=.18cm,fill,color=black] at (-1,.8) (x4) {};    
    \node[ball,text width=.18cm,fill,color=black] at (-1,-.8) (x5) {};    
    \node[ball,text width=.18cm,fill,color=black] at (-1.7,-2) (x6) {};    
    \node[ball,text width=.18cm,fill,color=black] at (-.3,-2) (x7) {};

    \node[above=.35cm of x5.north] (ref2) {};
    \coordinate (Int2) at (intersection of x5--x1 and ref2--x2);  

    \def\r{.225}
    \pgfmathsetmacro\x{\r*cos{60}};
    \pgfmathsetmacro\y{\r*sin{60}};
    \coordinate (c1u) at ($(x1)+(\x,\y)$);
    \coordinate (c1l) at ($(x1)!-0.175!(x2)$);
    \coordinate (c1r) at ($(x1)+(\x,-\y)$);
    \coordinate (c2u) at ($(x2)+(0,.2cm)$);
    \coordinate (c2d) at ($(x2)-(0,.2cm)$);
    \coordinate (c3) at ($(x3)!-0.15!(x2)$);
    \coordinate (c4) at ($(x4)!-0.15!(x1)$);
    \coordinate (c5u) at ($(x5)+(-\x,\y)$);
    \coordinate (c5r) at ($(x5)+(.3cm, 0)$);
    \coordinate (c5d) at ($(x5)+(0,-.3cm)$);
    \coordinate (c6) at ($(x6)!-0.15!(x5)$);
    \coordinate (c7) at ($(x7)!-0.15!(x5)$);

    \coordinate (c1xu) at ($(c1u)+(\x,\y)$);
    \coordinate (c1xl) at ($(x1)!-0.35!(x2)$);
    \coordinate (c1xr) at ($(x1)+(\x,-\y)+(\x,-\y)$);
    \coordinate (c2xu) at ($(x2)+(0,.4cm)$);
    \coordinate (c2xd) at ($(x2)-(0,.4cm)$);
    \coordinate (c3x) at ($(x3)+(.4cm,0)$);
    \coordinate (c4x) at ($(x4)!-.3!(x1)$);
    \coordinate (c5xu) at ($(c5u)+(-\x,\y)$);
    \coordinate (c5xr) at ($(c5r)+(.25cm,0)$);
    \coordinate (c5xd) at ($(c5d)+(0,-.25cm)$);
    \coordinate (c6x) at ($(x6)!-0.3!(x5)$);
    \coordinate (c7x) at ($(x7)!-0.3!(x5)$);

    \draw[-,thick,color=black] (x1) -- (x2) -- (x3); 
    \draw[-,thick,color=black] (x1) -- (x4);
    \draw[-,thick,color=black] (x5) -- (x1);
    \draw[-,thick,color=black] (x5) -- (x7);   
    \draw[-,thick,color=black] (x5) -- (x6); 

    \draw[thick] (c1u) circle (.2cm);
    \draw[thick] (c1l) ellipse (.25cm and .15cm);
    \draw[thick] (c1r) circle (.2cm);
    \draw[thick] (c2u) circle (.2cm);    
    \draw[thick] (c2d) circle (.2cm);
    \draw[thick] (c3) circle (.2cm);
    \draw[thick] (c4) circle (.2cm);
    \draw[thick] (c5u) circle (.2cm);
    \draw[thick] (c5r) ellipse (.25cm and .15cm);
    \draw[thick] (c5d) ellipse (.15cm and .25cm);
    \draw[thick] (c6) circle (.2cm);    
    \draw[thick] (c7) circle (.2cm);

    \def\rr{.113}
    \pgfmathsetmacro\xx{\rr*cos{60}};
    \pgfmathsetmacro\yy{\rr*sin{60}};
    \coordinate (a1u) at ($(c1xu)+(\xx,\yy)$);
    \coordinate (a1l) at ($(x1)!-0.7!(x2)$);
    \coordinate (a1r) at ($(c1xr)+(\xx,-\yy)$);
    \coordinate (a2u) at ($(c2xu)+(0, .25cm)$);
    \coordinate (a2d) at ($(c2xd)-(0, .25cm)$);
    \coordinate (a3u) at ($(x3)+(0, .5cm)$);
    \coordinate (a3r) at ($(x3)!-0.5!(x2)$);
    \coordinate (a3d) at ($(x3)-(0,.5cm)$);
    \coordinate (a4ur) at ($(c4x)+(\x,\y)$);
    \coordinate (a4ul) at ($(x4)!-.6!(x1)$);
    \coordinate (a4dl) at ($(c4x)-(\x,\y)$);
    \coordinate (a5u) at ($(c5xu)+(-\x,\y)$);
    \coordinate (a5r) at ($(c5xr)+(.1cm,0)$);
    \coordinate (a6ul) at ($(c6x)+(-.25,0)$);
    \coordinate (a6d) at ($(x6)!-.45!(x5)$);
    \coordinate (a6r) at ($(c6x)+(.25,0)$);
    \coordinate (a67) at ($(x6)!0.5!(x7)$);
    \coordinate (a7l) at ($(c7x)-(.25,0)$);
    \coordinate (a7d) at ($(x7)!-0.45!(x5)$);
    \coordinate (a7r) at ($(c7x)+(.25,0)$);
     
    \draw[red!50!black, thick] plot [smooth cycle] coordinates {(a3r) (a3u) (a2u) (a1u) (a4ur) (a4ul) (a4dl) (a1l) (a5u) (a6ul) (a6d) (a6r) (a67) (a7l) (a7d) (a7r) (a5r) (a1r) (a2d) (a3d)};
    \node[color=red!50!black,] at ($(x5)+(3,0)$) {\large $\mathfrak{g}_{\mathfrak{t}}\,=\,\mathcal{G}_{\mathfrak{t}}$}; 

    \coordinate (m1) at ($(x1)!0.5!(x4)$);
    \coordinate (m2) at ($(x1)!0.5!(x2)$);
    \coordinate (m3) at ($(x2)!0.5!(x3)$);
    \coordinate (m4) at ($(x1)!0.5!(x5)$);
    \coordinate (m5) at ($(x5)!0.5!(x6)$);
    \coordinate (m6) at ($(x5)!0.5!(x7)$);

    \node[very thick, cross=4pt, rotate=0, color=blue] at (m1) {};
    \node[very thick, cross=4pt, rotate=0, color=blue] at (m2) {};
    \node[very thick, cross=4pt, rotate=0, color=blue] at (m3) {};  
    \node[very thick, cross=4pt, rotate=0, color=blue] at (m4) {};  
    \node[very thick, cross=4pt, rotate=0, color=blue] at (m5) {};  
    \node[very thick, cross=4pt, rotate=0, color=blue] at (m6) {};

    \node[very thick, cross=4pt, rotate=0, color=blue] at (c1xu) {};
    \node[very thick, cross=4pt, rotate=0, color=blue] at (c1xl) {};
    \node[very thick, cross=4pt, rotate=0, color=blue] at (c1xr) {};
    \node[very thick, cross=4pt, rotate=0, color=blue] at (c2xu) {};
    \node[very thick, cross=4pt, rotate=0, color=blue] at (c2xd) {};
    \node[very thick, cross=4pt, rotate=0, color=blue] at (c3x) {};
    \node[very thick, cross=4pt, rotate=0, color=blue] at (c4x) {};
    \node[very thick, cross=4pt, rotate=0, color=blue] at (c5xu) {};
    \node[very thick, cross=4pt, rotate=0, color=blue] at (c5xr) {};
    \node[very thick, cross=4pt, rotate=0, color=blue] at (c5xd) {};
    \node[very thick, cross=4pt, rotate=0, color=blue] at (c6x) {};
    \node[very thick, cross=4pt, rotate=0, color=blue] at (c7x) {};
   \end{scope}
 \end{tikzpicture}
\end{wrapfigure}
This counting is easily done. The total number of vertices of a polytope $\mathcal{P}_{\mathcal{G}_{\mathfrak{t}}}$ is $3n_e+2n_h$, {\it i.e.} three for each straight edge of the associated graph $\mathcal{G}_{\mathfrak{t}}$ and two for each tadpole; furthermore by construction $n_h\,=\,2n_e$ and, therefore, the total number of vertices of $\mathcal{P}_{\mathcal{G}_{\mathfrak{t}}}$ is $7n_e$. When we consider any subgraph which contains all the straight edges, the related facet will have two vertices for each straight edge and one for each tadpole, so that the total number of vertices in it is $4n_e$. Thus, such a facet is a polytope which lives in $\mathbb{P}^{4n_e-L-1}$ and has $4n_e$ vertices. Consequently, for graphs with no internal loops ($L\,=\,0$), such codimension-$1$ facets are simplices -- an example is given in the picture above.

A second important observation is that the scattering amplitude is encoded in higher codimension faces. Concretely, the {\it scattering face} of a polytope $\mathcal{P}_{\mathcal{G}_{\mathfrak{t}}}$, is the face of codimension $n_h+1\,\equiv\,2n_e+1,$ identified on the associated graph $\mathcal{G}_{\mathfrak{t}}$ by the subgraph $\mathfrak{g}_{\mathfrak{t}}\,=\,\mathcal{G}_t$ and {\it all} the $2n_e$ subgraphs which exclude one tadpole at a time. Interestingly, there is an isomorphic face which is identified by the subgraph which includes {\it none} of the tadpoles and, again, {\it all} the $2n_e$ subgraphs which exclude one tadpole at a time. We will see explicit examples of them in the next subsection. What is worth to emphasise now is that the {\it scattering face} we have been discussed has the very same structure of the {\it scattering facet} which arises for the standard cosmological polytopes, being it identified by the set of vertices $\{\mathbf{x}_i+\mathbf{y}_i-\mathbf{x'}_i,\,-\mathbf{x}_i+\mathbf{y}_i+\mathbf{x'}_i\}$ ($i\,=\,1,\ldots,n_e$) associated to the sides of each edge which connects two different sites of $\mathcal{G}_{\mathfrak{t}}$
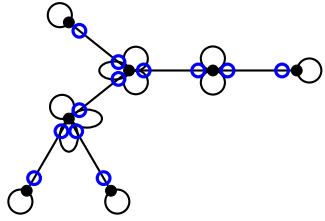
\begin{equation*}
 \begin{tikzpicture}[ball/.style = {circle, draw, align=center, anchor=north, inner sep=0}, cross/.style={cross out, draw, minimum size=2*(#1-\pgflinewidth), inner sep=0pt, outer sep=0pt}, scale=.8, transform shape]
   \begin{scope}
    \node[ball,text width=.18cm,fill,color=black] at (0,0) (x1) {};    
    \node[ball,text width=.18cm,fill,color=black,right=1.2cm of x1.east] (x2) {};    
    \node[ball,text width=.18cm,fill,color=black,right=1.2cm of x2.east] (x3) {};
    \node[ball,text width=.18cm,fill,color=black] at (-1,.8) (x4) {};    
    \node[ball,text width=.18cm,fill,color=black] at (-1,-.8) (x5) {};    
    \node[ball,text width=.18cm,fill,color=black] at (-1.7,-2) (x6) {};    
    \node[ball,text width=.18cm,fill,color=black] at (-.3,-2) (x7) {};

    \node[above=.35cm of x5.north] (ref2) {};
    \coordinate (Int2) at (intersection of x5--x1 and ref2--x2);  

    \def\r{.225}
    \pgfmathsetmacro\x{\r*cos{60}};
    \pgfmathsetmacro\y{\r*sin{60}};
    \coordinate (c1u) at ($(x1)+(\x,\y)$);
    \coordinate (c1l) at ($(x1)!-0.175!(x2)$);
    \coordinate (c1r) at ($(x1)+(\x,-\y)$);
    \coordinate (c2u) at ($(x2)+(0,.2cm)$);
    \coordinate (c2d) at ($(x2)-(0,.2cm)$);
    \coordinate (c3) at ($(x3)!-0.15!(x2)$);
    \coordinate (c4) at ($(x4)!-0.15!(x1)$);
    \coordinate (c5u) at ($(x5)+(-\x,\y)$);
    \coordinate (c5r) at ($(x5)+(.3cm, 0)$);
    \coordinate (c5d) at ($(x5)+(0,-.3cm)$);
    \coordinate (c6) at ($(x6)!-0.15!(x5)$);
    \coordinate (c7) at ($(x7)!-0.15!(x5)$);

    \coordinate (c1xu) at ($(c1u)+(\x,\y)$);
    \coordinate (c1xl) at ($(x1)!-0.35!(x2)$);
    \coordinate (c1xr) at ($(x1)+(\x,-\y)+(\x,-\y)$);
    \coordinate (c2xu) at ($(x2)+(0,.4cm)$);
    \coordinate (c2xd) at ($(x2)-(0,.4cm)$);
    \coordinate (c3x) at ($(x3)+(.4cm,0)$);
    \coordinate (c4x) at ($(x4)!-.3!(x1)$);
    \coordinate (c5xu) at ($(c5u)+(-\x,\y)$);
    \coordinate (c5xr) at ($(c5r)+(.25cm,0)$);
    \coordinate (c5xd) at ($(c5d)+(0,-.25cm)$);
    \coordinate (c6x) at ($(x6)!-0.3!(x5)$);
    \coordinate (c7x) at ($(x7)!-0.3!(x5)$);

    \draw[-,thick,color=black] (x1) -- (x2) -- (x3); 
    \draw[-,thick,color=black] (x1) -- (x4);
    \draw[-,thick,color=black] (x5) -- (x1);
    \draw[-,thick,color=black] (x5) -- (x7);   
    \draw[-,thick,color=black] (x5) -- (x6); 

    \draw[thick] (c1u) circle (.2cm);
    \draw[thick] (c1l) ellipse (.25cm and .15cm);
    \draw[thick] (c1r) circle (.2cm);
    \draw[thick] (c2u) circle (.2cm);    
    \draw[thick] (c2d) circle (.2cm);
    \draw[thick] (c3) circle (.2cm);
    \draw[thick] (c4) circle (.2cm);
    \draw[thick] (c5u) circle (.2cm);
    \draw[thick] (c5r) ellipse (.25cm and .15cm);
    \draw[thick] (c5d) ellipse (.15cm and .25cm);
    \draw[thick] (c6) circle (.2cm);    
    \draw[thick] (c7) circle (.2cm);

    \coordinate (c01) at ($(x1)!0.175!(x4)$);
    \coordinate (c02) at ($(x1)!0.825!(x4)$);
    \coordinate (c03) at ($(x1)!0.175!(x2)$);
    \coordinate (c04) at ($(x1)!0.825!(x2)$);
    \coordinate (c05) at ($(x2)!0.175!(x3)$);
    \coordinate (c06) at ($(x2)!0.825!(x3)$);
    \coordinate (c07) at ($(x1)!0.175!(x5)$);
    \coordinate (c08) at ($(x1)!0.825!(x5)$);
    \coordinate (c09) at ($(x5)!0.175!(x6)$);
    \coordinate (c10) at ($(x5)!0.825!(x6)$);
    \coordinate (c11) at ($(x5)!0.175!(x7)$);
    \coordinate (c12) at ($(x5)!0.825!(x7)$);

    \draw[very thick, color=blue] (c01) circle (3pt);
    \draw[very thick, color=blue] (c02) circle (3pt);
    \draw[very thick, color=blue] (c03) circle (3pt);
    \draw[very thick, color=blue] (c04) circle (3pt);
    \draw[very thick, color=blue] (c05) circle (3pt);
    \draw[very thick, color=blue] (c06) circle (3pt);
    \draw[very thick, color=blue] (c07) circle (3pt);
    \draw[very thick, color=blue] (c08) circle (3pt);
    \draw[very thick, color=blue] (c09) circle (3pt);
    \draw[very thick, color=blue] (c10) circle (3pt);
    \draw[very thick, color=blue] (c11) circle (3pt);
    \draw[very thick, color=blue] (c12) circle (3pt);

   \end{scope}
 \end{tikzpicture}
\end{equation*}
-- the open circles indicate the vertices of $\mathcal{P}_{\mathcal{G}_{\mathfrak{t}}}$  {\it belonging} to the face. Notice that this facet is identified by the intersection of the hyperplane $\mathcal{W}^{\mbox{\tiny (T)}}\,\equiv\,\sum_{v\in\mathcal{V}}\mathbf{x}_v$ with all the hyperplanes of the form $\mathcal{W}^{\mbox{\tiny ($\mathfrak{h}$)}}\equiv\,\sum_{v\in\mathcal{V}}\mathbf{x}_v+\sum_{e\in\mathcal{E}_{\mathfrak{h}}}\mathbf{h}_{e}$, where $\mathcal{E}_{\mathfrak{h}}$ is one of the subset of $\mathcal{E}$ containing only the edges of the tadpoles but one. The appearance of the scattering face as a codimension $2n_e+1$ face is the beautiful avatar of the fact that the high energy limit of the scattering amplitude is encoded into the leading coefficient of the Laurent expansion of the wavefunction in the neighbourhood of the total energy pole! In fact, in the limit $h_i\,\longrightarrow\,0$ ($\forall\,i\,=\,1,\ldots,2n_e$), the poles in the coefficient of the canonical form identified by these hyperplanes collapse to form a $(2n_e+1)$-order pole. This indeed the same order we would read off from the related edge.weighted graph. According \eqref{eq:OrdPol}, the order of the total energy pole is $\sum_{e\in\mathcal{E}_{\mathfrak{g}}^{\mbox{\tiny int}}}2l_e+1$, given that the total energy pole corresponds to the subgraph $\mathfrak{g}\,=\,\mathcal{G}$, with thus no external edges. Given that in our case all the edges have weight $l_e\,=\,1$, the sum in the counting formula returns the number $n_e$ of the edges of the graph. Consequently, the order of the pole is $2n_e+1$ as from the polytope analysis.


\subsection{An illustrative example}\label{subsec:ExCF}

For the sake of clarity, let us illustrate the construction just described with some example. The simplest case is given by the polytope $\mathcal{P}_{\mathcal{G}_{\mathfrak{t}}}$ itself, which lives in $\mathbb{P}^4$ and is identified by the vertices \eqref{eq:PB2}, which we write here again for convenience:
\begin{equation*}
 \left\{
  \mathbf{x}_1-\mathbf{y}_{12}+\mathbf{x}_2,\quad\mathbf{x}_1+\mathbf{y}_{12}-\mathbf{x}_2,\quad-\mathbf{x}_1+\mathbf{y}_{12}+\mathbf{x}_2,\quad2\mathbf{x}_1-\mathbf{h}_1,\quad\mathbf{h}_1,\quad2\mathbf{x}_2-\mathbf{h}_2,\quad\mathbf{h}_2
 \right\}.
\end{equation*}
It is in $1-1$ correspondence with the graph $\mathfrak{t}$:
\begin{equation*}
 \begin{tikzpicture}[line join = round, line cap = round, ball/.style = {circle, draw, align=center, anchor=north, inner sep=0}, 
                     axis/.style={very thick, ->, >=stealth'}, pile/.style={thick, ->, >=stealth', shorten <=2pt, shorten>=2pt}, every node/.style={color=black}, scale={1.25}]
  \begin{scope}[scale={.75}]
   \coordinate (A) at (0,0);
   \coordinate (B) at (-1.75,-2.25);
   \coordinate (C) at (+1.75,-2.25);
   \coordinate [label=left:{\footnotesize $\displaystyle x_1$}] (m1) at ($(A)!0.5!(B)$);
   \coordinate [label=right:{\footnotesize $\displaystyle x_2$}] (m2) at ($(A)!0.5!(C)$);
   \coordinate [label=below:{\footnotesize $\displaystyle y_{12}$}] (m3) at ($(m1)!0.5!(m2)$);
   \coordinate [label=right:{\footnotesize $\displaystyle h_2$}] (m4) at ($(m2)+(.5,0)$);
   \coordinate [label=left:{\footnotesize $\displaystyle h_1$}] (m5) at ($(m1)-(.5,0)$);   
   \tikzset{point/.style={insert path={ node[scale=2.5*sqrt(\pgflinewidth)]{.} }}} 

   \draw[-, very thick, color=red] (m1) -- (m2);    
   \draw[very thick, color=red] ($(m2)!0.5!(m4)$) circle (.25cm);
   \draw[very thick, color=red] ($(m1)!0.5!(m5)$) circle (.25cm);

   \draw[color=blue,fill=blue] (m1) circle (2pt);
   \draw[color=blue,fill=blue] (m2) circle (2pt);
  \end{scope}
 \end{tikzpicture}
\end{equation*}
Let us analyse in detail its facet structure. As mentioned in the previous section, some of the facets are simplices: they are identified by any subgraph containing the lowest codimension subgraph without tadpoles. In this case, the lowest codimension subgraph without tadpoles is the two-site line graph.
 
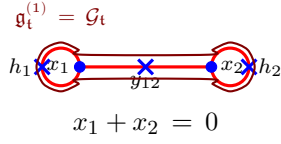
\begin{wrapfigure}{l}{4cm}
 \centering
 \begin{tikzpicture}[ball/.style = {circle, draw, align=center, anchor=north, inner sep=0}, cross/.style={cross out, draw, minimum size=2*(#1-\pgflinewidth), inner sep=0pt, outer sep=0pt}, transform shape]
  \begin{scope}[scale={1}]
   \coordinate (A) at (0,0);
   \coordinate (B) at (-1.75,-2.25);
   \coordinate (C) at (+1.75,-2.25);
   \coordinate [label=left:{\footnotesize $\displaystyle x_1$}] (m1) at ($(A)!0.5!(B)$);
   \coordinate [label=right:{\footnotesize $\displaystyle x_2$}] (m2) at ($(A)!0.5!(C)$);
   \coordinate [label=below:{\footnotesize $\displaystyle y_{12}$}] (m3) at ($(m1)!0.5!(m2)$);
   \coordinate [label=right:{\footnotesize $\displaystyle h_2$}] (m4) at ($(m2)+(.5,0)$);
   \coordinate [label=left:{\footnotesize $\displaystyle h_1$}] (m5) at ($(m1)-(.5,0)$);   
   \tikzset{point/.style={insert path={ node[scale=2.5*sqrt(\pgflinewidth)]{.} }}} 

   \draw[-, very thick, color=red] (m1) -- (m2);   
   \coordinate (cA) at ($(m1)!0.5!(m5)$);
   \coordinate (cB) at ($(m2)!0.5!(m4)$);
   \draw[very thick, color=red] ($(m2)!0.5!(m4)$) circle (.25cm);
   \draw[very thick, color=red] ($(m1)!0.5!(m5)$) circle (.25cm);

   \draw[color=blue,fill=blue] (m1) circle (2pt);
   \draw[color=blue,fill=blue] (m2) circle (2pt);

   \coordinate (c12) at ($(m1)!0.5!(m2)$);
   \coordinate (cL) at ($(m1)-(.5,0)$);
   \coordinate (cR) at ($(m2)+(.5,0)$); 

   \node[very thick, cross=4pt, rotate=0, color=blue] at (c12) {};
   \node[very thick, cross=4pt, rotate=0, color=blue] at (cL) {};
   \node[very thick, cross=4pt, rotate=0, color=blue] at (cR) {};

   \node[below=.5cm of c12] (W1) {$\displaystyle x_1+x_2\,=\,0$};

   \def\r{.325}
   \pgfmathsetmacro\bcx{\r*cos{45}};
   \pgfmathsetmacro\bcy{\r*sin{45}};
   \pgfmathsetmacro\cx{\r*cos{30}};
   \pgfmathsetmacro\cy{\r*sin{30}};

   \coordinate (a) at ($(cA)+(-\r,0)$);
   \coordinate (ab) at ($(cA)+(-\bcx,\bcy)$);
   \coordinate (b) at ($(cA)+(0,+\r)$);
   \coordinate (bc) at ($(cA)+(\bcx,\bcy)$);
   \coordinate (c) at ($(cA)+(\cx,\cy)$); 
   \coordinate (d) at ($(cB)+(-\cx,\cy)$);
   \coordinate (de) at ($(cB)+(-\bcx,\bcy)$);
   \coordinate (e) at ($(cB)+(0,\r)$);
   \coordinate (ef) at ($(cB)+(\bcx,\bcy)$);
   \coordinate (f) at ($(cB)+(\r,0)$);
   \coordinate (fg) at ($(cB)+(\bcx,-\bcy)$);
   \coordinate (g) at ($(cB)+(0,-\r)$);
   \coordinate (gh) at ($(cB)+(-\bcx,-\bcy)$);
   \coordinate (h) at ($(cB)+(-\cx,-\cy)$);
   \coordinate (i) at ($(cA)+(\cx,-\cy)$);
   \coordinate (ij) at ($(cA)+(\bcx,-\bcy)$);
   \coordinate (j) at ($(cA)+(0,-\r)$); 
   \coordinate (ja) at ($(cA)+(-\bcx,-\bcy)$);

   \draw[thick, red!50!black] plot [smooth cycle] coordinates {(a) (ab) (b) (bc) (c) (d) (de) (e) (ef) (f) (fg) (g) (gh) (h) (i) (ij) (j) (ja)};
   \node[above=.05cm of b, color=red!50!black] {\footnotesize $\displaystyle\mathfrak{g}_{\mathfrak{t}}^{\mbox{\tiny $(1)$}}\,=\,\mathcal{G}_{\mathfrak{t}}$};   
  \end{scope}
 \end{tikzpicture}
\end{wrapfigure}

There are four of such facets. The first one is identified by the graph itself and corresponds to the hyperplane $\mathcal{W}\,=\,\mathbf{x}_1+\mathbf{x}_2$. The four vertices on such a facet are
\begin{equation}\eqlabel{eq:fver1}
 \left\{
  \mathbf{x}_1+\mathbf{y}_{12}-\mathbf{x}_2,\; -\mathbf{x}_1+\mathbf{y}_{12}+\mathbf{x}_2,\;\mathbf{h}_1,\;\mathbf{h}_2
 \right\}
\end{equation}
which is indeed a tetrahedron in $\mathbb{P}^3$ with canonical form coefficient
\begin{equation}\eqlabel{eq:CFr1}
 \Omega^{\mbox{\tiny $(1)$}}\:=\:\frac{1}{(2h_1)(2h_2)(y_{12}^2-x_1^2)}.
\end{equation}

The other three facets of this type are identified by the hyperplanes related to the subgraphs which exclude either of the tadpoles as well as both, while containing the two-site line subgraph:
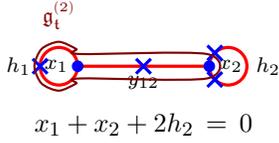
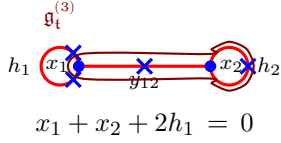
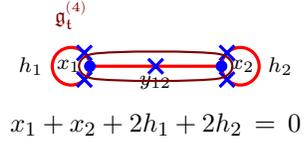
\begin{equation*}
 \begin{tikzpicture}[ball/.style = {circle, draw, align=center, anchor=north, inner sep=0}, cross/.style={cross out, draw, minimum size=2*(#1-\pgflinewidth), inner sep=0pt, outer sep=0pt}, transform shape]
  \begin{scope}
   \coordinate (A) at (0,0);
   \coordinate (B) at (-1.75,-2.25);
   \coordinate (C) at (+1.75,-2.25);
   \coordinate [label=left:{\footnotesize $\displaystyle x_1$}] (m1) at ($(A)!0.5!(B)$);
   \coordinate [label=right:{\footnotesize $\displaystyle x_2$}] (m2) at ($(A)!0.5!(C)$);
   \coordinate [label=below:{\footnotesize $\displaystyle y_{12}$}] (m3) at ($(m1)!0.5!(m2)$);
   \coordinate [label=right:{\footnotesize $\displaystyle h_2$}] (m4) at ($(m2)+(.5,0)$);
   \coordinate [label=left:{\footnotesize $\displaystyle h_1$}] (m5) at ($(m1)-(.5,0)$);   
   \tikzset{point/.style={insert path={ node[scale=2.5*sqrt(\pgflinewidth)]{.} }}} 

   \draw[-, very thick, color=red] (m1) -- (m2);   
   \coordinate (cA) at ($(m1)!0.5!(m5)$);
   \coordinate (cB) at ($(m2)!0.5!(m4)$);
   \draw[very thick, color=red] ($(m2)!0.5!(m4)$) circle (.25cm);
   \draw[very thick, color=red] ($(m1)!0.5!(m5)$) circle (.25cm);

   \draw[color=blue,fill=blue] (m1) circle (2pt);
   \draw[color=blue,fill=blue] (m2) circle (2pt);

   \coordinate (c12) at ($(m1)!0.5!(m2)$);
   \coordinate (cL) at ($(m1)-(.5,0)$);
   \pgfmathsetmacro\ax{.25*cos{135}};
   \pgfmathsetmacro\ay{.25*sin{135}};
   \coordinate (cR1) at ($(cB)+(\ax,\ay)$); 
   \coordinate (cR2) at ($(cB)+(\ax,-\ay)$);

   \node[very thick, cross=4pt, rotate=0, color=blue] at (c12) {};
   \node[very thick, cross=4pt, rotate=0, color=blue] at (cL) {};
   \node[very thick, cross=4pt, rotate=0, color=blue] at (cR1) {};
   \node[very thick, cross=4pt, rotate=0, color=blue] at (cR2) {};   

   \node[below=.5cm of c12] (W1) {$\displaystyle x_1+x_2+2h_2\,=\,0$};

   \def\r{.325}
   \pgfmathsetmacro\bcx{\r*cos{45}};
   \pgfmathsetmacro\bcy{\r*sin{45}};
   \pgfmathsetmacro\cx{\r*cos{30}};
   \pgfmathsetmacro\cy{\r*sin{30}};

   \coordinate (a) at ($(cA)+(-\r,0)$);
   \coordinate (ab) at ($(cA)+(-\bcx,\bcy)$);
   \coordinate (b) at ($(cA)+(0,+\r)$);
   \coordinate (bc) at ($(cA)+(\bcx,\bcy)$);
   \coordinate (c) at ($(cA)+(\cx,\cy)$); 
   \coordinate (d) at ($(cB)+(-\cx,\cy)$);
   \coordinate (h) at ($(cB)+(-\cx,-\cy)$);
   \coordinate (i) at ($(cA)+(\cx,-\cy)$);
   \coordinate (ij) at ($(cA)+(\bcx,-\bcy)$);
   \coordinate (j) at ($(cA)+(0,-\r)$); 
   \coordinate (ja) at ($(cA)+(-\bcx,-\bcy)$);

   \draw[thick, red!50!black] plot [smooth cycle] coordinates {(a) (ab) (b) (bc) (c) (d) (h) (i) (ij) (j) (ja)};
   \node[above=.05cm of b, color=red!50!black] {\footnotesize $\displaystyle\mathfrak{g}_{\mathfrak{t}}^{\mbox{\tiny $(2)$}}$};   
  \end{scope}
  \begin{scope}[shift={(5,0)}, transform shape]
   \coordinate (A) at (0,0);
   \coordinate (B) at (-1.75,-2.25);
   \coordinate (C) at (+1.75,-2.25);
   \coordinate [label=left:{\footnotesize $\displaystyle x_1$}] (m1) at ($(A)!0.5!(B)$);
   \coordinate [label=right:{\footnotesize $\displaystyle x_2$}] (m2) at ($(A)!0.5!(C)$);
   \coordinate [label=below:{\footnotesize $\displaystyle y_{12}$}] (m3) at ($(m1)!0.5!(m2)$);
   \coordinate [label=right:{\footnotesize $\displaystyle h_2$}] (m4) at ($(m2)+(.5,0)$);
   \coordinate [label=left:{\footnotesize $\displaystyle h_1$}] (m5) at ($(m1)-(.5,0)$);   
   \tikzset{point/.style={insert path={ node[scale=2.5*sqrt(\pgflinewidth)]{.} }}} 

   \draw[-, very thick, color=red] (m1) -- (m2);   
   \coordinate (cA) at ($(m1)!0.5!(m5)$);
   \coordinate (cB) at ($(m2)!0.5!(m4)$);
   \draw[very thick, color=red] ($(m2)!0.5!(m4)$) circle (.25cm);
   \draw[very thick, color=red] ($(m1)!0.5!(m5)$) circle (.25cm);

   \draw[color=blue,fill=blue] (m1) circle (2pt);
   \draw[color=blue,fill=blue] (m2) circle (2pt);

   \coordinate (c12) at ($(m1)!0.5!(m2)$);
   \coordinate (cR) at ($(m2)+(.5,0)$);
   \pgfmathsetmacro\ax{.25*cos{45}};
   \pgfmathsetmacro\ay{.25*sin{45}};
   \coordinate (cL1) at ($(cA)+(\ax,\ay)$); 
   \coordinate (cL2) at ($(cA)+(\ax,-\ay)$);

   \node[very thick, cross=4pt, rotate=0, color=blue] at (c12) {};
   \node[very thick, cross=4pt, rotate=0, color=blue] at (cR) {};
   \node[very thick, cross=4pt, rotate=0, color=blue] at (cL1) {};
   \node[very thick, cross=4pt, rotate=0, color=blue] at (cL2) {};   

   \node[below=.5cm of c12] (W1) {$\displaystyle x_1+x_2+2h_1\,=\,0$};

   \def\r{.325}
   \pgfmathsetmacro\bcx{\r*cos{45}};
   \pgfmathsetmacro\bcy{\r*sin{45}};
   \pgfmathsetmacro\cx{\r*cos{30}};
   \pgfmathsetmacro\cy{\r*sin{30}};

   \coordinate (b) at ($(cA)+(0,+\r)$);
   \coordinate (c) at ($(cA)+(\cx,\cy)$); 
   \coordinate (d) at ($(cB)+(-\cx,\cy)$);
   \coordinate (de) at ($(cB)+(-\bcx,\bcy)$);
   \coordinate (e) at ($(cB)+(0,\r)$);
   \coordinate (ef) at ($(cB)+(\bcx,\bcy)$);
   \coordinate (f) at ($(cB)+(\r,0)$);
   \coordinate (fg) at ($(cB)+(\bcx,-\bcy)$);
   \coordinate (g) at ($(cB)+(0,-\r)$);
   \coordinate (gh) at ($(cB)+(-\bcx,-\bcy)$);
   \coordinate (h) at ($(cB)+(-\cx,-\cy)$);
   \coordinate (i) at ($(cA)+(\cx,-\cy)$);
   
   \draw[thick, red!50!black] plot [smooth cycle] coordinates {(c) (d) (de) (e) (ef) (f) (fg) (g) (gh) (h) (i)};
   \node[above=.05cm of b, color=red!50!black] {\footnotesize $\displaystyle\mathfrak{g}_{\mathfrak{t}}^{\mbox{\tiny $(3)$}}$};   
  \end{scope}
  \begin{scope}[shift={(10,0)}, transform shape]
   \coordinate (A) at (0,0);
   \coordinate (B) at (-1.75,-2.25);
   \coordinate (C) at (+1.75,-2.25);
   \coordinate [label=left:{\footnotesize $\displaystyle x_1$}] (m1) at ($(A)!0.5!(B)$);
   \coordinate [label=right:{\footnotesize $\displaystyle x_2$}] (m2) at ($(A)!0.5!(C)$);
   \coordinate [label=below:{\footnotesize $\displaystyle y_{12}$}] (m3) at ($(m1)!0.5!(m2)$);
   \coordinate [label=right:{\footnotesize $\displaystyle h_2$}] (m4) at ($(m2)+(.5,0)$);
   \coordinate [label=left:{\footnotesize $\displaystyle h_1$}] (m5) at ($(m1)-(.5,0)$);   
   \tikzset{point/.style={insert path={ node[scale=2.5*sqrt(\pgflinewidth)]{.} }}} 

   \draw[-, very thick, color=red] (m1) -- (m2);   
   \coordinate (cA) at ($(m1)!0.5!(m5)$);
   \coordinate (cB) at ($(m2)!0.5!(m4)$);
   \draw[very thick, color=red] ($(m2)!0.5!(m4)$) circle (.25cm);
   \draw[very thick, color=red] ($(m1)!0.5!(m5)$) circle (.25cm);

   \draw[color=blue,fill=blue] (m1) circle (2pt);
   \draw[color=blue,fill=blue] (m2) circle (2pt);

   \coordinate (c12) at ($(m1)!0.5!(m2)$);
   \coordinate (cR) at ($(m2)+(.5,0)$);
   \pgfmathsetmacro\ax{.25*cos{45}};
   \pgfmathsetmacro\ay{.25*sin{45}};
   \coordinate (cL1) at ($(cA)+(\ax,\ay)$); 
   \coordinate (cL2) at ($(cA)+(\ax,-\ay)$);
   \coordinate (cR1) at ($(cB)+(-\ax,\ay)$); 
   \coordinate (cR2) at ($(cB)+(-\ax,-\ay)$);

   \node[very thick, cross=4pt, rotate=0, color=blue] at (c12) {};
   \node[very thick, cross=4pt, rotate=0, color=blue] at (cL1) {};
   \node[very thick, cross=4pt, rotate=0, color=blue] at (cL2) {};   
   \node[very thick, cross=4pt, rotate=0, color=blue] at (cR1) {};
   \node[very thick, cross=4pt, rotate=0, color=blue] at (cR2) {};

   \node[below=.5cm of c12] (W1) {$\displaystyle x_1+x_2+2h_1+2h_2\,=\,0$};

   \def\r{.325}
   \pgfmathsetmacro\bcx{\r*cos{45}};
   \pgfmathsetmacro\bcy{\r*sin{45}};
   \pgfmathsetmacro\cx{\r*cos{30}};
   \pgfmathsetmacro\cy{\r*sin{30}};

   \coordinate (b) at ($(cA)+(0,+\r)$);
   \coordinate (c) at ($(cA)+(\cx,\cy)$); 
   \coordinate (d) at ($(cB)+(-\cx,\cy)$);
   \coordinate (h) at ($(cB)+(-\cx,-\cy)$);
   \coordinate (i) at ($(cA)+(\cx,-\cy)$);
   
   \draw[thick, red!50!black] plot [smooth cycle] coordinates {(c) (d) (h) (i)};
   \node[above=.05cm of b, color=red!50!black] {\footnotesize $\displaystyle\mathfrak{g}_{\mathfrak{t}}^{\mbox{\tiny $(4)$}}$};   
  \end{scope}
 \end{tikzpicture}
\end{equation*}
whose respective vertices and canonical form coefficients are given by
\begin{equation}\eqlabel{eq:fver2}
 \begin{split}
  & \{\mathbf{x}_1+\mathbf{y}_{12}-\mathbf{x}_2,\; -\mathbf{x}_1+\mathbf{y}_{12}+\mathbf{x}_2,\;\mathbf{h}_1,\;2\mathbf{x}_2-\mathbf{h}_2\},
    \hspace{1.625cm}
    \Omega^{\mbox{\tiny $(2)$}}\:=\:-\frac{1}{(2h_1)(2h_2)(y_{12}^2-x_1^2)}\\
  & \{\mathbf{x}_1+\mathbf{y}_{12}-\mathbf{x}_2,\; -\mathbf{x}_1+\mathbf{y}_{12}+\mathbf{x}_2,\;2\mathbf{x}_1-\mathbf{h}_1,\;\mathbf{h}_2\},
    \hspace{1.625cm}
    \Omega^{\mbox{\tiny $(3)$}}\:=\:-\frac{1}{(2h_1)(2h_2)(y_{12}^2-x_2^2)}\\
  & \{\mathbf{x}_1+\mathbf{y}_{12}-\mathbf{x}_2,\; -\mathbf{x}_1+\mathbf{y}_{12}+\mathbf{x}_2,\;2\mathbf{x}_1-\mathbf{h}_1,\;2\mathbf{x}_2-\mathbf{h}_2\}, 
    \qquad
     \Omega^{\mbox{\tiny $(4)$}}\:=\:\frac{1}{(2h_1)(2h_2)(y_{12}^2-(x_1+2h_1)^2)}
 \end{split}
\end{equation}
These facets share a codimension $2$ (codimension $3$ with respect to the original polytope) face, which is identified by the intersection between the hyperplanes identified by the lowest codimension subgraph of $\mathcal{G}_{\mathfrak{t}}$ excluding only and only one tadpole at a time
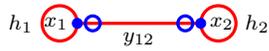
\begin{equation*}
 \begin{tikzpicture}[line join = round, line cap = round, ball/.style = {circle, draw, align=center, anchor=north, inner sep=0}, 
                     axis/.style={very thick, ->, >=stealth'}, pile/.style={thick, ->, >=stealth', shorten <=2pt, shorten>=2pt}, every node/.style={color=black}, scale={1.25}]
  \begin{scope}[scale={.75}]
   \coordinate (A) at (0,0);
   \coordinate (B) at (-1.75,-2.25);
   \coordinate (C) at (+1.75,-2.25);
   \coordinate [label=left:{\footnotesize $\displaystyle x_1$}] (m1) at ($(A)!0.5!(B)$);
   \coordinate [label=right:{\footnotesize $\displaystyle x_2$}] (m2) at ($(A)!0.5!(C)$);
   \coordinate [label=below:{\footnotesize $\displaystyle y_{12}$}] (m3) at ($(m1)!0.5!(m2)$);
   \coordinate [label=right:{\footnotesize $\displaystyle h_2$}] (m4) at ($(m2)+(.5,0)$);
   \coordinate [label=left:{\footnotesize $\displaystyle h_1$}] (m5) at ($(m1)-(.5,0)$);   
   \tikzset{point/.style={insert path={ node[scale=2.5*sqrt(\pgflinewidth)]{.} }}} 

   \draw[-, very thick, color=red] (m1) -- (m2);    
   \draw[very thick, color=red] ($(m2)!0.5!(m4)$) circle (.25cm);
   \draw[very thick, color=red] ($(m1)!0.5!(m5)$) circle (.25cm);
 
   \draw[color=blue,fill=blue] (m1) circle (2pt);
   \draw[color=blue,fill=blue] (m2) circle (2pt);

   \coordinate (cL) at ($(m1)!.125!(m2)$);
   \coordinate (cR) at ($(m1)!.875!(m2)$);
   \draw[very thick, color=blue] (cL) circle (3pt);   
   \draw[very thick, color=blue] (cR) circle (3pt);      
  \end{scope}
 \end{tikzpicture}
\end{equation*}
where, as in the previous section, the open circles indicates the two vertices {\it belonging} to the face. This is equivalent to taking the residues in $h_1\,=\,0$ and $h_2\,=\,0$ in each of the four canonical form coefficients $ \Omega^{\mbox{\tiny $(j)$}}$ ($j\,=\,1,\ldots,4$), returning the Lorentz-invariant flat-space scattering amplitude. In this concrete example we are just seeing what we discussed in full generality at the end of the previous section: the scattering face is a higher codimension face of the polytope, and its codimension corresponds to the order of the pole when the limits $h_j\,\longrightarrow\,0$ ($j\,=\,1,\,2$) are taken.

The polytope in question has four more facets: they are identified by any other subgraph which does not include the two-site line one, {\it i.e.} in our concrete example, they can include one tadpole and one site at a time or just one site:
\begin{equation*}
 \begin{tikzpicture}[line join = round, line cap = round, ball/.style = {circle, draw, align=center, anchor=north, inner sep=0}, cross/.style={cross out, draw, minimum size=2*(#1-\pgflinewidth), inner sep=0pt, outer sep=0pt}, 
                     axis/.style={very thick, ->, >=stealth'}, pile/.style={thick, ->, >=stealth', shorten <=2pt, shorten>=2pt}, every node/.style={color=black}, scale={1.25}]
  \begin{scope}[scale={.75}, transform shape]
   \coordinate (A) at (0,0);
   \coordinate (B) at (-1.75,-2.25);
   \coordinate (C) at (+1.75,-2.25);
   \coordinate [label=left:{\footnotesize $\displaystyle x_1$}] (m1) at ($(A)!0.5!(B)$);
   \coordinate [label=right:{\footnotesize $\displaystyle x_2$}] (m2) at ($(A)!0.5!(C)$);
   \coordinate [label=below:{\footnotesize $\displaystyle y_{12}$}] (m3) at ($(m1)!0.5!(m2)$);
   \coordinate [label=right:{\footnotesize $\displaystyle h_2$}] (m4) at ($(m2)+(.5,0)$);
   \coordinate [label=left:{\footnotesize $\displaystyle h_1$}] (m5) at ($(m1)-(.5,0)$);   
   \tikzset{point/.style={insert path={ node[scale=2.5*sqrt(\pgflinewidth)]{.} }}} 

   \draw[-, very thick, color=red] (m1) -- (m2);    
   \draw[very thick, color=red] ($(m2)!0.5!(m4)$) circle (.25cm);
   \draw[very thick, color=red] ($(m1)!0.5!(m5)$) circle (.25cm);

   \draw[color=blue,fill=blue] (m1) circle (2pt);
   \draw[color=blue,fill=blue] (m2) circle (2pt);

   \draw[thick, color=red!50!black] ($(m1)!0.5!(m5)$) circle (.325cm);

   \coordinate (cLL) at ($(m1)-(.5,0)$);
   \coordinate (cL) at ($(m1)!.125!(m2)$);
   \node[very thick, cross=4pt, rotate=0, color=blue] at (cLL) {};
   \node[very thick, cross=4pt, rotate=0, color=blue] at (cL) {};   

   \coordinate (c12) at ($(m1)!0.5!(m2)$);
   \node[below=.5cm of c12] (W1) {$\displaystyle x_1+y_{12}\,=\,0$};   

   \node[right=2cm of m2] (v1) {$\displaystyle\{\mathbf{x}_1-\mathbf{y}_{12}+\mathbf{x}_2,\;-\mathbf{x}_1+\mathbf{y}_{12}+\mathbf{x}_2,\;\mathbf{h}_1,\;2\mathbf{x}_2-\mathbf{h}_2,\;\mathbf{h}_2\}$};
  \end{scope}
  \begin{scope}[scale={.75}, shift={(0,-2)}, transform shape]
   \coordinate (A) at (0,0);
   \coordinate (B) at (-1.75,-2.25);
   \coordinate (C) at (+1.75,-2.25);
   \coordinate [label=left:{\footnotesize $\displaystyle x_1$}] (m1) at ($(A)!0.5!(B)$);
   \coordinate [label=right:{\footnotesize $\displaystyle x_2$}] (m2) at ($(A)!0.5!(C)$);
   \coordinate [label=below:{\footnotesize $\displaystyle y_{12}$}] (m3) at ($(m1)!0.5!(m2)$);
   \coordinate [label=right:{\footnotesize $\displaystyle h_2$}] (m4) at ($(m2)+(.5,0)$);
   \coordinate [label=left:{\footnotesize $\displaystyle h_1$}] (m5) at ($(m1)-(.5,0)$);   
   \tikzset{point/.style={insert path={ node[scale=2.5*sqrt(\pgflinewidth)]{.} }}} 

   \draw[-, very thick, color=red] (m1) -- (m2);    
   \draw[very thick, color=red] ($(m2)!0.5!(m4)$) circle (.25cm);
   \draw[very thick, color=red] ($(m1)!0.5!(m5)$) circle (.25cm);

   \draw[color=blue,fill=blue] (m1) circle (2pt);
   \draw[color=blue,fill=blue] (m2) circle (2pt);

   \draw[thick, color=red!50!black] (m1) circle (4pt);

   \coordinate (cA) at ($(m1)!0.5!(m5)$);
   \coordinate (cB) at ($(m2)!0.5!(m4)$);
   \coordinate (cL) at ($(m1)!.125!(m2)$);
   \pgfmathsetmacro\ax{.25*cos{45}};
   \pgfmathsetmacro\ay{.25*sin{45}};
   \coordinate (cL1) at ($(cA)+(\ax,\ay)$); 
   \coordinate (cL2) at ($(cA)+(\ax,-\ay)$);
   \node[very thick, cross=4pt, rotate=0, color=blue] at (cL1) {};
   \node[very thick, cross=4pt, rotate=0, color=blue] at (cL2) {};   
   \node[very thick, cross=4pt, rotate=0, color=blue] at (cL) {}; 

   \coordinate (c12) at ($(m1)!0.5!(m2)$);
   \node[below=.5cm of c12] (W1) {$\displaystyle x_1+y_{12}+2h_1\,=\,0$};  

   \node[right=2cm of m2] (v1) {$\displaystyle\{\mathbf{x}_1-\mathbf{y}_{12}+\mathbf{x}_2,\;-\mathbf{x}_1+\mathbf{y}_{12}+\mathbf{x}_2,\;2\mathbf{x}_1-\mathbf{h}_1,\;2\mathbf{x}_2-\mathbf{h}_2,\;\mathbf{h}_2\}$};
  \end{scope}
  \begin{scope}[scale={.75}, shift={(0,-4)}, transform shape]
   \coordinate (A) at (0,0);
   \coordinate (B) at (-1.75,-2.25);
   \coordinate (C) at (+1.75,-2.25);
   \coordinate [label=left:{\footnotesize $\displaystyle x_1$}] (m1) at ($(A)!0.5!(B)$);
   \coordinate [label=right:{\footnotesize $\displaystyle x_2$}] (m2) at ($(A)!0.5!(C)$);
   \coordinate [label=below:{\footnotesize $\displaystyle y_{12}$}] (m3) at ($(m1)!0.5!(m2)$);
   \coordinate [label=right:{\footnotesize $\displaystyle h_2$}] (m4) at ($(m2)+(.5,0)$);
   \coordinate [label=left:{\footnotesize $\displaystyle h_1$}] (m5) at ($(m1)-(.5,0)$);   
   \tikzset{point/.style={insert path={ node[scale=2.5*sqrt(\pgflinewidth)]{.} }}} 

   \draw[-, very thick, color=red] (m1) -- (m2);    
   \draw[very thick, color=red] ($(m2)!0.5!(m4)$) circle (.25cm);
   \draw[very thick, color=red] ($(m1)!0.5!(m5)$) circle (.25cm);

   \draw[color=blue,fill=blue] (m1) circle (2pt);
   \draw[color=blue,fill=blue] (m2) circle (2pt);

   \draw[thick, color=red!50!black] ($(m2)!0.5!(m4)$) circle (.325cm);

   \coordinate (cRR) at ($(m2)+(.5,0)$);
   \coordinate (cR) at ($(m1)!.875!(m2)$);
   \node[very thick, cross=4pt, rotate=0, color=blue] at (cRR) {};
   \node[very thick, cross=4pt, rotate=0, color=blue] at (cR) {};   

   \coordinate (c12) at ($(m1)!0.5!(m2)$);
   \node[below=.5cm of c12] (W1) {$\displaystyle y_{12}+x_2\,=\,0$};   

   \node[right=2cm of m2] (v1) {$\displaystyle\{\mathbf{x}_1-\mathbf{y}_{12}+\mathbf{x}_2,\;\mathbf{x}_1+\mathbf{y}_{12}-\mathbf{x}_2,\;2\mathbf{x}_1-\mathbf{h}_1,\;\mathbf{h}_1,\;\mathbf{h}_2\}$};
  \end{scope}
  \begin{scope}[scale={.75}, shift={(0,-6)}, transform shape]
   \coordinate (A) at (0,0);
   \coordinate (B) at (-1.75,-2.25);
   \coordinate (C) at (+1.75,-2.25);
   \coordinate [label=left:{\footnotesize $\displaystyle x_1$}] (m1) at ($(A)!0.5!(B)$);
   \coordinate [label=right:{\footnotesize $\displaystyle x_2$}] (m2) at ($(A)!0.5!(C)$);
   \coordinate [label=below:{\footnotesize $\displaystyle y_{12}$}] (m3) at ($(m1)!0.5!(m2)$);
   \coordinate [label=right:{\footnotesize $\displaystyle h_2$}] (m4) at ($(m2)+(.5,0)$);
   \coordinate [label=left:{\footnotesize $\displaystyle h_1$}] (m5) at ($(m1)-(.5,0)$);   
   \tikzset{point/.style={insert path={ node[scale=2.5*sqrt(\pgflinewidth)]{.} }}} 

   \draw[-, very thick, color=red] (m1) -- (m2);    
   \draw[very thick, color=red] ($(m2)!0.5!(m4)$) circle (.25cm);
   \draw[very thick, color=red] ($(m1)!0.5!(m5)$) circle (.25cm);

   \draw[color=blue,fill=blue] (m1) circle (2pt);
   \draw[color=blue,fill=blue] (m2) circle (2pt);

   \draw[thick, color=red!50!black] (m2) circle (4pt);

   \coordinate (cA) at ($(m1)!0.5!(m5)$);
   \coordinate (cB) at ($(m2)!0.5!(m4)$);
   \coordinate (cR) at ($(m1)!.875!(m2)$);
   \pgfmathsetmacro\ax{.25*cos{45}};
   \pgfmathsetmacro\ay{.25*sin{45}};
   \coordinate (cR1) at ($(cB)+(-\ax,\ay)$); 
   \coordinate (cR2) at ($(cB)+(-\ax,-\ay)$);
   \node[very thick, cross=4pt, rotate=0, color=blue] at (cR1) {};
   \node[very thick, cross=4pt, rotate=0, color=blue] at (cR2) {};   
   \node[very thick, cross=4pt, rotate=0, color=blue] at (cR) {}; 

   \coordinate (c12) at ($(m1)!0.5!(m2)$);
   \node[below=.5cm of c12] (W1) {$\displaystyle x_1+y_{12}+2h_1\,=\,0$};  

   \node[right=2cm of m2] (v1) {$\displaystyle\{\mathbf{x}_1-\mathbf{y}_{12}+\mathbf{x}_2,\;-\mathbf{x}_1+\mathbf{y}_{12}+\mathbf{x}_2,\;2\mathbf{x}_1-\mathbf{h}_1,\;\mathbf{h}_1\;2\mathbf{x}_2-\mathbf{h}_2\}$};
  \end{scope}
 \end{tikzpicture}
\end{equation*}
These facets are just square pyramids, whose square face has $\mathbf{x}_i$ related to the site outside of the subgraph as a midpoint
\begin{equation*}
 \begin{tikzpicture}[line join = round, line cap = round, ball/.style = {circle, draw, align=center, anchor=north, inner sep=0}, 
                     axis/.style={very thick, ->, >=stealth'}, pile/.style={thick, ->, >=stealth', shorten <=2pt, shorten>=2pt}, every node/.style={color=black}, scale={1.25}]
  \begin{scope}[scale={.5}, transform shape]
   \pgfmathsetmacro{\factor}{1/sqrt(2)};  
   \coordinate[label=right:{$\mathbf{4}$}] (B2) at (1.5,-3,-1.5*\factor);
   \coordinate[label=left:{$\mathbf{1}$}] (A1) at (-1.5,-3,-1.5*\factor);
   \coordinate[label=right:{$\mathbf{3}$}] (B1) at (1.5,-3.75,1.5*\factor);
   \coordinate[label=left:{$\mathbf{5}$}] (A2) at (-1.5,-3.75,1.5*\factor);  
   \coordinate[label=above:{$\mathbf{2}$}] (C1) at (0.75,-.65,.75*\factor);
   \coordinate  (C2) at (0.4,-6.05,.75*\factor);
   \coordinate (Int) at (intersection of A2--B2 and B1--C1);
   \coordinate (Int2) at (intersection of A1--B1 and A2--B2);

   \tikzstyle{interrupt}=[
    postaction={
        decorate,
        decoration={markings,
                    mark= at position 0.5 
                          with
                          {
                            \node[rectangle, color=white, fill=white, below=-.1 of Int] {};
                          }}}
   ]
  
   \draw[draw=none,fill=green!80,opacity=.3] (A2) -- (B1) -- (B2) -- (A1) -- cycle;
   \draw[draw=none,fill=blue!60, opacity=.45] (C1) -- (B2) -- (A1) -- cycle;
   \draw[draw=none,fill=blue!80, opacity=.7] (C1) -- (A2) -- (B1) -- cycle;
   \draw[draw=none,fill=blue!70, opacity=.5] (C1) -- (A1) -- (A2) -- cycle;
   \draw[draw=none,fill=blue!70, opacity=.5] (C1) -- (B2) -- (B1) -- cycle;  

   \node[right=1.75cm of B2, scale=1.5] (eq) {$\displaystyle\Omega^{\mbox{\tiny $(a)$}}\:=\:(-1)^{\sigma_i}\frac{2(x_j+h_j)}{2h_i(y_{12}^2-x_j^2)(y_{12}^2-(x_j+2h_j)^2)}$};
  \end{scope}
 \end{tikzpicture}
\end{equation*}
which is identified by the hyperplane corresponding to the equation $x_i+y_{12}+\sigma_i2h_i$, with $\sigma_i\,=\,0,1$, and $(i\,=\,1,\,2)$. If we now go on the facet of this square pyramid identified by $h_i\,=\,0$, {\it i.e.} its square base, which is a codimension $2$ face of the original polytope, the related canonical form is such that, in the degenerate limit $h_j\,\longrightarrow\,0$, reduces to the derivative of the scattering amplitude with respect to $x_j$: 
\begin{equation}\eqlabel{eq:deglimCF}
 \Omega\:=\:(-1)^{\sigma_i}\frac{2(x_j+h_j)}{(y_{12}^2-x_j^2)(y_{12}^2-(x_j+2h_j)^2)}\:\xrightarrow{h_j\longrightarrow0}\:(-1)^{\sigma_i}\frac{2x_j}{(y_{12}^2-x_j^2)^2}
       \:\equiv\:(-1)^{\sigma_i}\frac{\partial}{\partial x_j}\frac{1}{y_{12}^2-x_j^2}
\end{equation}

Hence, the degenerate limit of the canonical form of the codimension two face of the original polytope $\mathcal{P}_{\mathfrak{t}}$ is exactly the coefficient of the expected double pole in $x_i+y_{12}$. Thus, with this simple example we have provided an illustration of the beautiful fact, proved in general in the previous section, that the coefficient of the highest order poles of the wavefunction of the universe emerge as the degenerate limit ({\it i.e.} $h_i\,\longrightarrow\,0$ for all $i$'s) of the canonical form of higher codimension faces, which are easily identified via the associated graph. Beautifully, the codimension corresponds to the order of the pole in the wavefunction.


\subsection{Cosmological polytopes and perturbative mass}\label{subsec:CPpert}

Let us now discuss the combinatorics for the contribution to the wavefunction considering two-point couplings, which has been discussed in Section \ref{sec:Pert}. One of the key features of the graphs with two point vertices is the fact the same $y$ is associated to the two edges joined by such a vertex, which, as already shown, implies the presence of a high order pole: taking residues of the wavefunction associated to the graph with respect to the variables associated to its sites, then if $n_e$ edges are connected via $n_e-1$ white sites, then one gets an $n_e$-order pole in $2y_e$, {\it i.e.} $\prod_{e\in\bar{\mathcal{E}}}(2y_e)^{-n_e}$ -- where $\bar{\mathcal{E}}$ is the subset of edges which differs among each other for the associated $y_e$. Consequently, it is not in principle possible to associated a canonical positive geometry to these type of graphs, given that the canonical form associated to them are characterised by having single poles only.

However, the discussion on the halohedron and the one-loop bi-adjoint scalar \cite{Salvatori:2018aha} as well as the discussion of the previous section, taught us that functions with high order poles can be thought of as a degenerate limit of some canonical form. 

In the case of graphs with black and white vertices introduced in Section \ref{sec:Pert}, it is straightforward to identify the positive geometric which we should take the degenerate limit of: given that any graph with black and white sites satisfies the same combinatorial rules as the standard reduced graphs and can be thought of as a limit of them\footnote{Despite in \eqref{eq:WFdeg} we draw a line graph, such a relation hold generally: it is enough to substitute the two black nodes with arbitrary complicated graphs.}
\begin{equation}\eqlabel{eq:WFdeg}
 \begin{tikzpicture}[line join = round, line cap = round, ball/.style = {circle, draw, align=center, anchor=north, inner sep=0}, 
                     axis/.style={very thick, ->, >=stealth'}, pile/.style={thick, ->, >=stealth', shorten <=2pt, shorten>=2pt}, every node/.style={color=black}]
  \begin{scope}
   \coordinate[label=below:{\footnotesize $x_1$}] (x1) at (0,0);
   \coordinate[label=below:{\footnotesize $\omega_1$}] (w1) at ($(x1)+(1,0)$);
   \coordinate (t1) at ($(w1)+(.25,0)$);
   \coordinate (t2) at ($(t1)+(1,0)$);
   \coordinate[label=below:{\footnotesize $\omega_a$}] (w2) at ($(t2)+(.25,0)$);
   \coordinate[label=below:{\footnotesize $x_2$}] (x2) at ($(w2)+(1,0)$);
   \draw[-,thick] (x1) -- node[above] {\footnotesize $y$} (w1) -- (t1);
   \draw[-,dotted] (t1) -- (t2);
   \draw[-,thick] (t2) -- (w2) -- node[above] {\footnotesize $y$} (x2);
   \draw[fill,black] (x1) circle (2pt);
   \draw[fill=white] (w1) circle (2pt);
   \draw[fill=white] (w2) circle (2pt);
   \draw[fill,black] (x2) circle (2pt);
   \node[right=.25cm of x2] (eq) {$\displaystyle=\:\int\prod_{e\in\mathcal{E}}dy_e\delta(y_e-y)$};
  \end{scope}
  \begin{scope}[shift={(7.5,0)}, transform shape]
   \coordinate[label=below:{\footnotesize $x_1$}] (x1) at (0,0);
   \coordinate[label=below:{\footnotesize $\omega_1$}] (w1) at ($(x1)+(1,0)$);
   \coordinate (t1) at ($(w1)+(.25,0)$);
   \coordinate (t2) at ($(t1)+(1,0)$);
   \coordinate[label=below:{\footnotesize $\omega_a$}] (w2) at ($(t2)+(.25,0)$);
   \coordinate[label=below:{\footnotesize $x_2$}] (x2) at ($(w2)+(1,0)$);
   \draw[-,thick] (x1) -- node[above] {\footnotesize $y_{e_{11}}$} (w1) -- (t1);
   \draw[-,dotted] (t1) -- (t2);
   \draw[-,thick] (t2) -- (w2) -- node[above] {\footnotesize $y_{e_{a2}}$} (x2);
   \draw[fill,black] (x1) circle (2pt);
   \draw[fill=black] (w1) circle (2pt);
   \draw[fill=black] (w2) circle (2pt);
   \draw[fill,black] (x2) circle (2pt);
  \end{scope}
 \end{tikzpicture}
\end{equation}
Hence, the graphs with black and white graphs are related to a degenerate limit of the canonical form of the standard cosmological polytopes.

As a final comment, we can also consider the mass insertion on the edge-weighted graphs, as already discussed in Section \ref{sec:Pert}. Then, the mass correction corresponding to a graph with $n_m$ mass insertions is obtained as a double degenerate limit of the polytopes introduced in this paper: one class of limits, $h_j\,\longrightarrow\,0$, which makes poles collapse into higher order ones, and the other, $y_e\,\longrightarrow y$ for each $e$ connected by two point vertices.


\section{Conclusion}\label{sec:Concl}

In the last two years we started to scratch the surface of what the general features of cosmological observables, equivalently the wavefunction of the universe and the spatial correlators, may be and how fundamental physics is encoded into them. Contrasting to the status of the physics at sufficiently high energies, the requirements of unitarity, locality and Lorentz invariance are extremely constraining, determining which interactions are allowed, and fixing the basic structure of scattering processes. It would be ideal to reach a similar understanding in cosmology, but we are still pretty far from achieving it. This happens for a good reason: those principles which are basic in flat-space become approximate, and thus it is no longer clear which are the fundamental rules governing the physics at cosmological scales.

There are two complementary approaches -- see \cite{Arkani-Hamed:2018kmz} and  \cite{Arkani-Hamed:2017fdk, Arkani-Hamed:2018ahb, Benincasa:2018ssx} -- which recently have been undertaken to make progress in this direction, both of which are inspired by the most recent developments in the context of scattering amplitudes.  The present paper is a generalisation of the second one, and has started the detailed analysis of the wavefunction of the universe for more general scalars, which are described by a scalar in flat-space with time-dependent mass as well as time dependent couplings. Treating the time-dependent mass in its Fourier space, the wavefunction integrands which get defined in this way satisfy novel recursion relations, connecting states with different masses and involving certain differential operators. For certain masses, these recursion relations have the flat-space massless case -- {\it i.e.} the conformally coupled scalar in cosmology -- as a seed and thus the full structure for the wavefunction with these internal states is just inherited from the seed via these differential operators. This means that the very same combinatorial rules holding for the conformally coupled scalar can be translated to such a case via differential operators. However, it has an additional implication. From \cite{Benincasa:2018ssx} we learnt that the residues of all the poles of (the integrand of the) wavefunction of the universe for a massless scalar with time-dependent coupling constants can be interpreted as scattering processes or can be expressed in terms of scattering processes. Consequently, via the differential operators in the recursion relations for these more general states, also all the coefficients in a Laurent expansion around any of the singular points of the wavefunction integrand can be expressed in terms of scattering amplitudes. Indeed, what was observed so far was that the leading coefficient of the Laurent expansion around the total energy pole was (proportional to) the flat-space scattering amplitudes. For more generic, but still light states, we can perform a perturbative treatment for the mass corrections: we obtain a diagrammatics which is a straightforward generalisation of the one we have been using for the conformally-coupled case and it is related to it via a limit which imposes the energy conservation between two edge joined by a mass insertion. Now, if we were interested in computing the wavefunction with a given internal massive state, treating the mass perturbatively one realises that there is a very specific class of graphs which contributes: {\it e.g.} given a two-site graph, the perturbative mass corrections to it are all line graphs with internal two-point vertices, representing the mass insertions. While it is indeed not trivial to re-sum them because of the time dependence of the mass (one does not end up having a trivial geometric series), the fact that there is a very specific class of graphs involved leaves some hope for the possibility of re-summing it. We have not faced this issue explicitly, leaving it for future work.

One of the aims of the approach \cite{Arkani-Hamed:2017fdk, Arkani-Hamed:2018ahb, Benincasa:2018ssx} is to find an underlying first-principle mathematical structure which the wavefunction of the universe arises from. The cosmological polytopes, which encodes the wavefunction of the universe for the conformally coupled scalar in FRW cosmologies, are characterised -- as any other positive geometries -- by a canonical differential form having logarithmic singularities only in correspondence of the boundaries of the polytope. Its coefficient returns the wavefunction of the universe. Now, when we treat massive states, it is no longer true that even the wavefunction integrand has simple poles ({\it i.e.} logarithmic singularities) only. This would in principle suggest that the wavefunctions for these states should not be describable in terms of positive geometries, at least according to our current understanding. In this paper, we show that it is not the case for some specific values of the mass: starting from the very same building blocks as the cosmological polytope, {\it i.e.} the space of triangles which can be intersected in the midpoints of two out of its three sides to form an actual cosmological polytope, we can define a generalisation of this construction requiring to intersect the triangles in one of the intersectable midpoints and {\it the vertex opposite to it}. Or equivalently, we can enlarge the set of building blocks considering both triangles and segments, and intersecting them in their midpoints (holding the distinction between intersectable and non-intersectable sides for the triangles). Taking this last point of view, the result of such a prescription is a polytope whose canonical form $\Omega$ is nothing but the Newton's difference quotient of the canonical form $\hat{\Omega}$ that one would obtain by intersecting the very same number of triangles. Thus, a degenerate limit of $\Omega$ returns the derivative of $\hat{\Omega}$ with respect to the energies associated to the midpoints where the triangles have been intersected with the segments. All the polytopes constructed considering two segments for each triangle, intersected one for each intersectable side, are characterised by a canonical form that, in the degenerate limit, returns the wavefunction integrand for the $l\,=\,1$ states. These polytopes are still in a $1-1$ correspondence with graphs, with the triangles still associated to two-site graphs, while the segments to tadpoles ({\it i.e.} one-site one-loop graphs).

The faces of these polytopes beautifully encode the information of the wavefunction, {\it before} the degenerate limit. In particular, the scattering amplitude emerges from a higher codimension face: this is just the statement that the scattering amplitude is associated with the leading term in the Laurent expansion around the total energy pole. Beautifully, the codimension of this face {\it is} the order of such a pole. This is  more generally true for other faces.

One of the important points of \cite{Benincasa:2018ssx} was that the tree wavefunction can be reconstructed from the flat space scattering amplitude and requiring the absence of some unphysical singularities, with the loop ones which can be obtained via a particular projection. This is also true in the case discussed in the present paper, and it is manifest in the polytope/graph picture. We can consider a cosmological polytope constructed in the standard way, which a tree graph is associated to. We can then project it through cones with origin in ${\bf x}_i-{\bf x'}_i$, ${\bf x}_i$ and ${\bf x'}_i$ being associated with the sites of the $i$-th most external two-site subgraph: such a projection produces a polytope whose associated graph has an external tadpole for each $i$. Thus, the polytopes introduced in this paper can be reconstructed from the knowledge of the flat-space scattering amplitudes via the class of projections just discussed.

As mentioned at the beginning of this section, we have been just scratching the surface and there are a large number of questions. Let us mention two of them, which are more immediately inherent to the discussion presented in this paper. Our analysis holds for light states only, {\it i.e.} in dS${}_{d+1}$ for masses $m\,\in\,[0,\,d/2]$ (the complementary series in $d\,=\,3$), and in cosmologies $a(\eta)\,=\,\eta^{-\alpha}$ for $m\,=\,0$. So, together with investigating the possible resummation of the graphs contributing when the mass is treated perturbatively, it is indeed interesting to explore the structure for heavier states. Notice that the recursion relation we proved, it is valid in general. However, for heavier states, it does not have a clear seed and, importantly, it seems that it introduces states which are out of the Hilbert space.

Secondly, we are now in the position of treating states with spin different than zero. The natural first candidate is looking at the wavefunction for spin-$1$ states. However, what we have been accustomed to do so far, is to perform a graph by graph analysis. If on one side it has been illuminating in the cases of scalars, when we move to spin-$1$ states considering the sum of graphs becomes compulsory because we need to have a gauge invariant observable -- while a single graph is always gauge-dependent. This goes in parallel with the issue of finding a picture for the sum of graphs even in the scalar case: we expect that this picture is bounded to exist because we already know that the relevant underlying combinatorial structure for scattering amplitude of scalar interactions \cite{Arkani-Hamed:2017mur, Frost:2018djd, Salvatori:2018aha, Banerjee:2018tun, Raman:2019utu} which are all included in the cosmological polytope description.


\section*{Acknowledgements}

It is a pleasure to thank Humberto Gomez, Enrico Pajer and Cristian Vergu for insightful discussions. I am especially in debt with Enrico Pajer for comments on the paper. I would also like to thank the developers of SageMath \cite{sagemath}, Maxima \cite{maxima} and Tikz \cite{tantau:2013a}. I am supported in part by a grant from the Villum Fonden, an ERC-StG grant (N. 757978) and the Danish National Research Foundation (DNRF91).

\appendix

\bibliographystyle{utphys}
\bibliography{cprefs}	

\end{document}